\documentclass[acmlarge]{acmart}

\makeatletter
\newcommand{\myconfshort}{\acmConference@shortname}
\newcommand{\myconffull}{\acmConference@name}
\newcommand{\myconfdate}{\acmConference@date}
\newcommand{\myconfloc}{\acmConference@venue}
\AtBeginDocument{
  \fancypagestyle{firstpagestyle}{
    \fancyhead{}%
    \fancyfoot[C]{}%
  }
  \fancyhf{}
  \fancyhead[LO]{\@headfootfont\shorttitle}%
  \fancyhead[RE]{\@headfootfont\@shortauthors}%
  \fancyhead[LE]{\@headfootfont\footnotesize \myconfshort, \myconfdate, \myconfloc}%
  \fancyhead[RO]{\@headfootfont\footnotesize \myconfshort, \myconfdate, \myconfloc}%
  \fancyfoot[C]{}%
}
\makeatother
\acmBooktitle{\conffull\@ (\confshort), \confdate, \confloc}
\AtBeginDocument{%
  }

\copyrightyear{2026}
\acmYear{2026}
\setcopyright{cc}
\setcctype{by}
\acmConference[FAccT '26]{The 2026 ACM Conference on Fairness, Accountability, and Transparency}{June 25--28, 2026}{Montreal, QC, Canada}
\acmBooktitle{The 2026 ACM Conference on Fairness, Accountability, and Transparency (FAccT '26), June 25--28, 2026, Montreal, QC, Canada}
\acmDOI{10.1145/3805689.3812222}
\acmISBN{979-8-4007-2596-8/2026/06}
\usepackage{tabularx}
\usepackage{booktabs}
\usepackage{color-edits}

\usepackage{titletoc}

\usepackage{multirow}   % optional, for grouped rows
\usepackage{siunitx}    % optional, for number alignment

\usepackage[most]{tcolorbox} % for colored boxes
\usepackage{colortbl}  % provides \arrayrulecolor

\newtcolorbox{systemprompt}{
  colback=blue!5!white,    % light blue background
  colframe=blue!60!black,  % darker blue border
  coltitle=black,
  fonttitle=\bfseries,
  boxrule=0.8pt,           % border thickness
  arc=4pt,                 % rounded corners
  left=6pt, right=6pt,     % padding
  top=6pt, bottom=6pt,
  breakable                % allow page breaks if prompt is long
}

\newtcolorbox{rubricprompt}{
  colback=blue!5!white,    % light blue background
  colframe=blue!60!black,  % darker blue border
  coltitle=black,
  fonttitle=\bfseries,
  boxrule=0.8pt,           % border thickness
  arc=4pt,                 % rounded corners
  left=6pt, right=6pt,     % padding
  top=6pt, bottom=6pt,
  breakable                % allow page breaks if prompt is long
}

\newcolumntype{g}{!{\color{gray!35}\vrule width 0.4pt}}
\newcolumntype{C}[1]{>{\arraybackslash}p{#1}}
\newcommand{\lightrule}{\arrayrulecolor{gray!35}\midrule\arrayrulecolor{black}}

\newcommand{\xhdr}[1]{\vspace{1.7mm}\noindent{{\bf #1.}}}

\newcommand{\specialcell}[2][l]{%
  \begin{tabular}[#1]{@{}p{5cm}@{}}#2\end{tabular}}
\newcommand{\wrapcell}[2][t]{%
  \begin{tabular}[#1]{@{}p{\linewidth}@{}}#2\end{tabular}}

\newcommand{\ie}{i.e.,}
\newcommand{\eg}{e.g.,}
\newcommand{\dalle}{DALL·E}

\title[Evaluating AI-Generated Images of Cultural Artifacts]{Evaluating AI-Generated Images of Cultural Artifacts with Community-Informed Rubrics}

\author{Nari Johnson}
\email{narij@andrew.cmu.edu}
\affiliation{%
  \institution{Microsoft Research}
  \city{Cambridge}
  \country{UK}
}
\affiliation{%
  \institution{Carnegie Mellon University}
  \city{Pittsburgh}
  \state{PA}
  \country{USA}
}

\author{Deepthi Sudharsan}
\authornote{Second authors, equal contribution.}
\authornote{Work completed while at Microsoft Research.}
\email{deepthi.sudharsan@gmail.com}
\affiliation{%
  \institution{RiskSpan}
  \city{Bengaluru}
  \country{India}
}

\author{Hamna}
\authornotemark[1]
\email{hamnaabid@gmail.com}
\affiliation{%
  \institution{Microsoft Research}
  \city{Bengaluru}
  \country{India}
}

\author{Samantha Dalal}
\email{Sd3765@princeton.edu}
\affiliation{%
  \institution{Microsoft Research}
  \city{Cambridge}
  \country{UK}
}
\affiliation{%
  \institution{Princeton University}
  \city{Princeton}
  \state{NJ}
  \country{USA}
}

\author{Theo Holroyd}
\email{theo@theoholroyd.com}
\affiliation{%
  \institution{The Stephen Perse Foundation}
  \city{Cambridge}
  \country{UK}
}

\author{Anja Thieme}
\email{anthie@microsoft.com}
\affiliation{%
  \institution{Microsoft Research}
  \city{Cambridge}
  \country{UK}
}

\author{Hoda Heidari}
\email{hheidari@cmu.edu}
\affiliation{%
  \institution{Carnegie Mellon University}
  \city{Pittsburgh}
  \state{PA}
  \country{USA}
}

\author{Daniela Massiceti}
\email{daniela.massiceti@gmail.com}
\affiliation{%
  \institution{Microsoft Research}
  \city{Sydney}
  \country{Australia}
}

\author{Jennifer Wortman Vaughan}
\email{jenn@microsoft.com}
\affiliation{%
  \institution{Microsoft Research}
  \city{New York}
  \state{NY}
  \country{USA}
}

\author{Cecily Morrison}
\email{cecilym@microsoft.com}
\affiliation{%
  \institution{Microsoft Research}
  \city{Cambridge}
  \country{UK}
}

\renewcommand{\shortauthors}{Johnson et al.}

\begin{abstract} 
  Measurement is essential to improving AI performance and mitigating
  harms for marginalized groups. As generative AI systems are rapidly
  deployed across geographies and contexts, AI measurement practices
  must be designed to support repeatable, automatable application
  across different models, datasets, and evaluation settings. But the
  drive to automate measurement can be in tension with the ability for
  measurement instruments to capture the expertise and perspectives of
  communities impacted by AI. Recent work advocates for breaking
  measurement into several key stages: first moving from an abstract
  concept to be measured into a precise systematized concept;
  next operationalizing the systematized concept into a concrete
  measurement instrument; and finally applying the measurement
  instrument on data to produce measurements. This opens up an
  opportunity to incorporate community engagement in the
  systematization phase before operationalizing and applying
  measurement instruments. In this paper, we explore how to involve
  communities in systematizing the concept of ``cultural
  appropriateness'' in text-to-image models’ representation of
  culturally significant artifacts through case studies with three
  communities: blind and low vision individuals residing in the UK,
  residents of Kerala, and residents of Tamil Nadu. Our systematized
  concepts reflect community members' lived experiences interacting
  with each artifact and how they want their material culture to be
  depicted. We explore how these systematized concepts can be
  operationalized into automated measurement instruments that could be
  applied using a multimodal LLM-as-a-judge approach. We reflect on
  the benefits and limitations of such approaches. \looseness=-1
\end{abstract} 



\begin{document}

\maketitle

\section{Introduction} 

People across the world have adopted generative AI tools to automate the creation of images for design and ideation, marketing and advertising, illustration, and beyond ~\cite{jiang2023art,dehouche2023whats,Gillespie2024,exner2025ai}. But not all cultures are rendered appropriately. HCI and AI researchers have documented the many ways in which state-of-the-art generative AI systems systematically underperform at depicting historically marginalized cultures, including replicating historical biases in media \cite{Gillespie2024,mack2023towards}, reinforcing stereotypes \cite{bianchi2023easily,jha2024visage,hall2024towards,gautam2024melting}, and contributing to cultural erasure \cite{ghosh2024do,johnson2025position,magomere2025world,qadri2025case}.   \looseness=-1

Effective model evaluation is necessary to make these failures visible and measure progress toward addressing them \cite{weidinger2023sociotechnical,wallach2025position,harvey2025understanding}.  Today, many existing approaches to generative AI evaluation apply off-the-shelf benchmarks and metrics \cite{eriksson2025trust,kapania2025examining,harvey2025understanding}.  However, an emerging line of research suggests that these existing approaches to evaluation suffer from significant validity issues~\citep{wallach2025position,coston2023validity}, and in particular, break down when applied in marginalized, low-data contexts \cite{kreiss2022context,johnson2025position,hada2024large}.  More generally, these approaches do not attempt to incorporate the expertise or input of the people whose cultures are depicted in the thousands of AI-generated images created each day~\cite{mack2023towards,qadri2023ai,qadri2025case}. \looseness=-1

\begin{figure}[t] 
\centering 
\includegraphics[width=0.9\linewidth]{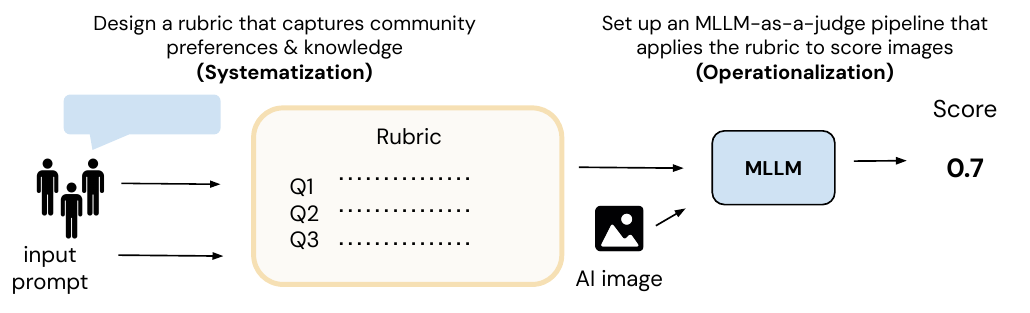} 
\caption{\textbf{Scaffolding community engagement to develop community-centered measures of cultural representation}.  Given an input prompt (\eg{} ``\emph{a photo of a guide cane}''), we invited community members to participate in designing a rubric that captures their expertise and preferences for each cultural artifact (systematization). Our research team then explored the use of this rubric within an automated  multimodal LLM-as-a-judge pipeline (operationalization).} 
\Description[Visualization of the measurement design process]{The figure shows an illustration of two key steps in the measurement design process: systematization and operationalization. The systematization process involves eliciting community members' feedback to construct an evaluation rubric for a given input prompt. The operationalization process then applies the created rubric to a new image to produce a numeric score.}
\label{fig:teaser} 
\end{figure}

\color{black}An emerging body of human-centered literature has explored ways to involve humans that hold relevant ``lived-experience expertise'' ~\citep{matias2025how} (\eg{} due to their knowledge, standpoint, or cultural identity) in the process of evaluating generative AI. 
This ranges from efforts that crowdsource culturally specific datasets \cite{orife2020masakhane,kunchukuttan2020bharat,seth2024dosa,magomere2025world,romero2024cvqa} to those that invite participants to evaluate model outputs~\cite{mack2023towards,aroyo2023dices,hall2024towards,qadri2025case}. At the same time, there is a rapidly growing literature on ways to develop practical, repeatable, and automatable evaluation practices by using MLLM-as-a-judge evaluations, which delegate evaluative judgments to multimodal language models (MLLMs) in lieu of human judges ~\cite{shankar2024who,szymanski2024limitations}. 
This paper aims to connect these two literatures by engaging people with lived-experience expertise in developing rubrics that can be used within the MLLM-as-a-judge paradigm (Figure \ref{fig:teaser}). \looseness=-1

Recent work by \citet{wallach2025position} calls for AI researchers to reimagine their approach to measurement by adopting an established measurement framework from the social sciences~\cite{adcock2001measurement}. Social scientists have long grappled with how abstract concepts, like political ideology or job satisfaction, can be captured in precise measurements~\cite{adcock2001measurement, cronbach1955construct}. Similar questions arise in the evaluation of generative AI systems, where one might wish to measure abstract concepts such as whether a language model can ``reason''~\cite{salaudeen2025measurement}, or whether an AI image depicts a "stereotype'' about a social group~\cite{bianchi2023easily}.  The framework breaks the process of measurement into three concrete steps: (1) \emph{systematizing} an abstract concept to be measured into a concrete definition; (2) \emph{operationalizing} the systematized concept into a measurement instrument; and (3) \emph{applying} the measurement instrument on data to produce measurements. 
The authors point to the opportunity for participants with different forms of expertise to be included in the systematization process, but stop short of offering practical guidance on how to facilitate such engagements. In this work, we ask: how can we scaffold community engagement in the systematization process to develop community-informed measurement instruments? 

To answer this question,
we facilitate research workshops with members of three
communities: blind and low vision individuals residing in the UK and residents of two distinct South Indian states, Kerala and Tamil Nadu.
We note that the term ``community'' has been used in a range of different ways across different academic disciplines ~\citep{bogardus1942fundamentals,wenger1998communities}.
In this work, we follow past HCI studies ~\cite{bergman2024stela,hamna2025building,matias2025how} and use ``community'' to refer to a group of individuals who hold a shared cultural identity, recognizing the shared ``lived-experience expertise that individuals hold that extends beyond their formal training or credentials'' ~\citep{matias2025how}.
Our goal is to identify ways for non-technical people (people who are not AI developers) to contribute their lived-experience expertise in shaping AI evaluation design. 
 In each context, we develop community-informed systematized concepts of \emph{cultural appropriateness} in depicting cultural artifacts~\cite{newmark1988translation}. 
In this work, we view cultural appropriateness loosely as the degree to which AI images of artifacts align with and respect the values, norms, and knowledge systems of the relevant community \citep{qadri2025case}, but leave it to communities to determine what this means. 
In studying culturally appropriate depictions, our goal is to capture community members’ understandings of how each artifact should---and critically, should not---be represented in AI media~\cite{qadri2023ai,mack2023towards,qadri2025case}. 
We structure our inquiry around the following research questions: \looseness=-1

\begin{itemize}
    \item \textbf{RQ1 (Systematization):} How can practitioners work with community members to systematize the cultural appropriateness of important cultural artifacts? How do the rubrics elicited through a community-centered approach differ from those generated by LLMs?

    \item \textbf{RQ2 (Operationalization):} How might practitioners automate the application of community-informed rubrics to score new AI images?

\end{itemize}

We contribute an empirical account of how practitioners can engage marginalized communities in the systematization process. 
We highlight how community members’ expertise, lived and embodied experiences with each artifact, and subjective preferences for representation shape our systematized concepts, demonstrating the potential value of involving communities in measurement.
We explore whether rubric application can be automated using an MLLM judge, surface several limitations of MLLM judges, and contribute methodological guidance for how practitioners can assess the feasibility of automating measurement.
We conclude by reflecting on open challenges and limitations of our exploratory work, both of which pose opportunities for future research to bring community expertise into the design of AI evaluation metrics.

\section{Related Work} 

\xhdr{Cultural Representation and AI-Generated Media} 
AI-generated images have the power to shape how communities are understood and perceived by others, as well as how they perceive themselves~\cite{qadri2023ai,hall1997representation}. 
Media representation of diverse peoples, especially those who hold historically marginalized identities, is an essential building block for social change and equality~\cite{Gillespie2024,walters2003all}. 
Marginalized communities have used AI tools to create visual media that aligns with their goals for representation~\cite{das2024provenance,huh2023genassist,adnin2024look,he2025dreamstory,hamna2025kahani}. However, these same AI tools have been shown to reproduce normative identities and narratives that contribute to the erasure of already marginalized communities~\cite{bianchi2023easily,Gillespie2024,mack2024they}. This can result in potential representational harm through unfavorable or demeaning depictions that can invoke psychological distress or frustration for users who aim to create media about their own culture \citep{mack2023towards,wenzel2025invisible}.\noindent

In this paper, we focus on the issue of appropriate  cultural representation---an often contested concept~\cite{chasalow2021representativeness}---in AI-generated images, as defined by community members. We adopt a broad conceptualization of culture as based on shared identity, recognizing that individuals can inhabit multiple cultural identities simultaneously \citep{zhou2025culture}. Culture may be grounded in place (\eg{} shared nationality~\cite{adilazuarda2024measuring}), or in other identity aspects including race or ethnicity~\cite{egede2025exploring}, disability~\cite{mack2024they}, sexuality~\cite{taylor2024cruising}, profession~\cite{tseng2025ownership}, or relational roles~\cite{viswanathan2025interaction}.
Cultural representations---those intended to capture tangible objects (artifacts, monuments) and intangible practices (traditions, rituals) salient to communities' cultural identity~\cite{hall2025human,adilazuarda2024culturesurvey,blake2000defining}---are not objective constructs with a definitive ground truth. Instead, they involve an articulation of  how community members would like their community to be depicted~\cite{qadri2025case}.  
While community members may have varying perspectives on which aspects of their culture should be highlighted, their knowledge and viewpoint are critical to deciding what makes a representation “appropriate”~\cite{hall2024towards,qadri2025case}.  \looseness=-1

In this paper, we focus on designing quantitative measures that reflect participants' lived understandings of their culture.
Prior work in AI and culture has emphasized the need for methodologies that ``translate qualitative insights into algorithmic interventions'' \cite{biega2025towards}. 
Our work responds to this call by bringing qualitative methods into the design of quantitative metrics by engaging communities in the design of rubrics that can be programmatically applied with MLLM-as-a-judge systems.

\xhdr{Measuring the Performance of Generative Text-to-Image Models} 
Our work contributes to a growing body of interdisciplinary scholarship focused on developing and critiquing performance measures used to evaluate generative AI systems.  Drawing from the quantitative social sciences, we define measurement as the assignment of quantitative or qualitative values to specific concepts or properties of a system~\cite{salaudeen2025measurement}. In text-to-image generation, concepts that are often measured include generated images' faithfulness to the user's generation prompt ~\cite{hu2023tifa,saxon2024who}, photorealism ~\cite{heusel2018gans,jayasumana2024rethinking}, and aesthetic value ~\cite{lee2023heim}. These measurements are applied to tasks like helping practitioners select deployment-appropriate systems~\cite{weidinger2023sociotechnical,salaudeen2025measurement,raji2021wide} or filtering content below quality thresholds~\cite{weidinger2023sociotechnical, kirstain2023pick,hong2024s}. \looseness=-1 

Despite the recognized importance of evaluation in generative AI development, significant challenges remain. Given the speed at which the AI industry is moving, there is a need for evaluation practices to be repeatable, automatable, and applicable across different models, datasets, and settings \cite{shankar2024who,weidinger2023sociotechnical,harvey2025understanding}. 
The dominant approach to AI evaluation in industry involves calculating standardized metrics that capture model performance on test datasets---a practice known as ``benchmarking''~\cite{raji2021wide,orr2024sport,wallach2025position}.  However, AI benchmarks are often decontextualized and divorced from real-world deployment contexts, leading to validity issues~\cite{raji2021wide,eriksson2025trust,deviyani2025contextual}.  This decontextualization manifests in the systematic exclusion of content from minoritized communities, including non-English language content and culturally specific imagery ~\cite{mcintosh2025inadequacies,rottger2025safety}. Recognizing these limitations of benchmarks, researchers have called for more localized, disaggregated, contextual evaluations~\cite{barocas2021designing,deviyani2025contextual}. Many existing efforts expand the scope of generative AI evaluations by curating globally diverse datasets of input prompts~\cite{seth2024dosa,kirk2024prism,winata2024worldcuisines,hall2025human,magomere2025world}.

In this work, we shift the focus from datasets to \emph{measures}.  It is increasingly common to automate evaluations using model-mediated evaluation approaches, where an auxiliary, pretrained model’s internal representations or judgments are used to stand in for human evaluative judgments as a form of ``ground truth''~\cite{kapania2025examining}. 
Drawing on the affordances for scale of LLM-as-a-judge pipelines, practitioners have increasingly turned to multimodal language models (MLLMs) as judges for evaluating generated images~\cite{hu2023tifa,zhang2023gpt4v,lin2024evaluating,chen2025multimodal}. In these pipelines, a multimodal model scores each image using a rubric, which in some cases is generated by another model.  Such approaches, however, introduce a fundamental dependency: the quality of the measurements becomes limited by the capabilities and biases of the auxiliary models~\cite{magomere2025world,kreiss2022context,seth2024dosa, kapania2025examining,massiceti2024explaining}. Since these models are trained on historical datasets that may lack diversity or contain societal biases, they can systematically undervalue or misrepresent content from marginalized communities. For example, if an auxiliary model was trained primarily on Western imagery, it may poorly evaluate AI-generated images of non-Western cultural practices, foods, or aesthetics. We ask whether this limitation might be overcome by the design of more thoughtful rubrics.

\begin{figure}[t] 
    \centering 
    \includegraphics[width=0.85\textwidth]{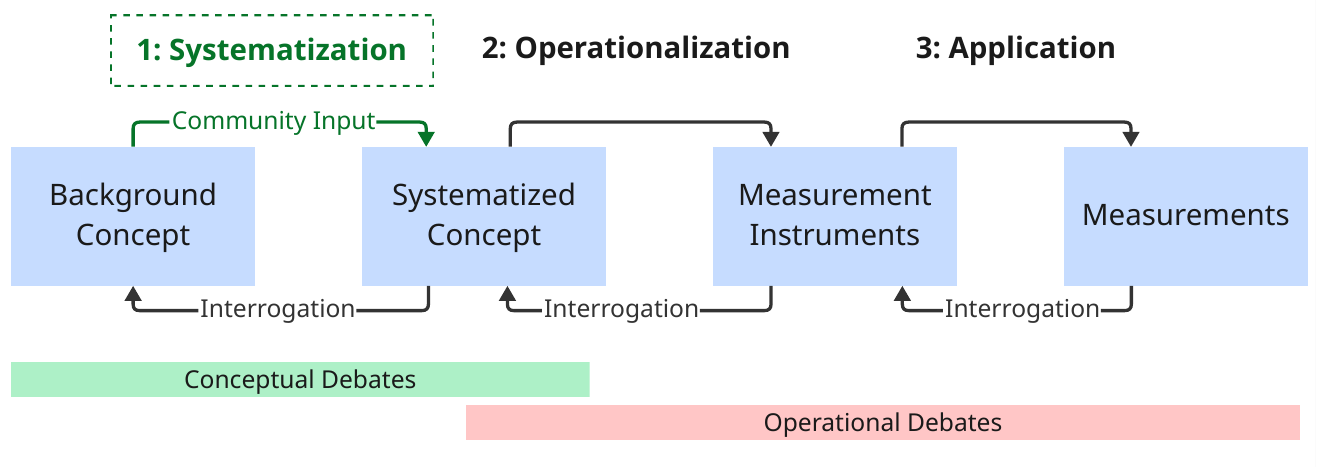} 
    \caption{\textbf{Measurement framework from the social sciences \cite{adcock2001measurement,wallach2025position}.} We study how to center community expertise in the systematization process before operationalizing the systematized concept as an automated MLLM-as-a-judge system.} 
    \label{fig:framework} 
    \Description[Visualization of Adcock and Collier's measurement framework]{A flowchart illustration of Adcock and Collier's measurement framework. The first step, systematization, involves moving from a "background concept" to a "systematized concept" by incorporating community input. The second step, operationalization, moves from a "systematized concept" to a "measurement instrument". The third step, application, applies the "measurement instrument" to produce "measurements". "Interrogation" arrows pointing backwards from later steps of the measurement process, back towards earlier steps, to illustrate the iterative nature of measurement design.}
\end{figure} 
 
\xhdr{Applying Measurement Theory to AI Evaluations}
Recent work by ~\citet{wallach2025position} advocates for researchers to rethink how they evaluate generative AI systems, drawing on the tradition of measurement theory from the social sciences. They build on the framework of \citet{adcock2001measurement}, pictured in Figure~\ref{fig:framework}, which shows how abstract concepts can be systematically transformed into concrete, quantifiable measurements. First, during \emph{systematization}, a vague background concept
is transformed into a more concrete, well-defined \emph{systematized concept}. Next, during \emph{operationalization}, the systematized concept is developed into a \emph{measurement instrument} (e.g., using metrics or code). Finally, there is \emph{application}, in which the measurement instrument is applied to a set of instances (e.g., outputs of an AI model), resulting in the creation of measurements. This process is intended to be iterative, with iteration and refinement at each stage. As \citet{wallach2025position} note, separating systematization from operationalization has the benefit that \emph{conceptual debates} about what is being measured and why can be separated from \emph{operational debates} about \textit{how} a concept will be measured in practice. This allows people with different backgrounds and situated knowledge to be involved in conceptual debates without needing to understand the technical details of how the measurement instrument will be developed, standing in contrast with dominant evaluation practices, where participants are typically only involved at the end of the process, tasked with evaluating the quality of model outputs~\cite{otani2023verifiable,hu2023tifa,johnson2025position}.

While prior work has drawn on disciplinary expertise in systematizing generative AI performance measures---\eg{} by consulting psychologists or linguists~\cite{corvi-etal-2025-taxonomizing}---there is little work on how to engage community members in translating their lived experience and knowledge into systematized concepts for evaluation. One notable exception is concurrent work by \citet{nguyen2026validating}, who engage community members to iteratively validate and refine a systematized concept of erasure.  In this work, we explore how we can scaffold community engagement in the systematization process to develop community-informed systematized concepts that can be operationalized using MLLM-as-a-judge approaches and ultimately applied across different datasets and models.

\section{Methods}
\label{sec:all-methods}

\subsection{Selecting communities and cultural artifacts}
\label{sec:workshop_background}

We collaborated with members of three distinct communities: blind and low-vision (BLV) people located in the United Kingdom, and current and former residents of two states in South India: Tamil Nadu and Kerala.
We chose these specific communities because members of the research team belong to them. This close researcher–community collaboration supported activities such as artifact selection, participant recruitment, and interpretation of our findings. 
Studying cultural representation across communities varying in ability, geography, and ethnicity allows us to identify both shared principles and context-specific needs in how communities want to be represented.
Recognizing the rich diversity and breadth of experiences that exists within each of these communities, we follow past work ~\citep{bergman2024stela} to engage community members not as representatives of the entire population with which they share one facet of their cultural identity, but instead as individual experts in their experience as people who hold those identities.

For the BLV community, we focused our study on a single region (the UK).
We chose to geographically situate our study because the material culture of blindness is shaped by local factors, such as the public funding of assistive technology use in schools~\cite{afabbraille}.
This broad scoping of eligibility also enabled us to recruit from community organizations with membership bases that were similarly scoped to the UK, such as an email list of UK-based assistive technology users. \looseness=-1
We also conducted two separate engagements with residents of two Indian states: Tamil Nadu and Kerala.
Cultural studies scholars have discussed how India's $28$ region-states function not merely as administrative regions, but also each carry their own unique histories and cultural practices that play an important role in creating a shared cultural identity ~\citep{tambiah1967social,Indianculturalmosaic2009}.
For example, residents of Kerala and Tamil Nadu speak different languages (and write in different scripts), and participate in both cultural ceremonies (\eg{} festivals like Pongal) and everyday activities (\eg{} dance forms like Bharatanatyam) that are unique to their region.
We follow past work ~\citep{seth2024dosa} and center community engagements on cultural artifacts that are \emph{unique to each state}, and do not have a direct equivalent in neighboring geographic states.
Each engagement was facilitated by a co-author who was from the respective state.

\color{black}

\begin{figure}[t] 
\centering 
\includegraphics[width=\linewidth]{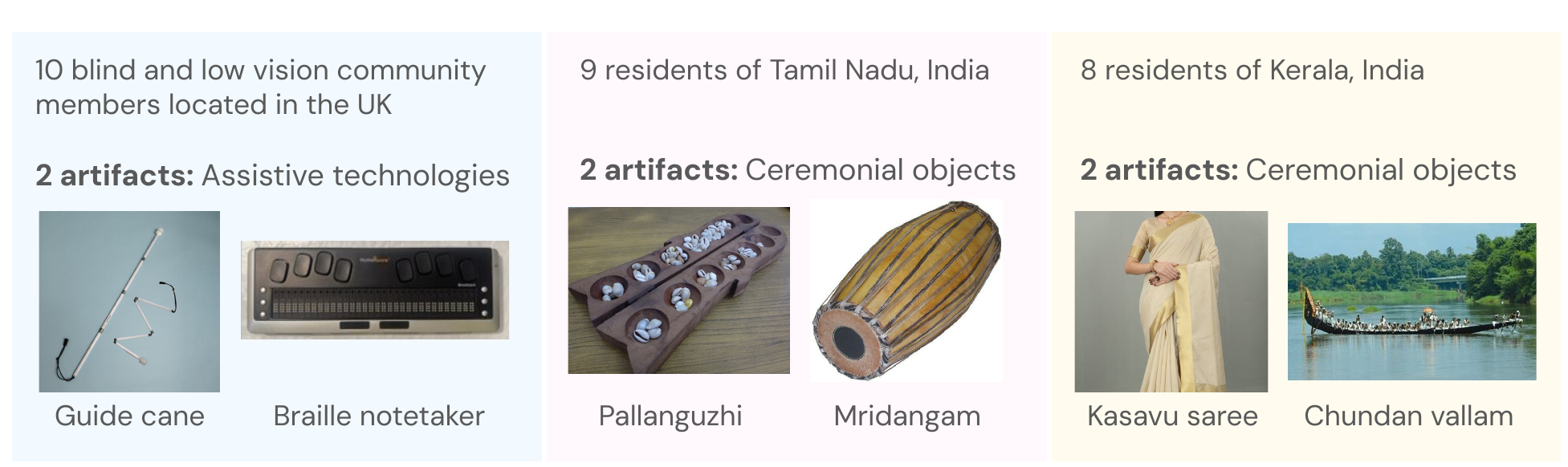} 
\caption{\textbf{Selected culturally significant artifacts.} From right to left: (1) With the blind and low vision community, we selected a \emph{guide cane} (a mobility aid that is held diagonally across one's body) and a \emph{braille notetaker} (an electronic device that can be used to read and write notes in tactile braille). (2) With residents of Tamil Nadu, we selected  \emph{Pallanguzhi} (a two-player
mancala game where players compete to collect cowry shells or seeds) and \emph{Mridangam} (a percussion instrument widely
used in South Asian classical music). (3) With residents of Kerala, we selected a \emph{Kasavu saree} (a handwoven saree
from Kerala, known for its off-white body with a gold border), and \emph{Chundan Vallam} (a traditional boat
from Kerala with a raised prow commonly used in festival races).} 
\Description[Reference photos of the six selected cultural artifacts.]{The figure shows a single reference photograph selected for each of the six cultural artifacts. We include a short description of each artifact in the figure's caption.}
\label{fig:community-summary} 
\end{figure}

Guided by the members of our research team who belong to each community, we selected salient cultural artifacts~\cite{newmark1988translation}. These artifacts are not intended to represent the richness and breadth of each community's material culture~\cite{zhou2025culture}, but to serve as a starting point for our exploration of rubric design.
For the BLV community, we selected two assistive technologies: a \emph{guide cane}, a mobility device that helps its user detect obstacles and navigate their surroundings, and a \emph{braille notetaker}, an electronic device used to write and read notes in tactile braille. Prior work has shown that people with disabilities value accurate depictions of assistive technologies in visual media~\cite{mack2023towards,zhang2022just} and AI-generated images~\cite{mack2024they}, but text-to-image models fail to accurately depict them~\cite{mack2023towards,mack2024they}. The particular technologies were chosen because of their wide use by and cultural significance to the BLV community in the UK, and are pictured in Figure~\ref{fig:community-summary}. \looseness=-1

For the two Indian communities, we selected objects regarded as meaningful to their state's cultural identity and likely to be known by residents. From Kerala, we selected (1) the \emph{Kasavu saree}, a traditional white and gold garment, and (2) the \emph{Chundan Vallam}, a snake boat used in traditional boat racing festivals. From Tamil Nadu, we selected (1) the \emph{Pallanguzhi}, a traditional two-player mancala board game, and (2) the \emph{Mridangam}, a percussion instrument that is widely used in South Indian classical music. 
Prior work has shown failures of text-to-image at depicting both scenes and objects from South Asian cultures~\cite{qadri2023ai,qadri2025case}.
These objects, pictured in Figure \ref{fig:community-summary}, carry historic and symbolic value for their roles in each state's unique cultural ceremonies, festivals, and traditions ~\cite{newmark1988translation,yao2024benchmarking}. We organize workshops by state as residents of neighboring states are less likely to be familiar with these objects. For instance, the Chundan Vallam is unique to the Keralan water festival tradition, and would not be widely known by residents of neighboring Tamil Nadu.

\subsection{Eliciting culturally appropriate artifact depictions in workshops}
To create the rubrics, we conducted a series of synchronous, hour-long online workshops with community members to elicit their knowledge of and desires for representation of the selected cultural artifacts. Workshops were facilitated by members of our research team, and took place between December 2024 and April 2025. To recruit participants, we used a purposive sampling approach~\cite{palinkas2012purposeful}, drawing on the research team’s existing networks across each community. Our final sample included 10 BLV participants in the UK, 9 participants from Tamil Nadu, and 8 participants from Kerala. Participants were compensated either £75 (UK) or 500 INR (India) based on their location, and all interviews were conducted in English. All workshop studies were approved by our institution's ethics review board. We detail the structure of the workshops, including activity guides, participant demographics, and steps we took to ensure the workshops were accessible for members of the BLV community in Appendix~\ref{apdx:blv-accessibility}. \looseness=-1

For each artifact, we curated a small set of AI-generated images and real photographs to facilitate discussions of cultural representation. We generated images using two state-of-the-art text-to-image models at the time of the study, Stable Diffusion 3 \cite{sd3} and DALL·E 3 \cite{dalle3}, prompting each model with a simple template (\eg{} “A photo of a \{artifact name\}”). For each artifact, we generated an initial pool of images and then grouped images by shared visual characteristics, sampling across groups to create a manageable but meaningfully diverse set for participants to review~\cite{mack2023towards}. In addition, we curated real photographs of each artifact from trusted sources (\eg{} the Royal National Institute of Blind People \cite{rnib-about}). These photographs helped ground discussion by providing clear examples of culturally appropriate depictions. For instance, we could more easily locate a photograph of a braille notetaker that depicted a QWERTY input keyboard correctly than generate one. We could also curate photographs that sat near the boundary of what participants might consider acceptable to facilitate discussion. The final set of images used in the workshops is included in Appendix \ref{apdx:complete-protocols}.  \looseness=-1

For each artifact, we conducted two study activities. First, 
after presenting each image (by screen-sharing and/or presenting an alt-text description for BLV participants), the facilitator asked participants to share if they felt that the image could or could not be shown to the general public
to represent the artifact and why. The facilitator then prompted each participant to elaborate on what made a particular image a good, bad, offensive, or incorrect depiction of the artifact. We discussed $10$ total images of each assistive technology, and $16$ total images of each Indian artifact.

Second, we invited participants to reflect on what they had seen so far to create a list of the most important criteria for a culturally appropriate portrayal of the object and provide reasons for each of their responses. This activity encouraged participants to articulate concrete visual criteria that shaped their decisions. The study facilitator encouraged participants to reflect on the degree of acceptable variability for objects and how visual features should be prioritized. 

\subsection{Systematizing cultural representation into rubrics}
\label{sec:syst}

We used the data collected from workshops to develop systematized concepts of ``cultural appropriateness'' for each artifact. These took the form of simple rubrics made up of binary yes/no criteria that community members identified as integral to an appropriate representation of the artifact; see Figure \ref{fig:example-rubric-cane} for an example of the rubric produced for a guide cane. We adopted this structure because rubrics are a common measurement approach (\eg{} in domains beyond AI, such as the social sciences \cite{adcock2001measurement}), 
and they are compatible with existing MLLM-as-a-judge pipelines \cite{chen2025multimodal}. \looseness=-1

To distill criteria for the rubrics, we first organized participants’ statements from both activities into two categories: (1) \emph{criteria}, or the concrete visual features or physical characteristics that shaped participants’ decision-making, and (2) \emph{justifications} that participants provided for their decisions.
In deciding which criteria to include in the rubrics, we made two key decisions. First, we excluded criteria that were highly contested across participants. 
For instance, community members disagreed about whether a guide cane with red reflective tape on its body was an appropriate representation. As such, we did not include a criterion about red tape in our rubric. 
We also excluded criteria that participants consistently described as less essential to depicting the essence of each artifact.  Consequently, rather than offering a complete description of each artifact, our rubrics focus on the 5 to 10 core features that participants agreed were most essential to a culturally appropriate depiction.
We acknowledge that this choice is value-laden, and discuss alternative ways that practitioners might navigate and resolve contested criteria in Section \ref{sec:discussion-limitations}. 
Importantly, the resulting rubric criteria for each cultural artifact should be interpreted as being \emph{particular to} (rather than portable across) the specific cultural contexts that they are created in collaboration with.\footnote{For instance, a guide cane rubric created based on engagements facilitated in a different part of the world (\eg{} with residents of Tanzania) may result in a different set of scoring criteria that reflect a different set of participant experiences and desires for cultural representation \citep{hall2024towards}.} \looseness=-1

Finally, we grouped the criteria into higher-level themes around the motivation for their inclusion. These themes reflect community-specific understandings of what constitutes (in)appropriate cultural representation. For example, BLV community members emphasized that assistive technologies should be usable and accessible to blind users, clarifying why particular visual features were prioritized. We structured each rubric by these themes. \looseness=-1

Using the rubric, we say that an image is ``culturally appropriate,'' receiving a label of 1, only if it satisfies all criteria.
Otherwise, we say that an image is culturally inappropriate (label 0).
While these final binary cultural appropriateness labels do not capture the subjective and contested nature of cultural representation, we adopt this aggregation function for its simplicity to calculate, and its direct comparability with participants' judgments.

As a point of comparison, we also generated rubrics automatically, in line with common AI industry practices \cite{szymanski2024comparing,shankar2024who,chen2025multimodal}.
Specifically, we prompted GPT-4o to produce evaluation criteria describing the ``most important visual characteristics that should be present or absent'' in culturally appropriate depictions of each artifact (details in Appendix \ref{apdx:rubric-comparisons}).
To compare our community-informed rubrics with the automatically generated rubrics,
the first author worked with community members on the research team to annotate each rubric, with the aim of identifying overlaps, divergences, and omissions relative to both our community-informed rubrics and our own knowledge of the artifacts. 
We then assessed each criterion for its alignment with community perspectives, and identified broader patterns in what LLM-generated rubrics captured or overlooked.  \looseness=-1

\subsection{Operationalizing the rubrics through MLLM-as-a-judge}\label{sec:methods-operationalization}

To operationalize our systematized concepts, we explore whether our community-elicited rubrics can be automatically applied using an MLLM-as-a-judge pipeline. 
We adopt a simple system prompt, following past work \cite{zheng2023judging,chen2025multimodal}, that instructs the MLLM to ``\emph{determine whether the provided image meets each criterion.}''
We adopt GPT 4-o~\cite{openai2024gpt4o} as our judge model for its demonstrated performance on MLLM-as-a-judge tasks \cite{zhang2023gpt4v}
and report all results averaged over five random seeds. We provide additional details in Appendix~\ref{apdx:gpt4o-judge}.

To validate the operationalization---that is, whether the MLLM-as-a-judge measurement instrument captures the systematized concept---we compare judgments of the rubric criteria from an MLLM judge and human annotators on a new dataset of AI-generated images.  We create a dataset containing 50 images of each artifact generated from 5 different models (\dalle{} 3, Stable Diffusion 3 Medium, Stable Diffusion 3.5 Large, GPT Image-1, and Flux.1 Dev) using simple prompts (like ``\emph{a photo of a guide cane}'').  We include image generation details in Appendix \ref{apdx:new-daataset-of-images}. These images depict the types of representations and errors that state-of-the-art models produce today. For each image, a member of our research team manually annotated whether each rubric criterion is met, which we compared against annotations assigned by the MLLM judge. Each rubric was applied by the research team member who facilitated workshop engagements and led the creation of the rubric, to ensure its consistent application across the dataset. We include an extended discussion of manual (human) application of our rubrics and inter-annotator agreement in Appendix \ref{apdx:iaa}. 
\section{Results}
In this section, we present our community-informed rubrics (Section \ref{sec:results-rubrics}) and an exploration of the feasibility of applying these rubrics automatically using an MLLM-as-a-judge approach (Section \ref{sec:results-operationalization}).
We reflect both on key trends that are common across the three communities we engaged with, and on key differences across communities and artifacts.\looseness=-1

\subsection{Systematizing community expertise into evaluation rubrics (RQ1)}\label{sec:results-rubrics}

\begin{figure}[t] 
\centering 
\includegraphics[width=\linewidth]{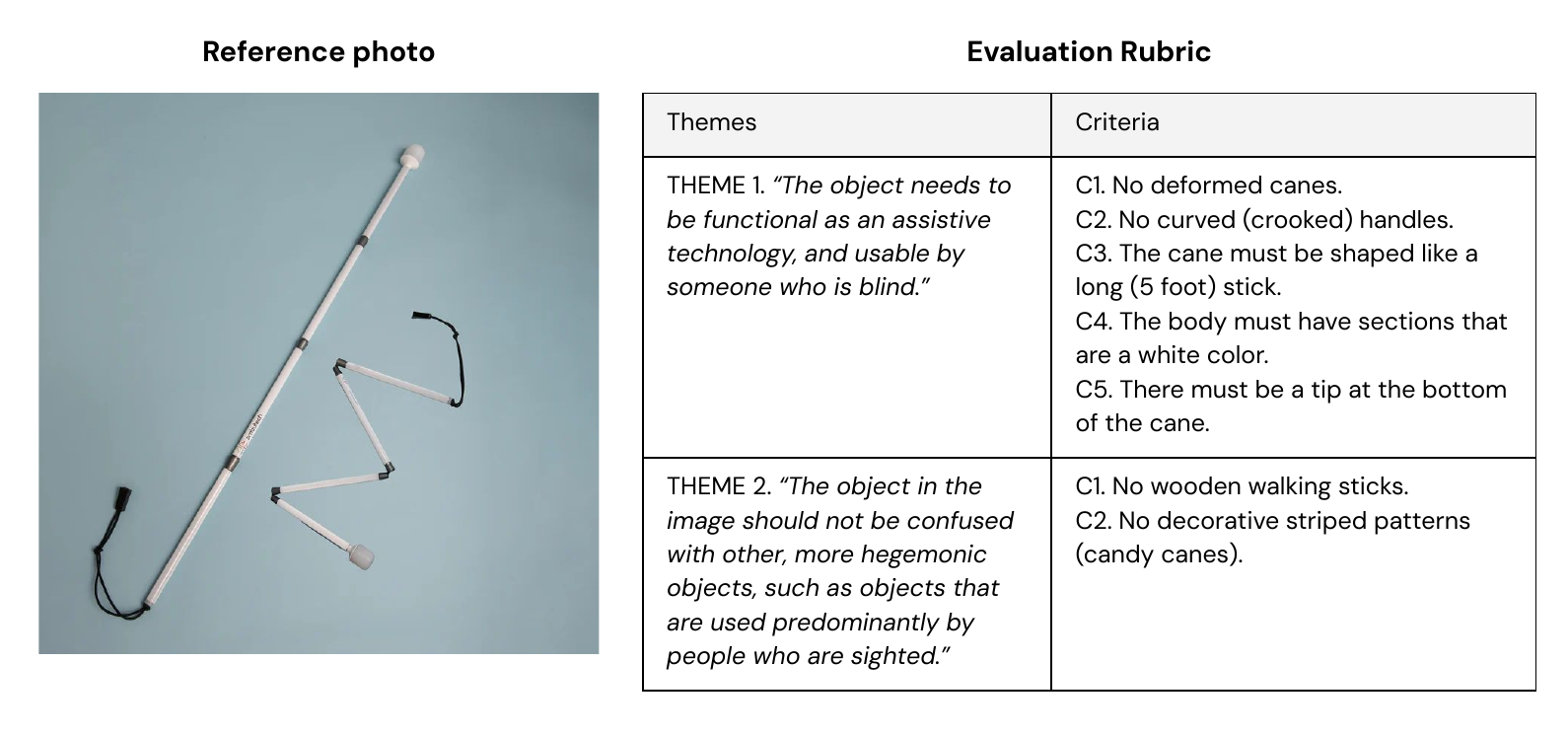} 
\caption{\textbf{A rubric to score images of a guide cane, designed with BLV community members.} Criteria that correspond to visual features in images are organized under two themes that describe participants' desires for cultural representation.} 
\Description[Reference photo and evaluation rubric for a guide cane.]{The left of the figure shows a reference photograph of a guide cane. The photo shows two guide canes: one cane is extended and ready for usage, and the other cane is folded along the joints on its body for convenient storage. Both canes have a long white body, a round white tip, and a straight grip with an attached wrist strap for convenient use. The right of the figure shows the community-informed evaluation rubric for a guide cane, which organizes evaluation criteria under two high-level themes.}
\label{fig:example-rubric-cane} 
\end{figure} 

In this section, we present the rubrics used to systematize the concept of ``culturally appropriate'' representation for each cultural artifact. The rubric for a guide cane is shown in Figure \ref{fig:example-rubric-cane}; the remainder can be found in Appendix \ref{apdx:systematization-results}.
For each community, we first identified high-level themes that capture key dimensions of how participants evaluated cultural representation.
We used these themes to organize a set of criteria that correspond to concrete, observable visual features. 
\looseness=-1

\subsubsection{Themes and criteria to assess cultural representation}
Across all three communities, participants consistently evaluated images along two core dimensions: \emph{functionality}, whether an artifact could plausibly serve its intended purpose, and \emph{recognizability}, whether the depiction preserves the key features that distinguish the artifact from related, but culturally distinct objects.  Within the BLV context, participants drew on their embodied experiences as users of each assistive technology to assess whether a depiction would be able to serve its intended function, \eg{} by imagining how one might hold a pictured cane or whether depictions of braille could be read by touch.  Participants repeatedly pointed out when generated images ``looked like'' other recognizable objects that did not serve the intended purpose of the assistive technology, such as a generated image of a guide cane that resembled a walking stick, a distinct mobility aid that differs in its function, material, and handle shape.  When asked to elaborate, participants explained that a walking stick is predominantly used by people who are sighted.  Community members felt that these ``confused'' depictions were not only misleading, but were disappointing in that they failed to communicate the unique material culture of blindness.

We found many key similarities in the higher-level themes that describe how residents of Kerala and Tamil Nadu evaluated their cultural artifacts.
Participants from both Kerala and Tamil Nadu emphasized evaluating each artifact's functionality by assessing whether the artifact pictured could be used for its intended purpose. For instance, participants from Tamil Nadu noted how the Mridangam drum’s leather straps running along its body are critical for tuning the instrument when played. Similarly, participants from Kerala described the unique shape of the Chundan Vallam racing boat (\eg{} its narrow, pointed stern) as central to its performance, and agreed that depictions should include the Amarakaran: the standing oarsmen who “orchestrates and steers the team’s focus during the race”. Similarly, participants from both Indian states repeatedly pointed out when generated images failed to capture what they felt were the most culturally distinct features of each artifact, such as the golden hand-woven border that distinguishes a Kasavu saree from sarees emblematic of other regions (``if it's not white with a gold border, then it's just a saree''), or confused depictions that resembled other board games (like checkers) instead of Pallanguzhi.

\looseness=-1

Taken together, these two shared dimensions of functionality and recognizability enable us to unpack what is shared across all three communities in their desires for cultural representation, while including rubric themes and criteria that remain context-specific and grounded within the unique perspectives of each community.

\subsubsection{Assessing the alignment of our rubrics with community preferences}\label{sec:results-llm-rubric-comparison}

We interrogate the validity of our rubrics by examining whether they align with preferences expressed by community members during workshops. Specifically, we compare (a) labels of cultural appropriateness obtained by having a member of our research team manually evaluate the rubric criteria to (b) participants' judgments of whether or not images were appropriate to be shown from the first workshop activity. We aggregate judgments across workshop participants by taking the majority judgment.\footnote{We note as one limitation of this comparison that these participant judgments were used to create the rubrics in the first place. Ideally we would interrogate validity using a new set of images and new participant judgments.} 
We find that for over 80\% of images, the label produced by manually applying the rubric matches the majority participant judgment (see Appendix Table \ref{table:validity-systematizations}).
Note that we may not expect perfect alignment because of the inherently plural and contested nature of cultural representation, which lacks a singular ground truth.
In our own workshops, participants' judgments of cultural appropriateness disagreed with each other for 33\% of the workshop images.

\subsubsection{Comparing our community-informed rubrics with LLM-generated rubrics}\label{sec:llm-rubric-comparison}

We observe several differences when examining how our rubrics differ from those generated by an off-the-shelf LLM, as described in Section~\ref{sec:syst}.
Many of the LLM-generated rubric criteria accurately capture surface-level properties of the artifacts, such as noting the ``distinct golden border'' of a Kasavu saree or describing a guide cane as ``a long, slender stick'' (Appendix \ref{apdx:rubric-comparisons}). However, several LLM-generated rubrics include criteria that are inaccurate, reflecting a fundamental misunderstanding of the objects. For example, the rubric for braille notetaker requires that the depiction ``resemble an electronic notebook'' and ``include a display for visual feedback,'' two features that are inaccessible to blind users and uncharacteristic of braille notetakers.
More generally, most LLM-generated rubrics omitted or underspecified features that community members viewed as essential for functionality or recognizability.
The Mridangam rubric, for example, omitted the drum's characteristic black circular membrane that is necessary to its tone. These omissions were not uniform across artifacts: some LLM-generated rubrics, such as that for the Kasavu saree, overlapped substantially with community rubrics, whereas others, most notably the braille notetaker, diverged sharply (see Appendix \ref{apdx:rubric-comparisons} for further discussion).
This suggests that community participation is essential for artifacts or concepts that are at present poorly captured by LLMs. \looseness=-1

\begin{table}[t]
\small
\centering
\caption{\textbf{Human vs. MLLM application of the rubrics.} Using 50 generated images per artifact, we report (i) the proportion of images labeled as culturally appropriate by humans versus an MLLM, and (ii) agreement between MLLM and human labels. We further disaggregate agreement by images the human labeled as appropriate versus inappropriate.}
\label{tab:operationalization-results-maintext}
\begin{tabular}{l c g c c c c}
\toprule
\textbf{Artifact} & \textbf{\shortstack{Human\\(\% Appropriate)}} & \textbf{\shortstack{MLLM\\(\% Appropriate)}} & \shortstack{\textbf{Agreement}\\Overall} & \shortstack{\textbf{Agreement}\\Appropriate images} & \shortstack{\textbf{Agreement}\\Inappropriate images} \\
\midrule
Guide cane & 0.40 & 0.44 & 0.84 & 0.84 & 0.83 \\ 
Braille notetaker & 0.08 & 0.20  & 0.82 & 0.65 & 0.83 \\
Pallanguzhi & 0.18 & 0.12 & 0.78 & 0.22 & 0.90 \\
Mridangam & 0.10 & 0.21  & 0.84 & 0.76 & 0.85 \\
Kasavu saree & 0.12 & 0.21 & 0.88 & 0.87 & 0.88 \\
Chundan Vallam & 0.00 & 0.17 & 0.83 & N/A & 0.83 \\
\bottomrule
\end{tabular}
\end{table}

\subsection{Operationalizing MLLM-as-a-judge metrics (RQ2)}\label{sec:results-operationalization}

We demonstrate how our community-informed rubrics could be operationalized into automated measurement instruments by using an MLLM to assess the rubric criteria, as described in Section \ref{sec:methods-operationalization}. We also consider a manual implementation, in which criteria are annotated by a member of our research team.
We use both of these measurement instruments (automated and manual) to label 300 images generated by five state-of-the-art text-to-image models as culturally appropriate or not.  Table \ref{tab:operationalization-results-maintext} summarizes the results.
In what follows, we first analyze the human annotations of criteria to examine how state-of-the-art models depict cultural artifacts, showing how the rubrics enable interpretable comparisons that reveal model-specific failure modes. We then compare MLLM-generated and human annotations. We emphasize that our goal in this work is not to optimize this implementation of the MLLM-as-a-judge system, but rather explore its feasibility, as best practices for MLLM-as-a-judge approaches remain an active area of research \cite{shankar2024who,li2025generation,chehbouni2025valid}.

\subsubsection{What rubrics reveal about cultural representation}\label{sec:interpreting-rubric-results}
We begin by examining the human annotations of the rubric criteria for the 50 images per artifact.
Across all artifacts, we find that state-of-the-art text-to-image models rarely produce depictions that satisfy all rubric criteria, resulting in low rates of images being labeled as culturally appropriate (first column in Table \ref{tab:operationalization-results-maintext}). The proportion of culturally appropriate images varies substantially across artifacts---40\% of guide cane images meet all criteria, compared to roughly 10\% of braille notetaker and Mridangam images, and no images of the Chundan Vallam. This pronounced class imbalance for all but one artifact reflects the current limitations of frontier models in representing cultural artifacts. This indicates that our community-informed rubrics capture meaningful errors of representation, even for models that we did not show to participants during our workshops, such as GPT Image-1 and Flux (discussed further in Appendix \ref{apdx:operationalization-interpretations}). 

Inspecting the specific criteria that are commonly violated, we find that different models exhibit distinct failure modes (see Section \ref{apdx:operationalization-interpretations} in the appendix). For example, the GPT Image-1 images of the Mridangam are primarily deemed inappropriate because they (incorrectly) add atypical decorative patterns on the drum’s body. In contrast, the other models violate a much wider range of rubric criteria: \eg{} applying our rubrics reveals how all of the \dalle{} 3, Flux.1 DEV, and Stable Diffusion images fail to depict the Mridangam’s black circular membrane. 
This analysis highlights the potential value of our rubric criteria in providing interpretable insights about the types of errors of representation made by different models, and the ability to track and measure them over time.\looseness=-1

\begin{figure}[t] 
\centering 
\includegraphics[width=0.9\linewidth]{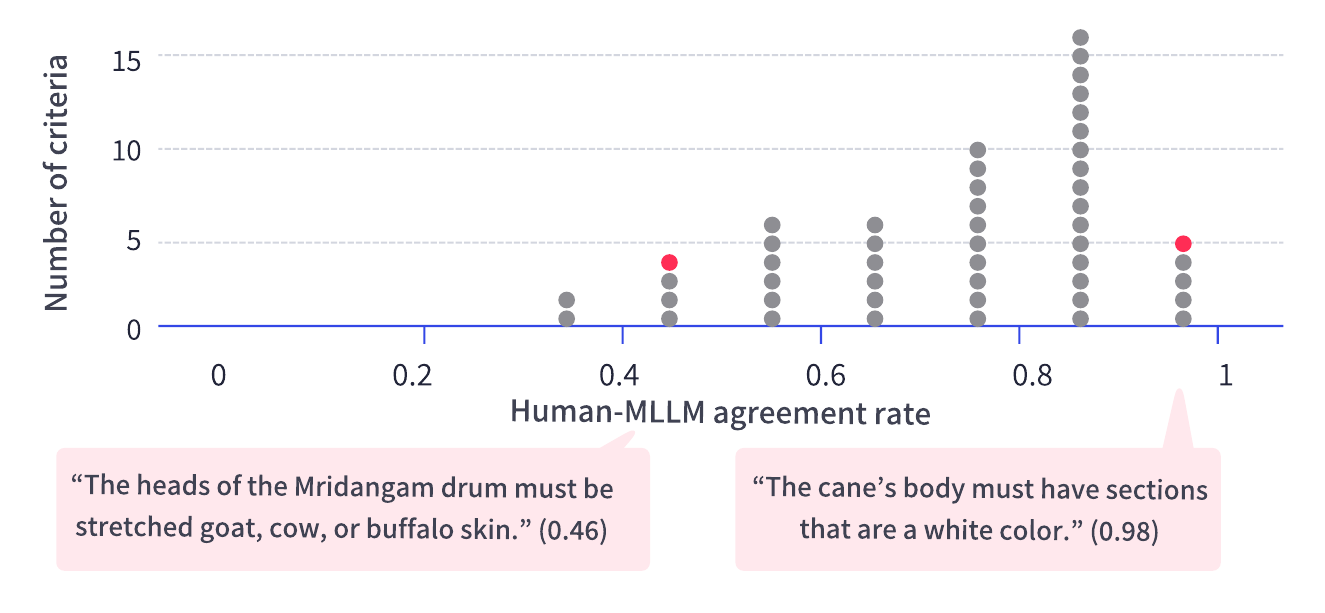} 
\caption{\textbf{Human-MLLM judge alignment for individual rubric criteria}. A histogram that shows the human-MLLM agreement rate for individual rubric criteria. We find that there is high variance in the MLLM's ability to annotate criteria accurately. For instance, GPT 4-o has low accuracy (agreement rate 0.46) annotating whether a drum's head is made of the correct material, but high accuracy (agreement rate 0.98) determining whether a cane is white in color. 
\Description[Histogram illustrating the human-MLLM agreement rate across all of the rubric criteria]{The histogram illustrates how the agreement rate varies substantially across criteria from ~30-90\%.}}
\label{fig:judge-histogram} 
\end{figure} 

\subsubsection{Interrogating the validity of automated application}\label{sec:result-mllm-human-comparison}
To assess the feasibility of automating rubric application, we compare MLLM-generated annotations to human annotations for each rubric criterion. Across artifacts, the MLLM’s labels of cultural appropriateness agree with human judgments for 78–88\% of the images in our dataset (fourth column in Table \ref{tab:operationalization-results-maintext}), with a consistent tendency to over-predict appropriateness for five of the six artifacts. Disaggregating by images labeled as appropriate versus inappropriate by humans reveals two types of MLLM errors.  Human-MLLM agreement for the \emph{inappropriate} images tells us the MLLM's ability to recognize violated criteria.  In contrast, human-MLLM agreement for the \emph{appropriate} images tells us whether the MLLM can recognize valid depictions.  Each has different implications for how practitioners might revise their rubric criteria to improve judge performance, as the best steps forward depend on the type of error being made---for instance, relaxing or clarifying criteria to address false negatives vs. strengthening or expanding criteria to address false positives.

To better understand the differences between the manual and automated implementations, we break down agreement by each of the individual rubric criteria. Figure \ref{fig:judge-histogram} shows the distribution of agreement scores across the 48 total criteria taken from our six rubrics. 
Agreement varies substantially: while many criteria are annotated accurately (44\% of criteria have agreement rates above 80\%), 10\% of criteria have agreement rates below 50\%, indicating performance no better or even worse than chance. 
Many of the criteria on which the MLLM-judge performed poorly (\ie{} the ``low-accuracy criteria'') were those that required the artifact to be depicted in a specific shape or spatial arrangement, such as a valid braille cell configuration (39\% agreement), the stern shape of a Chundan Vallam (43\%), or arrangement of pits on a Pallanguzhi board (64\%).
Other low-accuracy criteria are those based on features that are difficult (for both humans and machines) to infer from visual information alone ~\citep{farhadi2009describing}, such as the material of a drum's head (46\%), or the type of wood used to create a Pallanguzhi board (56\%). 
In contrast, high-accuracy criteria typically describe visually salient features, such as the color of a guide cane (98\%). Many low-accuracy criteria correspond to features that community members identify as highly important, highlighting opportunities to improve MLLM-based judging.  We include the agreement rate for each criterion and a further discussion of judge performance in Appendix \ref{apdx:human-mllm-agreement}. \looseness=-1

\section{Discussion}\label{sec:discussion}

In this section, we discuss our study’s contributions to and implications for the emerging interdisciplinary debate around how to bring stakeholder perspectives into the design of evaluations for generative AI, a topic that is gaining attention within the FAccT community \cite{kawakami2025translation,nguyen2026validating}.
Our goal in conducting this study is not to advocate for any one particular methodology as the ideal way for practitioners to engage stakeholders in measurement design. 
Instead, our goals are to highlight the key benefits and challenges that surfaced across engagements with three different communities, to offer reflections on the methodological limitations of our exploratory approach, and to identify opportunities for future work. To this end, 
we reflect on open research challenges in automating the application of rubrics (Section \ref{sec:discussion-automation}), situate our work within the larger discourse on tensions between participation and scale (Section \ref{sec:discussion-participation}), and discuss important limitations of our study (Section \ref{sec:discussion-limitations}).

\subsection{Towards the design of valid, community-centered MLLM-as-a-judge pipelines}\label{sec:discussion-automation} A gold-standard approach to centering community participation in measurement would involve sustained community engagement across all phases of the measurement lifecycle (Figure~\ref{fig:framework}), but we recognize that such approaches might be prohibitively costly for both communities and industry practitioners ~\cite{delgado2023participatory,ParticipationScaleTensions}. Accordingly, our methodology reflects a pragmatic design choice: we invite community members to participate in systematization through one-time workshops, and then examine the extent to which MLLM judges can operationalize these criteria effectively off-the-shelf, without further customization or experimentation.
Our findings, however, reveal substantial variation in how reliably MLLM judges apply different rubric criteria, raising important concerns about their resulting validity. Prior work has documented the brittleness of MLLM-based judges, including systematic biases \cite{ye2024justiceprejudice,lin2024evaluating} and sensitivity to prompt formulation \cite{li2025generation}. More broadly, both AI researchers and quantitative social scientists emphasize that the measurement design process is inherently iterative \cite{adcock2001measurement,shankar2024who,gabreegziabher2024metricmate,wallach2025position}. As \citet{adcock2001measurement} argue, systematized concepts often require revision once confronted with the practical challenges of operationalization, \eg{} the realization that a systematized concept may be too difficult to reliably measure. 

Iteratively refining rubric criteria to improve their legibility to automated judges is a critical next step for future work and a limitation of this current study that aimed to focus on systematization. Addressing these questions requires grappling with both technical challenges in the design of MLLM-as-a-judge pipelines \cite{li2025generation,gabreegziabher2024metricmate}, and human-centered challenges in thoughtfully engaging community members in a highly specialized and technical measurement design process \cite{suresh2023kaliedoscope}. 
Future work can also conduct more critical studies of the implicit limitations of automated approaches to measurement, \eg{} drawing upon literature from critical algorithm studies \citep{weizenbaum1976computer,kawakami2026ai}, and provide guidance for when alternative ways of evaluating AI systems (\eg{} direct engagement with stakeholders \citep{mack2023towards,weidinger2023sociotechnical}) may be preferred.

\subsection{Defining the scope and purpose of community-centered measurement}\label{sec:discussion-participation} Our goal in this work is to scaffold community participation in the evaluation of general-purpose, universally-scoped foundation models. Our methodology builds upon the framework put forward by \citet{suresh2024participation}, who pose that even though foundation model design is largely centralized, local communities can still pursue meaningful participatory efforts at the ``surface layer,'' which corresponds to a specific context (\eg{} generating culturally representative media that depicts the unique material culture of Kerala) and groups of stakeholders (\eg{} residents of Kerala whose culture is being depicted in those images).

Our findings show that community participation can shape measurement design by contributing lived, embodied expertise that is difficult to capture through reference images or automated approaches alone. Rather than providing a surface-level description of the visual attributes of each artifact, participants' descriptions spoke to how artifacts are used and experienced in practice. 
This insight affirms past work showing that anonymous crowd workers often lack the cultural expertise required for evaluative tasks, even when provided with reference images~\cite{hall2024towards,hall2025human,qadri2025case}. 
In contrast, many LLM-generated rubrics failed to capture these salient features and in some cases reflected confused or inaccurate understandings.
This finding extends those from past work, which describe how LLM-generated rubrics are often overly vague or underspecified ~\cite{szymanski2024comparing}, to reveal how these rubrics may reflect deeper cultural misalignment. Taken together, these results suggest that community participation in defining evaluation criteria can help bridge epistemic gaps in large pretrained models, particularly for low-data cultural artifacts \cite{massiceti2024explaining,hall2025human}, improving the validity of the resulting measurement instruments. \looseness=-1

Beyond asking community members to contribute their knowledge, community participation also enables communities to express their more normative desires and preferences for cultural representation. For example, while it is common for braille notetaker devices to have either a QWERTY or braille keyboard, blind community members consistently expressed a preference for braille keyboards, which better emphasize tactile culture and align with shared advocacy goals around increased public investment in accessible technologies. 
Thus, even in settings where multiple depictions may be technically accurate, participation enables communities to make critical decisions about how they desire to be represented. This more normative role reveals how participation is still valuable in a setting where an LLM can fully describe a cultural artifact, as it is still up to the community to make choices about which depictions they do, or do not, desire. \looseness=-1

While our findings illustrate the potential benefits of human-centered approaches to measurement through close collaborations with three communities and just a small set of artifacts, many open questions remain. 
In real-world settings, foundation model providers must prioritize engagement across thousands of communities and a wide range of content \cite{weidinger2023sociotechnical,ParticipationScaleTensions,young2025participatory}, a scale at which both our methods and related qualitative approaches to identifying representational harms (\eg{} \cite{mack2023towards,qadri2025case}) are not feasible to apply. 
Our findings highlight key considerations for identifying where community participation is most valuable in measurement design. 
We found that LLM-generated rubrics may serve as a reasonable starting point for some cultural artifacts (Section \ref{sec:llm-rubric-comparison}), but not others. 
Practitioners may consider inviting community members to assess the face validity of LLM-generated rubrics as a lightweight check. 
If the rubrics diverge significantly from community members' understandings of the content being evaluated, this is a signal that a deeper engagement (using methods like ours) may be warranted. \looseness=-1

In this study, systematization of knowledge into rubrics was done synchronously, and in detail. The depth of our exploratory engagement reveals several promising directions for adapting this process to reduce the potential time and labor expected of both communities and practitioners. One direction is to elicit community members' desires for cultural representation asynchronously, drawing upon known deliberation approaches (\eg{} the Delphi technique ~\cite{hsu2007delphi}) or emerging interfaces to support users in designing LLM-as-a-judge pipelines ~\cite{shankar2024who,pan2024human,gabreegziabher2024metricmate}. 
Practitioners can also avoid starting from scratch by instead asking communities to revise and critique LLM-generated rubrics, treating automated outputs as provisional baselines rather than authoritative representations. 
Another complementary direction for future work can explore how practitioners can ground rubric development in themes identified by prior qualitative research, \eg{} shared themes that articulate participants' desires for cultural representation, like those identified here and by \citet{qadri2025case}, or work that documents the representational harms experienced by marginalized communities \cite{bianchi2023easily,qadri2023ai,mack2024they,magomere2025world,bennett2025toward}. 
Practitioners can use these themes identified by past literature to scaffold more efficient elicitation of specific visual criteria that can be used to score images.
While such lower-touch approaches may improve scalability, our study suggests that synchronous engagement plays an important role in fostering participant buy-in and surfacing deeper insights into participants' more normative desires for cultural representation (the higher-level themes) \cite{qadri2025case}, highlighting an open question about how much of this depth can be preserved in more lightweight models of participation.

\subsection{Limitations}\label{sec:discussion-limitations}

There are many approaches to scaffolding engagements with impacted communities---for instance, pursuing sustained engagements with organizations with defined memberships and structures of communication~\citep{ParticipationScaleTensions,thieme2026engaging}.  In contrast, in this work, we chose to engage individuals who held shared identities and life experiences but otherwise did not know each other.  While this recruitment strategy is in line with past HCI research on representational harm~\citep{mack2024they,qadri2025case,bergman2024stela}, our broad conceptualization of ``community'' as a group of individuals who hold a shared cultural identity has limitations.  Our workshop data allowed us to identify preferences that were shared across the individuals we interviewed. However, our small sample of participants recruited through convenience sampling does not enable us to make claims that will generalize across the entire community: a common limitation of qualitative research ~\citep{kumar2019engaging}. \looseness=-1

We made the value-laden choice to scaffold participation as consultation ~\citep{arnstein1969ladder}---one-time workshops to elicit participants' preferences---rather than engaging the community as full collaborators with the power to shape and own research outputs~\citep{delgado2023participatory,suresh2024participation}.  While there are several potential benefits to pursuing lower-touch approaches to participation, such as respecting and attempting to minimize the labor required from participants ~\citep{ParticipationScaleTensions}, we acknowledge that such approaches run the risk of being exploitative if outputs are misused by researchers, or if community members are not adequately compensated for their labor ~\citep{delgado2023participatory,dalal2024provocation}.  There is more work needed to develop methods that meaningfully shift power to community members, \eg{} as part of a grassroots, community-led project where community members have full ownership over critical measurement decisions ~\citep{orife2020masakhane}. \looseness=-1

Our decision to prioritize criteria that were agreed upon across participants when designing rubrics, excluding those that were contested, has several limitations.
While our approach focuses on those criteria viewed as most essential to a culturally appropriate depiction, without engaging more deeply with contestedness and disagreement, it runs the risk of replicating dominant or hegemonic views ~\citep{dev2026unified}. Future research can build upon our methods, which \emph{surface} disagreement across participants, to pursue alternative approaches to \emph{reconcile} these disagreements through continued engagements, \eg{} using deliberative or discursive methods to reach group consensus ~\citep{bergman2024stela,qadri2025case}, or exploring innovative approaches to measurement that allow for more than one singular ``ground truth'' ~\citep{sorensen2024roadmap,guerdan2025validating}.
Finally, our rubrics systematize cultural appropriateness for depictions of cultural artifacts in isolation, and further work is needed to extend these methods to design rubrics for more realistic, complex cultural scenes \citep{qadri2025case,bennett2025toward,thieme2026engaging}.

\section{Conclusion}
Rich qualitative work from the FAccT community has demonstrated that AI models fail to represent people from marginalized communities as they wish to be seen. We explore one potential path forward: supporting participation in designing evaluations that can be automatically applied to score AI-generated outputs. 
Our findings open a broader discussion about the technical challenges of designing valid measures, and the value community input adds relative to fully automated evaluation approaches. 
We are hopeful that practitioners can adapt and extend the exploratory methods presented in this work to create more contextually grounded AI evaluations. \looseness=-1

\newpage
\section{Endmatter Statements}\label{sec:endmatter}

\subsection{Ethical Considerations Statement}

Our study protocol, recruitment material, and consent form were reviewed and approved by our institution’s ethics review board to ensure they followed best practices. Participants were compensated and informed that they could withdraw from the study at any time without consequence.

A central ethical consideration throughout the study was co-creating access with participants with differing needs. Drawing on best practices in accessibility research \cite{bennett2018interdependence,muehlbradt2022whats,das2024provenance}, we adapted our protocol to support non-visual access to AI images by creating carefully designed alt text, and inviting BLV community members to participate in cross-ability pairs, detailed in Appendix \ref{apdx:blv-accessibility}. We also regularly solicited participant feedback during workshops so that we could modify our facilitation approach to increase comfort for participants, \eg{} by repeating our alt text descriptions or offering more regular breaks to rest in between activities.

When designing our protocol, we carefully considered the risks of exposing participants to harmful or offensive AI-generated depictions, as prior work has shown that repeated exposure to representational harms can be disempowering \cite{wenzel2023can,zhang2025aura}.  To mitigate these risks, we pre-generated all images shown in workshops rather than generating them live. Pre-generation also allowed our research team to curate a diverse set of meaningfully distinct images that we could show participants, reducing the annotation burden expected from participants so that they would not need to give us feedback on the same types of errors.

% Reflecting on the potential adverse impacts of our methods, we acknowledge that

Our approach inherits familiar risks and limitations of related participatory AI efforts that are discussed in past scholarship \cite{sloane2020participation,delgado2023participatory,suresh2024participation,ParticipationScaleTensions}. First, we our proposed methods require labor from participants who may already be marginalized. To this end, we encourage practitioners to consider how participation can be meaningfully compensated, and discuss ways in which participation might otherwise be made less burdensome. While participation in measurement offers great potential to shape foundation model development, we also caution that participatory evaluation may not yield immediate or tangible benefits for community members, such as improved representations in deployed models \cite{sloane2020participation,dalal2024provocation}. Accordingly, we urge practitioners to be transparent about how the outputs of participation (\eg{} the resulting measurement instruments) will be used and what benefits, if any, participants can reasonably expect (\eg{} beyond contributing to research).

\subsection{Generative AI Usage Statement}

%All authors confirm that they did not use generative AI to generate text for the paper. 
Generative AI tools were not used to generate original content, analyses, arguments, findings, or interpretations. All academic contributions, including our literature review, workshop data collection, workshop data analysis, interpretation, and writing were produced by the authors. %The authors take full responsibility for all content in our manuscript.
The authors did make use of generative AI tools to support brainstorming throughout the paper writing process.  ChatGPT (based on GPT-5.2) was used to brainstorm alternative terminology choices, for instance, for describing measurement approaches that can be deployed at scale.%(landed on ``designed to support repeatable, automatable application across different models, datasets, and evaluation settings'' and variants of this term).
% and for describing the class of approaches that includes CLIPscore, VQAscore, and LLM-as-a-judge (landed on ``model-mediated evaluation'').
ChatGPT was also used to brainstorm ways to more clearly format results tables and figures.  For all uses, the authors did not copy the text directly output by LLMs, but instead reviewed what was generated and used their own judgment to make revisions to the paper.

% Additional sections we could potentially add AFTER the paper is accepted

%\subsection{Author Contributions}

%\subsection{Competing Interests}

\subsection{Acknowledgements}

We thank our study participants from each community who shared their knowledge and expertise.
We thank Hanna Wallach, Siobhan Mackenzie Hall, Xinnuo Xu, Melanie Pradier, Martin Grayson, Camilla Longden, Rita Marques, Shaily Bhatt, Shivani Kapania, Luke Guerdan, Fernando Diaz, and Anna Kawakami for helpful feedback and conversations that shaped this work.
NJ and HH acknowledge support from the NSF (IIS2229881) and the CMU Block Center for Technology and Society.
Any opinions, findings, conclusions, or recommendations expressed in this material are those of the authors and do not reflect the views of the National Science Foundation and other funding agencies.

\bibliographystyle{ACM-Reference-Format}
\bibliography{main}

@article{arnstein1969ladder,
  title={A Ladder of Citizen Participation},
  author={Arnstein, Sherry R.},
  journal={Journal of the American Institute of Planners},
  volume={35},
  number={4},
  pages={216--224},
  year={1969},
  publisher={Taylor \& Francis}
}

@misc{guerdan2025validating,
      title={Validating LLM-as-a-Judge Systems under Rating Indeterminacy}, 
      author={Luke Guerdan and Solon Barocas and Kenneth Holstein and Hanna Wallach and Zhiwei Steven Wu and Alexandra Chouldechova},
      year={2025},
      archivePrefix={arXiv},
      primaryClass={cs.LG},
      url={https://arxiv.org/abs/2503.05965}, 
}

@inproceedings{chang2017revolt,
author = {Chang, Joseph Chee and Amershi, Saleema and Kamar, Ece},
title = {Revolt: Collaborative Crowdsourcing for Labeling Machine Learning Datasets},
year = {2017},
isbn = {9781450346559},
publisher = {Association for Computing Machinery},
address = {New York, NY, USA},
url = {https://doi.org/10.1145/3025453.3026044},
doi = {10.1145/3025453.3026044},
booktitle = {Proceedings of the 2017 CHI Conference on Human Factors in Computing Systems},
pages = {2334–2346},
numpages = {13},
location = {Denver, Colorado, USA},
series = {CHI '17}
}

@inproceedings{callison2009fast,
    title = "Fast, Cheap, and Creative: Evaluating Translation Quality Using {A}mazon{'}s {M}echanical {T}urk",
    author = "Callison-Burch, Chris",
    editor = "Koehn, Philipp  and
      Mihalcea, Rada",
    booktitle = "Proceedings of the 2009 Conference on Empirical Methods in Natural Language Processing",
    month = aug,
    year = "2009",
    address = "Singapore",
    publisher = "Association for Computational Linguistics",
    url = "https://aclanthology.org/D09-1030/",
    pages = "286--295"
}

@misc{akyurek2025prbench,
      title={PRBench: Large-Scale Expert Rubrics for Evaluating High-Stakes Professional Reasoning}, 
      author={Afra Feyza Akyürek and Advait Gosai and Chen Bo Calvin Zhang and Vipul Gupta and Jaehwan Jeong and Anisha Gunjal and Tahseen Rabbani and Maria Mazzone and David Randolph and Mohammad Mahmoudi Meymand and Gurshaan Chattha and Paula Rodriguez and Diego Mares and Pavit Singh and Michael Liu and Subodh Chawla and Pete Cline and Lucy Ogaz and Ernesto Hernandez and Zihao Wang and Pavi Bhatter and Marcos Ayestaran and Bing Liu and Yunzhong He},
      year={2025},
      archivePrefix={arXiv},
      primaryClass={cs.CL},
      url={https://arxiv.org/abs/2511.11562}, 
}

@article{kawakami2026ai,
author = {Anna Kawakami and Jordan Taylor and Sarah Fox and Haiyi Zhu and Kenneth Holstein},
title ={AI failure loops in devalued work: The confluence of overconfidence in AI and underconfidence in worker expertise},

journal = {Big Data \& Society},
volume = {13},
number = {1},
pages = {20539517261424164},
year = {2026},
doi = {10.1177/20539517261424164},
URL = {https://doi.org/10.1177/20539517261424164},
}

@book{weizenbaum1976computer,
author = {Weizenbaum, Joseph},
title = {Computer Power and Human Reason: From Judgment to Calculation},
year = {1976},
isbn = {0716704641},
publisher = {W. H. Freeman \& Co.},
address = {USA}
}

@misc{sorensen2024roadmap,
      title={A Roadmap to Pluralistic Alignment}, 
      author={Taylor Sorensen and Jared Moore and Jillian Fisher and Mitchell Gordon and Niloofar Mireshghallah and Christopher Michael Rytting and Andre Ye and Liwei Jiang and Ximing Lu and Nouha Dziri and Tim Althoff and Yejin Choi},
      year={2024},
      archivePrefix={arXiv},
      primaryClass={cs.AI},
      url={https://arxiv.org/abs/2402.05070}, 
}

@misc{dev2026unified,
      title={A Unified Framework to Quantify Cultural Intelligence of AI}, 
      author={Sunipa Dev and Vinodkumar Prabhakaran and Rutledge Chin Feman and Aida Davani and Remi Denton and Charu Kalia and Piyawat L Kumjorn and Madhurima Maji and Rida Qadri and Negar Rostamzadeh and Renee Shelby and Romina Stella and Hayk Stepanyan and Erin van Liemt and Aishwarya Verma and Oscar Wahltinez and Edem Wornyo and Andrew Zaldivar and Saška Mojsilović},
      year={2026},
      archivePrefix={arXiv},
      primaryClass={cs.AI},
      url={https://arxiv.org/abs/2603.01211}, 
}

@inproceedings{thieme2026engaging,
  title={Engaging Communities Meaningfully in Defining Disability Representation for AI Image Generation},
  author={Thieme, Anja and Marques, Rita Faia and Grayson, Martin and Balachandar, Sidhika and Cassidy, Cameron Tyler and Choksi, Madiha Zahrah and Longden, Camilla and Huda, Reeda Shimaz and Kalovwe, Nicholas Ileve and Mallon, Christina and Mansperger, Courtney and Massiceti, Daniela and Mitra, Bhaskar and Nzioka, Ruth Mueni and Tanase, Ioana and You, Yuzhe and Morrison, Cecily},
  booktitle={Proceedings of the CHI Conference on Human Factors in Computing Systems (CHI '26)},
  year={2026},
  publisher={ACM},
  url={https://www.microsoft.com/en-us/research/wp-content/uploads/2026/01/CHI_paper_FINAL.pdf}
}

@inproceedings{coston2023validity,
  title={A validity perspective on evaluating the justified use of data-driven decision-making algorithms},
  author={Coston, Amanda and Kawakami, Anna and Zhu, Haiyi and Holstein, Ken and Heidari, Hoda},
  booktitle={2023 IEEE conference on secure and trustworthy machine learning (SaTML)},
  pages={690--704},
  year={2023},
  organization={IEEE}
}

@misc{raji2021wide,
      title={AI and the Everything in the Whole Wide World Benchmark}, 
      author={Inioluwa Deborah Raji and Emily M. Bender and Amandalynne Paullada and Emily Denton and Alex Hanna},
      year={2021},
      archivePrefix={arXiv},
      primaryClass={cs.LG},
      url={https://arxiv.org/abs/2111.15366}, 
}

@inproceedings{barocas2021designing,
  title={Designing disaggregated evaluations of ai systems: Choices, considerations, and tradeoffs},
  author={Barocas, Solon and Guo, Anhong and Kamar, Ece and Krones, Jacquelyn and Morris, Meredith Ringel and Vaughan, Jennifer Wortman and Wadsworth, W Duncan and Wallach, Hanna},
  booktitle={Proceedings of the 2021 AAAI/ACM Conference on AI, Ethics, and Society},
  pages={368--378},
  year={2021}
}

@inproceedings{orr2024sport,
author = {Orr, Will and Kang, Edward B.},
title = {AI as a Sport: On the Competitive Epistemologies of Benchmarking},
year = {2024},
isbn = {9798400704505},
publisher = {Association for Computing Machinery},
address = {New York, NY, USA},
url = {https://doi.org/10.1145/3630106.3659012},
doi = {10.1145/3630106.3659012},
booktitle = {Proceedings of the 2024 ACM Conference on Fairness, Accountability, and Transparency},
pages = {1875–1884},
numpages = {10},
location = {Rio de Janeiro, Brazil},
series = {FAccT '24}
}

@misc{eriksson2025trust,
      title={Can We Trust AI Benchmarks? An Interdisciplinary Review of Current Issues in AI Evaluation}, 
      author={Maria Eriksson and Erasmo Purificato and Arman Noroozian and Joao Vinagre and Guillaume Chaslot and Emilia Gomez and David Fernandez-Llorca},
      year={2025},
      archivePrefix={arXiv},
      primaryClass={cs.AI},
      url={https://arxiv.org/abs/2502.06559}, 
}

@misc{dalle3,
  author       = {{OpenAI}},
  title        = {{DALL·E 3}},
  howpublished = {\url{https://openai.com/index/dall-e-3/}},
  year         = {2024},
  note         = {Accessed: 2025-09-11}
}

@misc{gptimage1,
  author       = {{OpenAI}},
  title        = {GPT-Image-1},
  howpublished = {\url{https://platform.openai.com/docs/guides/image-generation?image-generation-model=gpt-image-1}},
  year         = {2025},
  note         = {Accessed: 2025-09-11}
}

@misc{flux2024,
    author={{Black Forest Labs}},
    title={FLUX},
    year={2024},
    howpublished={\url{https://github.com/black-forest-labs/flux}},
}

@online{sd35,
  author       = {{Stability AI}},
  title        = {Introducing Stable Diffusion 3.5},
  year         = {2024},
  url          = {https://stability.ai/news/introducing-stable-diffusion-3-5},
  note         = {Accessed: 2025-09-11}
}

@online{sd3,
  author       = {{Stability AI}},
  title        = {Announcing the Open Release of Stable Diffusion 3 Medium, Our Most Sophisticated Image Generation Model to Date },
  year         = {2024},
  url          = {https://stability.ai/news/stable-diffusion-3-medium},
  note         = {Accessed: 2025-09-11}
}

@misc{delgado2023participatory,
      title={The Participatory Turn in AI Design: Theoretical Foundations and the Current State of Practice}, 
      author={Fernando Delgado and Stephen Yang and Michael Madaio and Qian Yang},
      year={2023},
      archivePrefix={arXiv},
      primaryClass={cs.HC},
      url={https://arxiv.org/abs/2310.00907}, 
}

@article{young2025participatory,
  author       = {Meg Young},
  title        = {Participatory AI? Begin with the Most Affected People},
  journal      = {TechPolicy.Press},
  year         = {2025},
  month        = {February 19},
  url          = {https://www.techpolicy.press/participatory-ai-begin-with-the-most-affected-people/},
  note         = {Essay published as part of the Participatory AI Research \& Practice Symposium reflections series}
}

@misc{kirstain2023pick,
      title={Pick-a-Pic: An Open Dataset of User Preferences for Text-to-Image Generation}, 
      author={Yuval Kirstain and Adam Polyak and Uriel Singer and Shahbuland Matiana and Joe Penna and Omer Levy},
      year={2023},
      archivePrefix={arXiv},
      primaryClass={cs.CV},
      url={https://arxiv.org/abs/2305.01569}, 
}

@misc{heusel2018gans,
      title={GANs Trained by a Two Time-Scale Update Rule Converge to a Local Nash Equilibrium}, 
      author={Martin Heusel and Hubert Ramsauer and Thomas Unterthiner and Bernhard Nessler and Sepp Hochreiter},
      year={2018},
      archivePrefix={arXiv},
      primaryClass={cs.LG},
      url={https://arxiv.org/abs/1706.08500}, 
}

@misc{deviyani2025contextual,
      title={Contextual Metric Meta-Evaluation by Measuring Local Metric Accuracy}, 
      author={Athiya Deviyani and Fernando Diaz},
      year={2025},
      archivePrefix={arXiv},
      primaryClass={cs.CL},
      url={https://arxiv.org/abs/2503.19828}, 
}

@misc{rottger2025safety,
      title={SafetyPrompts: a Systematic Review of Open Datasets for Evaluating and Improving Large Language Model Safety}, 
      author={Paul Röttger and Fabio Pernisi and Bertie Vidgen and Dirk Hovy},
      year={2025},
      archivePrefix={arXiv},
      primaryClass={cs.CL},
      url={https://arxiv.org/abs/2404.05399}, 
}

@misc{jayasumana2024rethinking,
      title={Rethinking FID: Towards a Better Evaluation Metric for Image Generation}, 
      author={Sadeep Jayasumana and Srikumar Ramalingam and Andreas Veit and Daniel Glasner and Ayan Chakrabarti and Sanjiv Kumar},
      year={2024},
      archivePrefix={arXiv},
      primaryClass={cs.CV},
      url={https://arxiv.org/abs/2401.09603}, 
}

@article{mcintosh2025inadequacies,
   title={Inadequacies of Large Language Model Benchmarks in the Era of Generative Artificial Intelligence},
   ISSN={2691-4581},
   url={http://dx.doi.org/10.1109/TAI.2025.3569516},
   DOI={10.1109/tai.2025.3569516},
   journal={IEEE Transactions on Artificial Intelligence},
   publisher={Institute of Electrical and Electronics Engineers (IEEE)},
   author={McIntosh, Timothy R and Susnjak, Teo and Arachchilage, Nalin and Liu, Tong and Xu, Dan and Watters, Paul and Halgamuge, Malka N},
   year={2025},
   pages={1–18} }

@misc{hu2023tifa,
      title={TIFA: Accurate and Interpretable Text-to-Image Faithfulness Evaluation with Question Answering}, 
      author={Yushi Hu and Benlin Liu and Jungo Kasai and Yizhong Wang and Mari Ostendorf and Ranjay Krishna and Noah A Smith},
      year={2023},
      archivePrefix={arXiv},
      primaryClass={cs.CV},
      url={https://arxiv.org/abs/2303.11897}, 
}

@inproceedings{shankar2024who,
author = {Shankar, Shreya and Zamfirescu-Pereira, J.D. and Hartmann, Bjoern and Parameswaran, Aditya and Arawjo, Ian},
title = {Who Validates the Validators? Aligning LLM-Assisted Evaluation of LLM Outputs with Human Preferences},
year = {2024},
isbn = {9798400706288},
publisher = {Association for Computing Machinery},
address = {New York, NY, USA},
url = {https://doi.org/10.1145/3654777.3676450},
doi = {10.1145/3654777.3676450},
booktitle = {Proceedings of the 37th Annual ACM Symposium on User Interface Software and Technology},
articleno = {131},
numpages = {14},
location = {Pittsburgh, PA, USA},
series = {UIST '24}
}

@misc{ye2024justiceprejudice,
      title={Justice or Prejudice? Quantifying Biases in LLM-as-a-Judge}, 
      author={Jiayi Ye and Yanbo Wang and Yue Huang and Dongping Chen and Qihui Zhang and Nuno Moniz and Tian Gao and Werner Geyer and Chao Huang and Pin-Yu Chen and Nitesh V Chawla and Xiangliang Zhang},
      year={2024},
      archivePrefix={arXiv},
      primaryClass={cs.CL},
      url={https://arxiv.org/abs/2410.02736}, 
}

@misc{lin2024evaluating,
      title={Evaluating Text-to-Visual Generation with Image-to-Text Generation}, 
      author={Zhiqiu Lin and Deepak Pathak and Baiqi Li and Jiayao Li and Xide Xia and Graham Neubig and Pengchuan Zhang and Deva Ramanan},
      year={2024},
      archivePrefix={arXiv},
      primaryClass={cs.CV},
      url={https://arxiv.org/abs/2404.01291}, 
}

@misc{lee2023heim,
      title={Holistic Evaluation of Text-To-Image Models}, 
      author={Tony Lee and Michihiro Yasunaga and Chenlin Meng and Yifan Mai and Joon Sung Park and Agrim Gupta and Yunzhi Zhang and Deepak Narayanan and Hannah Benita Teufel and Marco Bellagente and Minguk Kang and Taesung Park and Jure Leskovec and Jun-Yan Zhu and Li Fei-Fei and Jiajun Wu and Stefano Ermon and Percy Liang},
      year={2023},
      archivePrefix={arXiv},
      primaryClass={cs.CV},
      url={https://arxiv.org/abs/2311.04287}, 
}

@article{adcock2001measurement,
  title={{Measurement validity: A shared standard for qualitative and quantitative research}},
  author={Adcock, Robert and Collier, David},
  journal={American Political Science Review},
  volume={95},
  number={3},
  pages={529--546},
  year={2001},
  publisher={Cambridge University Press}
}

@article{hada2024large,
  title={{Are Large Language Model-based Evaluators the Solution to Scaling Up Multilingual Evaluation?}},
  author={Hada, Rishav and Gumma, Varun and de Wynter, Adrian and Diddee, Harshita and Ahmed, Mohamed and Choudhury, Monojit and Bali, Kalika and Sitaram, Sunayana},
  journal={arXiv preprint arXiv:2309.07462},
  year={2024}
}

@article{kreiss2022context,
  title={Context Matters for Image Descriptions for Accessibility: Challenges for Referenceless Evaluation Metrics},
  author={Kreiss, Elisa and Bennett, Cynthia and Hooshmand, Shayan and Zelikman, Eric and Morris, Meredith Ringel and Potts, Christopher},
  journal={arXiv preprint arXiv:2205.10646},
  year={2022}
}

@inproceedings{saxon2024who,
 author = {Saxon, Michael and Jahara, Fatima and Khoshnoodi, Mahsa and Lu, Yujie and Sharma, Aditya and Wang, William Yang},
 booktitle = {Advances in Neural Information Processing Systems},
 editor = {A. Globerson and L. Mackey and D. Belgrave and A. Fan and U. Paquet and J. Tomczak and C. Zhang},
 pages = {85630--85657},
 publisher = {Curran Associates, Inc.},
 title = {Who Evaluates the Evaluations? Objectively Scoring Text-to-Image Prompt Coherence Metrics with T2IScoreScore (TS2)},
 url = {https://proceedings.neurips.cc/paper_files/paper/2024/file/9b9cfd5428153ccfbd4ba34b7e007305-Paper-Conference.pdf},
 volume = {37},
 year = {2024}
}

@misc{orife2020masakhane,
      title={Masakhane -- Machine Translation For Africa}, 
      author={Iroro Orife and Julia Kreutzer and Blessing Sibanda and Daniel Whitenack and Kathleen Siminyu and Laura Martinus and Jamiil Toure Ali and Jade Abbott and Vukosi Marivate and Salomon Kabongo and Musie Meressa and Espoir Murhabazi and Orevaoghene Ahia and Elan van Biljon and Arshath Ramkilowan and Adewale Akinfaderin and Alp Öktem and Wole Akin and Ghollah Kioko and Kevin Degila and Herman Kamper and Bonaventure Dossou and Chris Emezue and Kelechi Ogueji and Abdallah Bashir},
      year={2020},
      archivePrefix={arXiv},
      primaryClass={cs.CL},
      url={https://arxiv.org/abs/2003.11529}, 
}

@inproceedings{wenzel2025invisible,
  title     = {Invisible by Design? Generative AI and Mirrors of Misrepresentation},
  author    = {Wenzel, Kimi and Ghosh, Avijit and Pendse, Sachin and Milani, Stephanie and Singh, Ajeet and Dabbish, Laura and Kaufman, Geoff},
  booktitle = {Proceedings of the ACM Conference on Fairness, Accountability, and Transparency (FAccT) CRAFT},
  year      = {2025},
  note      = {Workshop (CRAFT)},
  url       = {https://earthling.my.canva.site/invisible}
}

@misc{kunchukuttan2020bharat,
      title={AI4Bharat-IndicNLP Corpus: Monolingual Corpora and Word Embeddings for Indic Languages}, 
      author={Anoop Kunchukuttan and Divyanshu Kakwani and Satish Golla and Gokul N. C. and Avik Bhattacharyya and Mitesh M. Khapra and Pratyush Kumar},
      year={2020},
      archivePrefix={arXiv},
      primaryClass={cs.CL},
      url={https://arxiv.org/abs/2005.00085}, 
}

@inproceedings{aroyo2023dices,
author = {Aroyo, Lora and Taylor, Alex S. and D\'{\i}az, Mark and Homan, Christopher M. and Parrish, Alicia and Serapio-Garc\'{\i}a, Greg and Prabhakaran, Vinodkumar and Wang, Ding},
title = {DICES dataset: diversity in conversational AI evaluation for safety},
year = {2023},
publisher = {Curran Associates Inc.},
address = {Red Hook, NY, USA},
booktitle = {Proceedings of the 37th International Conference on Neural Information Processing Systems},
articleno = {2321},
numpages = {13},
location = {New Orleans, LA, USA},
series = {NIPS '23}
}

@misc{salaudeen2025measurement,
      title={Measurement to Meaning: A Validity-Centered Framework for AI Evaluation}, 
      author={Olawale Salaudeen and Anka Reuel and Ahmed Ahmed and Suhana Bedi and Zachary Robertson and Sudharsan Sundar and Ben Domingue and Angelina Wang and Sanmi Koyejo},
      year={2025},
      archivePrefix={arXiv},
      primaryClass={cs.CY},
      url={https://arxiv.org/abs/2505.10573}, 
}

@misc{openai_dalle_2022,
  author       = {OpenAI},
  title        = {{DALL·E now available without waitlist}},
  year         = 2022,
  month        = sep,
  url          = {https://openai.com/index/dall-e-now-available-without-waitlist/},
  note         = {Accessed: 2024-05-28}
}

@article{kumar2019engaging,
author = {Kumar, Neha and Karusala, Naveena and Ismail, Azra and Wong-Villacres, Marisol and Vishwanath, Aditya},
title = {Engaging Feminist Solidarity for Comparative Research, Design, and Practice},
year = {2019},
issue_date = {November 2019},
publisher = {Association for Computing Machinery},
address = {New York, NY, USA},
volume = {3},
number = {CSCW},
url = {https://doi.org/10.1145/3359269},
doi = {10.1145/3359269},
journal = {Proc. ACM Hum.-Comput. Interact.},
month = nov,
articleno = {167},
numpages = {24}
}

@inproceedings{qadri2023ai,
  title={Ai’s regimes of representation: A community-centered study of text-to-image models in south asia},
  author={Qadri, Rida and Shelby, Renee and Bennett, Cynthia L and Denton, Remi},
  booktitle={Proceedings of the 2023 ACM Conference on Fairness, Accountability, and Transparency},
  pages={506--517},
  year={2023}
}

@article{qadri2025case,
  title={{The Case for "Thick Evaluations" of Cultural Representation in AI}},
  author={Qadri, Rida and Diaz, Mark and Wang, Ding and Madaio, Michael},
  journal={arXiv preprint arXiv:2503.19075},
  year={2025}
}

@article{Gillespie2024,
author = {Tarleton Gillespie},
title ={Generative AI and the politics of visibility},

journal = {Big Data \& Society},
volume = {11},
number = {2},
pages = {20539517241252131},
year = {2024},
doi = {10.1177/20539517241252131},

URL = { 
    
        https://doi.org/10.1177/20539517241252131
},
}

@book{hall1997representation,
  title={Representation: Cultural Representations and Signifying Practices},
  editor={Hall, Stuart},
  year={1997},
  publisher={Sage Publications},
  address={London}
}

@inproceedings{hamna2025kahani,
author = {Hamna and Sudharsan, Deepthi and Seth, Agrima and Budhiraja, Ritvik and Khullar, Deepika and Jain, Vyshak and Bali, Kalika and Vashistha, Aditya and Segal, Sameer},
title = {Kahani: Culturally-Nuanced Visual Storytelling Tool for Non-Western Cultures},
year = {2025},
isbn = {9798400714849},
publisher = {Association for Computing Machinery},
address = {New York, NY, USA},
url = {https://doi.org/10.1145/3715335.3735478},
doi = {10.1145/3715335.3735478},
booktitle = {Proceedings of the 2025 ACM SIGCAS/SIGCHI Conference on Computing and Sustainable Societies},
pages = {379–400},
numpages = {22},
location = {
},
series = {COMPASS '25}
}

@article{he2025dreamstory,
  title={Dreamstory: Open-domain story visualization by llm-guided multi-subject consistent diffusion},
  author={He, Huiguo and Yang, Huan and Tuo, Zixi and Zhou, Yuan and Wang, Qiuyue and Zhang, Yuhang and Liu, Zeyu and Huang, Wenhao and Chao, Hongyang and Yin, Jian},
  journal={IEEE Transactions on Pattern Analysis and Machine Intelligence},
  year={2025},
  publisher={IEEE}
}

@inproceedings{adnin2024look,
author = {Adnin, Rudaiba and Das, Maitraye},
title = {"I look at it as the king of knowledge": How Blind People Use and Understand Generative AI Tools},
year = {2024},
isbn = {9798400706776},
publisher = {Association for Computing Machinery},
address = {New York, NY, USA},
url = {https://doi.org/10.1145/3663548.3675631},
doi = {10.1145/3663548.3675631},
booktitle = {Proceedings of the 26th International ACM SIGACCESS Conference on Computers and Accessibility},
articleno = {64},
numpages = {14},
location = {St. John's, NL, Canada},
series = {ASSETS '24}
}

@article{cronbach1955construct,
  title={Construct validity in psychological tests.},
  author={Cronbach, Lee J and Meehl, Paul E},
  journal={Psychological bulletin},
  volume={52},
  number={4},
  pages={281},
  year={1955},
  publisher={American Psychological Association}
}

@inproceedings{bianchi2023easily,
author = {Bianchi, Federico and Kalluri, Pratyusha and Durmus, Esin and Ladhak, Faisal and Cheng, Myra and Nozza, Debora and Hashimoto, Tatsunori and Jurafsky, Dan and Zou, James and Caliskan, Aylin},
title = {Easily Accessible Text-to-Image Generation Amplifies Demographic Stereotypes at Large Scale},
year = {2023},
isbn = {9798400701924},
publisher = {Association for Computing Machinery},
address = {New York, NY, USA},
url = {https://doi.org/10.1145/3593013.3594095},
doi = {10.1145/3593013.3594095},
booktitle = {Proceedings of the 2023 ACM Conference on Fairness, Accountability, and Transparency},
pages = {1493–1504},
numpages = {12},
location = {Chicago, IL, USA},
series = {FAccT '23}
}

@article{ghosh2024do, title={Do Generative AI Models Output Harm while Representing Non-Western Cultures: Evidence from A Community-Centered Approach}, volume={7}, url={https://ojs.aaai.org/index.php/AIES/article/view/31651}, DOI={10.1609/aies.v7i1.31651}, abstractNote={Our research investigates the impact of Generative Artificial Intelligence (GAI) models, specifically text-to-image generators (T2Is), on the representation of non-Western cultures, with a focus on Indian contexts. Despite the transformative potential of T2Is in content creation, concerns have arisen regarding biases that may lead to misrepresentations and marginalizations. Through a Non-Western community-centered approach
and grounded theory analysis of 5 focus groups from diverse Indian subcultures, we explore how T2I outputs to English input prompts depict Indian culture and its subcultures, uncovering novel representational harms such as exoticism and cultural misappropriation. These findings highlight the urgent need for inclusive and culturally sensitive T2I systems. We propose design guidelines informed by a sociotechnical perspective, contributing to the development of more equitable and representative GAI technologies globally. Our work underscores the necessity of adopting a community-centered approach to comprehend the sociotechnical dynamics of these models, complementing existing work in this space while identifying and addressing the potential negative repercussions and harms that may arise as these models are deployed on a global scale.}, number={1}, journal={Proceedings of the AAAI/ACM Conference on AI, Ethics, and Society}, author={Ghosh, Sourojit and Narayanan Venkit, Pranav and Gautam, Sanjana and Wilson, Shomir and Caliskan, Aylin}, year={2024}, month={Oct.}, pages={476-489} }

@inproceedings{zhang2022just,
author = {Zhang, Kexin and Deldari, Elmira and Lu, Zhicong and Yao, Yaxing and Zhao, Yuhang},
title = {“It’s Just Part of Me:” Understanding Avatar Diversity and Self-presentation of People with Disabilities in Social Virtual Reality},
year = {2022},
isbn = {9781450392587},
publisher = {Association for Computing Machinery},
address = {New York, NY, USA},
url = {https://doi.org/10.1145/3517428.3544829},
doi = {10.1145/3517428.3544829},
booktitle = {Proceedings of the 24th International ACM SIGACCESS Conference on Computers and Accessibility},
articleno = {4},
numpages = {16},
location = {Athens, Greece},
series = {ASSETS '22}
}

@inproceedings{mack2023towards,
author = {Mack, Kelly and Hsu, Rai Ching Ling and Monroy-Hern\'{a}ndez, Andr\'{e}s and Smith, Brian A. and Liu, Fannie},
title = {Towards Inclusive Avatars: Disability Representation in Avatar Platforms},
year = {2023},
isbn = {9781450394215},
publisher = {Association for Computing Machinery},
address = {New York, NY, USA},
url = {https://doi.org/10.1145/3544548.3581481},
doi = {10.1145/3544548.3581481},
booktitle = {Proceedings of the 2023 CHI Conference on Human Factors in Computing Systems},
articleno = {607},
numpages = {13},
location = {Hamburg, Germany},
series = {CHI '23}
}

@book{newmark1988translation,
  author    = {Newmark, Peter},
  title     = {A Textbook of Translation},
  year      = {1988},
  publisher = {Prentice Hall},
  address   = {New York},
  volume    = {66},
}

@article{knight1998machine,
    title = "Machine Transliteration",
    author = "Knight, Kevin  and
      Graehl, Jonathan",
    editor = "Hirschberg, Julia",
    journal = "Computational Linguistics",
    volume = "24",
    number = "4",
    year = "1998",
    address = "Cambridge, MA",
    publisher = "MIT Press",
    url = "https://aclanthology.org/J98-4003/",
    pages = "599--612"
}

@inproceedings{gupta2012mining,
    title = "Mining {H}indi-{E}nglish Transliteration Pairs from Online {H}indi Lyrics",
    author = "Gupta, Kanika  and
      Choudhury, Monojit  and
      Bali, Kalika",
    editor = "Calzolari, Nicoletta  and
      Choukri, Khalid  and
      Declerck, Thierry  and
      Do{\u{g}}an, Mehmet U{\u{g}}ur  and
      Maegaard, Bente  and
      Mariani, Joseph  and
      Moreno, Asuncion  and
      Odijk, Jan  and
      Piperidis, Stelios",
    booktitle = "Proceedings of the Eighth International Conference on Language Resources and Evaluation ({LREC}'12)",
    month = may,
    year = "2012",
    address = "Istanbul, Turkey",
    publisher = "European Language Resources Association (ELRA)",
    url = "https://aclanthology.org/L12-1179/",
    pages = "2459--2465"
}

@inproceedings{yao2024benchmarking,
    title = "Benchmarking Machine Translation with Cultural Awareness",
    author = "Yao, Binwei  and
      Jiang, Ming  and
      Bobinac, Tara  and
      Yang, Diyi  and
      Hu, Junjie",
    editor = "Al-Onaizan, Yaser  and
      Bansal, Mohit  and
      Chen, Yun-Nung",
    booktitle = "Findings of the Association for Computational Linguistics: EMNLP 2024",
    month = nov,
    year = "2024",
    address = "Miami, Florida, USA",
    publisher = "Association for Computational Linguistics",
    url = "https://aclanthology.org/2024.findings-emnlp.765/",
    doi = "10.18653/v1/2024.findings-emnlp.765",
    pages = "13078--13096"
}

@inproceedings{braham2015invisible,
author = {Branham, Stacy M. and Kane, Shaun K.},
title = {The Invisible Work of Accessibility: How Blind Employees Manage Accessibility in Mixed-Ability Workplaces},
year = {2015},
isbn = {9781450334006},
publisher = {Association for Computing Machinery},
address = {New York, NY, USA},
url = {https://doi.org/10.1145/2700648.2809864},
doi = {10.1145/2700648.2809864},
booktitle = {Proceedings of the 17th International ACM SIGACCESS Conference on Computers \& Accessibility},
pages = {163–171},
numpages = {9},
location = {Lisbon, Portugal},
series = {ASSETS '15}
}

@article{das2019doesnt,
author = {Das, Maitraye and Gergle, Darren and Piper, Anne Marie},
title = {"It doesn't win you friends": Understanding Accessibility in Collaborative Writing for People with Vision Impairments},
year = {2019},
issue_date = {November 2019},
publisher = {Association for Computing Machinery},
address = {New York, NY, USA},
volume = {3},
number = {CSCW},
url = {https://doi.org/10.1145/3359293},
doi = {10.1145/3359293},
journal = {Proc. ACM Hum.-Comput. Interact.},
month = nov,
articleno = {191},
numpages = {26}
}

@article{muehlbradt2022whats,
author = {Muehlbradt, Annika and Kane, Shaun K.},
title = {What's in an ALT Tag? Exploring Caption Content Priorities through Collaborative Captioning},
year = {2022},
issue_date = {March 2022},
publisher = {Association for Computing Machinery},
address = {New York, NY, USA},
volume = {15},
number = {1},
issn = {1936-7228},
url = {https://doi.org/10.1145/3507659},
doi = {10.1145/3507659},
journal = {ACM Trans. Access. Comput.},
month = mar,
articleno = {6},
numpages = {32}
}

@inproceedings{bennett2018interdependence,
author = {Bennett, Cynthia L. and Brady, Erin and Branham, Stacy M.},
title = {Interdependence as a Frame for Assistive Technology Research and Design},
year = {2018},
isbn = {9781450356503},
publisher = {Association for Computing Machinery},
address = {New York, NY, USA},
url = {https://doi.org/10.1145/3234695.3236348},
doi = {10.1145/3234695.3236348},
booktitle = {Proceedings of the 20th International ACM SIGACCESS Conference on Computers and Accessibility},
pages = {161–173},
numpages = {13},
location = {Galway, Ireland},
series = {ASSETS '18}
}

@article{palinkas2012purposeful,
author = {Palinkas, Lawrence and Horwitz, Sarah and Green, Carla and Wisdom, Jennifer and Duan, Naihua and Hoagwood, Kimberly},
year = {2013},
month = {11},
pages = {},
title = {Purposeful Sampling for Qualitative Data Collection and Analysis in Mixed Method Implementation Research},
volume = {42},
journal = {Administration and policy in mental health},
doi = {10.1007/s10488-013-0528-y}
}

@misc{ openai-promptrevision,
  author = {{OpenAI Developer Community Forum}},
  title = "API Image Generation in Dall-E-3 changes my original prompt without my permission",
  year = "2023",
  url = "https://community.openai.com/t/api-image-generation-in-dall-e-3-changes-my-original-prompt-without-my-permission/476355/2",
  note = "[Online; accessed 6-September-2025]"
}

@misc{ rnib-about,
  author = {{The Royal National Institute of Blind People}},
  title = "About us",
  year = "2025",
  url = "https://www.rnib.org.uk/about-us/",
  note = "[Online; accessed 6-September-2025]"
}

@misc{afabbraille,
  author = {Tim Connell},
  title = "The Challenge of Assistive Technology and Braille Literacy",
  year = "2008",
  url = "https://www.afb.org/aw/9/1/14277",
  note = "[Online; accessed 6-September-2025]"
}

@inproceedings{das2024provenance,
  title={From provenance to aberrations: Image creator and screen reader user perspectives on alt text for AI-generated images},
  author={Das, Maitraye and Fiannaca, Alexander J and Morris, Meredith Ringel and Kane, Shaun K and Bennett, Cynthia L},
  booktitle={Proceedings of the 2024 CHI Conference on Human Factors in Computing Systems},
  pages={1--21},
  year={2024}
}

@inproceedings{wenzel2023can,
author = {Wenzel, Kimi and Devireddy, Nitya and Davison, Cam and Kaufman, Geoff},
title = {Can Voice Assistants Be Microaggressors? Cross-Race Psychological Responses to Failures of Automatic Speech Recognition},
year = {2023},
isbn = {9781450394215},
publisher = {Association for Computing Machinery},
address = {New York, NY, USA},
url = {https://doi.org/10.1145/3544548.3581357},
doi = {10.1145/3544548.3581357},
booktitle = {Proceedings of the 2023 CHI Conference on Human Factors in Computing Systems},
articleno = {109},
numpages = {14},
location = {Hamburg, Germany},
series = {CHI '23}
}

@book{wenger1998communities,
  author    = {Wenger, Etienne},
  title     = {Communities of Practice: Learning, Meaning, and Identity},
  year      = {1998},
  publisher = {Cambridge University Press}
}

@book{bogardus1942fundamentals,
  author    = {Bogardus, Emory S.},
  title     = {Fundamentals of Social Psychology},
  edition   = {3},
  year      = {1942},
  publisher = {D. Appleton-Century Company},
  address   = {New York and London}
}

@article{tambiah1967social, title={Social Change in Modern India. By M. N. Srinivas. University of California Press: Berkeley and Los Angeles, and Cambridge University Press: London, 1966. Pp. xv + 194, 40s.}, volume={1}, DOI={10.1017/S0026749X00002687}, number={4}, journal={Modern Asian Studies}, author={Tambiah, S. J.}, year={1967}, pages={404–405}}

@INPROCEEDINGS{farhadi2009describing,
  author={Farhadi, Ali and Endres, Ian and Hoiem, Derek and Forsyth, David},
  booktitle={2009 IEEE Conference on Computer Vision and Pattern Recognition}, 
  title={Describing objects by their attributes}, 
  year={2009},
  volume={},
  number={},
  pages={1778-1785},
  doi={10.1109/CVPR.2009.5206772}}

@inproceedings{huh2023genassist,
  title={GenAssist: Making image generation accessible},
  author={Huh, Mina and Peng, Yi-Hao and Pavel, Amy},
  booktitle={Proceedings of the 36th Annual ACM Symposium on User Interface Software and Technology},
  pages={1--17},
  year={2023}
}

@misc{hamna2025building,
      title={Building Benchmarks from the Ground Up: Community-Centered Evaluation of LLMs in Healthcare Chatbot Settings}, 
      author={Hamna and Gayatri Bhat and Sourabrata Mukherjee and Faisal Lalani and Evan Hadfield and Divya Siddarth and Kalika Bali and Sunayana Sitaram},
      year={2025},
      archivePrefix={arXiv},
      primaryClass={cs.CL},
      url={https://arxiv.org/abs/2509.24506}, 
}

@inproceedings{mack2024they,
  title={“They only care to show us the wheelchair”: disability representation in text-to-image AI models},
  author={Mack, Kelly Avery and Qadri, Rida and Denton, Remi and Kane, Shaun K and Bennett, Cynthia L},
  booktitle={Proceedings of the 2024 CHI Conference on Human Factors in Computing Systems},
  pages={1--23},
  year={2024}
}

@article{seth2024dosa,
  title={Dosa: A dataset of social artifacts from different indian geographical subcultures},
  author={Seth, Agrima and Ahuja, Sanchit and Bali, Kalika and Sitaram, Sunayana},
  journal={arXiv preprint arXiv:2403.14651},
  year={2024}
}

@article{jha2024visage,
  title={Visage: A global-scale analysis of visual stereotypes in text-to-image generation},
  author={Jha, Akshita and Prabhakaran, Vinodkumar and Denton, Remi and Laszlo, Sarah and Dave, Shachi and Qadri, Rida and Reddy, Chandan K and Dev, Sunipa},
  journal={arXiv preprint arXiv:2401.06310},
  year={2024}
}

@article{weidinger2023sociotechnical,
	title        = {Sociotechnical Safety Evaluation of Generative AI Systems},
	author       = {Weidinger, Laura and Rauh, Maribeth and Marchal, Nahema and Manzini, Arianna and Hendricks, Lisa Anne and Mateos-Garcia, Juan and Bergman, Stevie and Kay, Jackie and Griffin, Conor and Bariach, Ben and others},
	year         = 2023,
	journal      = {arXiv preprint arXiv:2310.11986}
}

@inproceedings{suresh2023kaliedoscope,
author = {Suresh, Harini and Shanmugam, Divya and Chen, Tiffany and Bryan, Annie G and D'Amour, Alexander and Guttag, John and Satyanarayan, Arvind},
title = {Kaleidoscope: Semantically-grounded, context-specific ML model evaluation},
year = {2023},
isbn = {9781450394215},
publisher = {Association for Computing Machinery},
address = {New York, NY, USA},
url = {https://doi.org/10.1145/3544548.3581482},
doi = {10.1145/3544548.3581482},
booktitle = {Proceedings of the 2023 CHI Conference on Human Factors in Computing Systems},
articleno = {775},
numpages = {13},
location = {Hamburg, Germany},
series = {CHI '23}
}

@inproceedings{hall2024towards,
author = {Hall, Melissa and Bell, Samuel J. and Ross, Candace and Williams, Adina and Drozdzal, Michal and Soriano, Adriana Romero},
title = {Towards Geographic Inclusion in the Evaluation of Text-to-Image Models},
year = {2024},
isbn = {9798400704505},
publisher = {Association for Computing Machinery},
address = {New York, NY, USA},
url = {https://doi.org/10.1145/3630106.3658927},
doi = {10.1145/3630106.3658927},
booktitle = {Proceedings of the 2024 ACM Conference on Fairness, Accountability, and Transparency},
pages = {585–601},
numpages = {17},
location = {Rio de Janeiro, Brazil},
series = {FAccT '24}
}

@inproceedings{kapania2025examining,
      title={Examining the Expanding Role of Synthetic Data Throughout the AI Development Pipeline}, 
      author={Shivani Kapania and Stephanie Ballard and Alex Kessler and Jennifer Wortman Vaughan},
      year={2025},
      booktitle = {Proceedings of the 2025 ACM Conference on Fairness, Accountability, and Transparency}
}

@misc{johnson2025position,
title	= {Position: To Make Text-to-Image Models that Work for Marginalized Communities, We Need New Measurement Practices for the Long Tail},author	= {Nari Johnson and Hamna Abid and Deepthi Sudharsan and Theo Holroyd and Samantha Dalal and {Siobhan Mackenzie Hall} and {Jennifer Wortman Vaughan} and Daniela Massiceti and Cecily Morrison},year	= {2025},url={https://www.microsoft.com/en-us/research/publication/position-to-make-text-to-image-models-that-work-for-marginalized-communities-we-need-new-measurement-practices-for-the-long-tail/}}

@misc{sloane2020participation,
      title={Participation is not a Design Fix for Machine Learning}, 
      author={Mona Sloane and Emanuel Moss and Olaitan Awomolo and Laura Forlano},
      year={2020},
      archivePrefix={arXiv},
      primaryClass={cs.CY},
      url={https://arxiv.org/abs/2007.02423}, 
}

@inproceedings{massiceti2024explaining,
  title={Explaining CLIP's performance disparities on data from blind/low vision users},
  author={Massiceti, Daniela and Longden, Camilla and Slowik, Agnieszka and Wills, Samuel and Grayson, Martin and Morrison, Cecily},
  booktitle={Proceedings of the IEEE/CVF Conference on Computer Vision and Pattern Recognition},
  pages={12172--12182},
  year={2024}
}

@article{gautam2024melting,
  title={{From melting pots to misrepresentations: Exploring harms in Generative AI}},
  author={Gautam, Sanjana and Venkit, Pranav Narayanan and Ghosh, Sourojit},
  journal={arXiv preprint arXiv:2403.10776},
  year={2024}
}

@misc{harvey2025understanding,
      title={Understanding and Meeting Practitioner Needs When Measuring Representational Harms Caused by LLM-Based Systems}, 
      author={Emma Harvey and Emily Sheng and Su Lin Blodgett and Alexandra Chouldechova and Jean Garcia-Gathright and Alexandra Olteanu and Hanna Wallach},
      year={2025},
      archivePrefix={arXiv},
      primaryClass={cs.CY},
      url={https://arxiv.org/abs/2506.04482}, 
}

@inproceedings{jiang2023art,
author = {Jiang, Harry H. and Brown, Lauren and Cheng, Jessica and Khan, Mehtab and Gupta, Abhishek and Workman, Deja and Hanna, Alex and Flowers, Johnathan and Gebru, Timnit},
title = {AI Art and its Impact on Artists},
year = {2023},
isbn = {9798400702310},
publisher = {Association for Computing Machinery},
address = {New York, NY, USA},
url = {https://doi.org/10.1145/3600211.3604681},
doi = {10.1145/3600211.3604681},
booktitle = {Proceedings of the 2023 AAAI/ACM Conference on AI, Ethics, and Society},
pages = {363–374},
numpages = {12},
location = {Montr\'{e}al, QC, Canada},
series = {AIES '23}
}

@article{dehouche2023whats,
title = {What’s in a text-to-image prompt? The potential of stable diffusion in visual arts education},
journal = {Heliyon},
volume = {9},
number = {6},
pages = {e16757},
year = {2023},
issn = {2405-8440},
doi = {https://doi.org/10.1016/j.heliyon.2023.e16757},
url = {https://www.sciencedirect.com/science/article/pii/S2405844023039646},
author = {Nassim Dehouche and Kullathida Dehouche}
}

@misc{exner2025ai,
author = {Exner, Yannick and Hartmann, Jochen and Netzer, Oded and Zhang, Shunyuan},
year = {2025},
month = {01},
pages = {},
title = {AI in Disguise - How AI-Generated Ads' Visual Cues Shape Consumer Perception and Performance},
doi = {10.2139/ssrn.5096969}
}

@misc{chen2025multimodal,
      title={Multi-Modal Language Models as Text-to-Image Model Evaluators}, 
      author={Jiahui Chen and Candace Ross and Reyhane Askari-Hemmat and Koustuv Sinha and Melissa Hall and Michal Drozdzal and Adriana Romero-Soriano},
      year={2025},
      archivePrefix={arXiv},
      primaryClass={cs.CV},
      url={https://arxiv.org/abs/2505.00759}, 
}

@misc{szymanski2024limitations,
      title={Limitations of the LLM-as-a-Judge Approach for Evaluating LLM Outputs in Expert Knowledge Tasks}, 
      author={Annalisa Szymanski and Noah Ziems and Heather A. Eicher-Miller and Toby Jia-Jun Li and Meng Jiang and Ronald A. Metoyer},
      year={2024},
      archivePrefix={arXiv},
      primaryClass={cs.HC},
      url={https://arxiv.org/abs/2410.20266}, 
}

@online{kawakami2025translation,
  author =       {Anna Kawakami and Su Lin Blodgett and Solon Barocas and Alex Chouldechova and Abigail Jacobs and Emily Sheng and Jenn Wortman Vaughan and Hanna Wallach and Amy Winecoff and Angelina Wang and Haiyi Zhu and Ken Holstein},
  year =         {2025},
  title =        {Translation Tutorial: AI Measurement as a Stakeholder-Engaged Design Practice},
  url =          {https://drive.google.com/file/d/12qQd6ROfacYAtoQ-iihgQxslNSIKF1Pu/view},
  lastaccessed = {January 10, 2026},
}

@Article{biega2025towards,
  author =	{Biega, Asia and Born, Georgina and Diaz, Fernando and Gray, Mary L. and Qadri, Rida},
  title =	{{Towards a Multidisciplinary Vision for Culturally Inclusive Generative AI (Dagstuhl Seminar 25022)}},
  pages =	{33--49},
  journal =	{Dagstuhl Reports},
  ISSN =	{2192-5283},
  year =	{2025},
  volume =	{15},
  number =	{1},
  editor =	{Biega, Asia and Born, Georgina and Diaz, Fernando and Gray, Mary L. and Qadri, Rida},
  publisher =	{Schloss Dagstuhl -- Leibniz-Zentrum f{\"u}r Informatik},
  address =	{Dagstuhl, Germany},
  URL =		{https://drops.dagstuhl.de/entities/document/10.4230/DagRep.15.1.33},
  URN =		{urn:nbn:de:0030-drops-236775},
  doi =		{10.4230/DagRep.15.1.33}
}

@inproceedings{gabreegziabher2024metricmate,
author = {Gebreegziabher, Simret Araya and Chiang, Charles and Wang, Zichu and Ashktorab, Zahra and Brachman, Michelle and Geyer, Werner and Li, Toby Jia-Jun and G\'{o}mez-Zar\'{a}, Diego},
title = {MetricMate: An Interactive Tool for Generating Evaluation Criteria for LLM-as-a-Judge Workflow},
year = {2025},
isbn = {9798400713842},
publisher = {Association for Computing Machinery},
address = {New York, NY, USA},
url = {https://doi.org/10.1145/3729176.3729199},
doi = {10.1145/3729176.3729199},
booktitle = {Proceedings of the 4th Annual Symposium on Human-Computer Interaction for Work},
articleno = {22},
numpages = {18},
location = {
},
series = {CHIWORK '25}
}

@inproceedings{hashemi2024rubric,
    title = "{LLM}-Rubric: A Multidimensional, Calibrated Approach to Automated Evaluation of Natural Language Texts",
    author = "Hashemi, Helia  and
      Eisner, Jason  and
      Rosset, Corby  and
      Van Durme, Benjamin  and
      Kedzie, Chris",
    editor = "Ku, Lun-Wei  and
      Martins, Andre  and
      Srikumar, Vivek",
    booktitle = "Proceedings of the 62nd Annual Meeting of the Association for Computational Linguistics (Volume 1: Long Papers)",
    month = aug,
    year = "2024",
    address = "Bangkok, Thailand",
    publisher = "Association for Computational Linguistics",
    url = "https://aclanthology.org/2024.acl-long.745/",
    doi = "10.18653/v1/2024.acl-long.745",
    pages = "13806--13834"
}

@misc{szymanski2024comparing,
      title={Comparing Criteria Development Across Domain Experts, Lay Users, and Models in Large Language Model Evaluation}, 
      author={Annalisa Szymanski and Simret Araya Gebreegziabher and Oghenemaro Anuyah and Ronald A. Metoyer and Toby Jia-Jun Li},
      year={2024},
      archivePrefix={arXiv},
      primaryClass={cs.HC},
      url={https://arxiv.org/abs/2410.02054}, 
}

@article{hsu2007delphi,
  title={The Delphi technique: Making sense of consensus},
  author={Hsu, Chien-Chi and Sandford, Brian A.},
  journal={Practical Assessment, Research, and Evaluation},
  year={2007},
  volume={12},
  number={10},
  pages={1--8},
  note={A widely cited methodological overview of the Delphi method},
  url={https://openpublishing.library.umass.edu/pare/article/id/1418/}
}

@inproceedings{pan2024human,
    title = "Human-Centered Design Recommendations for {LLM}-as-a-judge",
    author = "Pan, Qian  and
      Ashktorab, Zahra  and
      Desmond, Michael  and
      Santill{\'a}n Cooper, Mart{\'i}n  and
      Johnson, James  and
      Nair, Rahul  and
      Daly, Elizabeth  and
      Geyer, Werner",
    editor = "Soni, Nikita  and
      Flek, Lucie  and
      Sharma, Ashish  and
      Yang, Diyi  and
      Hooker, Sara  and
      Schwartz, H. Andrew",
    booktitle = "Proceedings of the 1st Human-Centered Large Language Modeling Workshop",
    month = aug,
    year = "2024",
    address = "TBD",
    publisher = "ACL",
    url = "https://aclanthology.org/2024.hucllm-1.2/",
    doi = "10.18653/v1/2024.hucllm-1.2",
    pages = "16--29"
}

@misc{li2025generation,
      title={From Generation to Judgment: Opportunities and Challenges of LLM-as-a-judge}, 
      author={Dawei Li and Bohan Jiang and Liangjie Huang and Alimohammad Beigi and Chengshuai Zhao and Zhen Tan and Amrita Bhattacharjee and Yuxuan Jiang and Canyu Chen and Tianhao Wu and Kai Shu and Lu Cheng and Huan Liu},
      year={2025},
      archivePrefix={arXiv},
      primaryClass={cs.AI},
      url={https://arxiv.org/abs/2411.16594}, 
}

@misc{chehbouni2025valid,
      title={Neither Valid nor Reliable? Investigating the Use of LLMs as Judges}, 
      author={Khaoula Chehbouni and Mohammed Haddou and Jackie Chi Kit Cheung and Golnoosh Farnadi},
      year={2025},
      archivePrefix={arXiv},
      primaryClass={cs.CL},
      url={https://arxiv.org/abs/2508.18076}, 
}

@misc{openai2024gpt4o,
      title={GPT-4o System Card}, 
      author={{OpenAI and others}},
      year={2024},
      archivePrefix={arXiv},
      primaryClass={cs.CL},
      url={https://arxiv.org/abs/2410.21276}, 
}

@misc{dalal2024provocation,
      title={Provocation: Who benefits from "inclusion" in Generative AI?}, 
      author={Samantha Dalal and Siobhan Mackenzie Hall and Nari Johnson},
      year={2024},
      archivePrefix={arXiv},
      primaryClass={cs.CY},
      url={https://arxiv.org/abs/2411.09102}, 
}

@article{zhang2025aura,
author = {Zhang, Alice Qian and Amores, Judith and Shen, Hong and Czerwinski, Mary and Gray, Mary L. and Suh, Jina},
title = {AURA: Amplifying Understanding, Resilience, and Awareness for Responsible AI Content Work},
year = {2025},
issue_date = {May 2025},
publisher = {Association for Computing Machinery},
address = {New York, NY, USA},
volume = {9},
number = {2},
url = {https://doi.org/10.1145/3710931},
doi = {10.1145/3710931},
journal = {Proc. ACM Hum.-Comput. Interact.},
month = may,
articleno = {CSCW033},
numpages = {45}
}

@inproceedings{bennett2025toward,
author = {Bennett, Cynthia L. and Kane, Shaun K. and Harrington, Christina N.},
title = {Toward Community-Led Evaluations of Text-to-Image AI Representations of Disability, Health, and Accessibility},
year = {2025},
isbn = {9798400721403},
publisher = {Association for Computing Machinery},
address = {New York, NY, USA},
url = {https://doi.org/10.1145/3757887.3763012},
doi = {10.1145/3757887.3763012},
booktitle = {Proceedings of the 5th ACM Conference on Equity and Access in Algorithms, Mechanisms, and Optimization},
pages = {256–270},
numpages = {15},
location = {
},
series = {EAAMO '25}
}

@misc{zhang2023gpt4v,
      title={GPT-4V(ision) as a Generalist Evaluator for Vision-Language Tasks}, 
      author={Xinlu Zhang and Yujie Lu and Weizhi Wang and An Yan and Jun Yan and Lianke Qin and Heng Wang and Xifeng Yan and William Yang Wang and Linda Ruth Petzold},
      year={2023},
      archivePrefix={arXiv},
      primaryClass={cs.CV},
      url={https://arxiv.org/abs/2311.01361}, 
}

@misc{zheng2023judging,
      title={Judging LLM-as-a-Judge with MT-Bench and Chatbot Arena}, 
      author={Lianmin Zheng and Wei-Lin Chiang and Ying Sheng and Siyuan Zhuang and Zhanghao Wu and Yonghao Zhuang and Zi Lin and Zhuohan Li and Dacheng Li and Eric P. Xing and Hao Zhang and Joseph E. Gonzalez and Ion Stoica},
      year={2023},
      archivePrefix={arXiv},
      primaryClass={cs.CL},
      url={https://arxiv.org/abs/2306.05685}, 
}

@inproceedings{otani2023verifiable,
  title={{Toward Verifiable and Reproducible Human Evaluation for Text-to-Image Generation}},
  author={Otani, Mayu and Togashi, Riku and Sawai, Yu and Ishigami, Ryosuke and Nakashima, Yuta and Rahtu, Esa and Heikkil{\"a}, Janne and Satoh, Shin’ichi},
  booktitle={Proceedings of the IEEE/CVF Conference on Computer Vision and Pattern Recognition},
  pages={14277--14286},
  year={2023}
}

@article{hall2025human,
  title={The Human Labour of Data Work: Capturing Cultural Diversity through World Wide Dishes},
  author={Hall, Siobhan Mackenzie and Dalal, Samantha and Sefala, Raesetje and Yuehgoh, Foutse and Alaagib, Aisha and Hamzaoui, Imane and Ishida, Shu and Magomere, Jabez and Crais, Lauren and Salama, Aya and others},
  journal={arXiv preprint arXiv:2502.05961},
  year={2025}
}

@inproceedings{hong2024s,
  title={Who's in and who's out? A case study of multimodal CLIP-filtering in DataComp},
  author={Hong, Rachel and Agnew, William and Kohno, Tadayoshi and Morgenstern, Jamie},
  booktitle={Proceedings of the 4th ACM Conference on Equity and Access in Algorithms, Mechanisms, and Optimization},
  pages={1--17},
  year={2024}
}

@inproceedings{magomere2025world,
  title={The World Wide recipe: A community-centred framework for fine-grained data collection and regional bias operationalisation},
  author={Magomere, Jabez and Ishida, Shu and Afonja, Tejumade and Salama, Aya and Kochin, Daniel and Foutse, Yuehgoh and Hamzaoui, Imane and Sefala, Raesetje and Alaagib, Aisha and Dalal, Samantha and others},
  booktitle={Proceedings of the 2025 ACM Conference on Fairness, Accountability, and Transparency},
  pages={246--282},
  year={2025}
}

@inproceedings{wallach2025position,
  title={Position: Evaluating generative ai systems is a social science measurement challenge},
  author={Hanna Wallach and Meera Desai and A. Feder Cooper and Angelina Wang and Chad Atalla and Solon Barocas and Su Lin Blodgett and Alexandra Chouldechova and Emily Corvi and P. Alex Dow and Jean Garcia-Gathright and Alexandra Olteanu and Nicholas Pangakis and Stefanie Reed and Emily Sheng and Dan Vann and Jennifer Wortman Vaughan and Matthew Vogel and Hannah Washington and Abigail Z. Jacobs},
  booktitle = {Proceedings of the 42nd International Conference on Machine Learning (ICML)},
  year={2025}
}

@inproceedings{corvi-etal-2025-taxonomizing,
    title = "Taxonomizing Representational Harms using Speech Act Theory",
    author = "Corvi, Emily  and
      Washington, Hannah  and
      Reed, Stefanie  and
      Atalla, Chad  and
      Chouldechova, Alexandra  and
      Dow, P. Alex  and
      Garcia-Gathright, Jean  and
      Pangakis, Nicholas J  and
      Sheng, Emily  and
      Vann, Dan  and
      Vogel, Matthew  and
      Wallach, Hanna",
    booktitle = "Findings of the Association for Computational Linguistics",
    year = "2025",
    url = "https://aclanthology.org/2025.findings-acl.202/",
    doi = "10.18653/v1/2025.findings-acl.202",
}

@inproceedings{viswanathan2025interaction,
author = {Viswanathan, Sruthi and Ibrahim, Seray and Shankar, Ravi and Binns, Reuben and Van Kleek, Max and Slovak, Petr},
title = {The Interaction Layer: An Exploration for Co-Designing User-LLM Interactions in Parental Wellbeing Support Systems},
year = {2025},
isbn = {9798400713941},
publisher = {Association for Computing Machinery},
address = {New York, NY, USA},
url = {https://doi.org/10.1145/3706598.3714088},
doi = {10.1145/3706598.3714088},
booktitle = {Proceedings of the 2025 CHI Conference on Human Factors in Computing Systems},
articleno = {310},
numpages = {25},
location = {
},
series = {CHI '25}
}

@inproceedings{tseng2025ownership,
author = {Tseng, Emily and Young, Meg and Le Qu\'{e}r\'{e}, Marianne Aubin and Rinehart, Aimee and Suresh, Harini},
title = {"Ownership, Not Just Happy Talk": Co-Designing a Participatory Large Language Model for Journalism},
year = {2025},
isbn = {9798400714825},
publisher = {Association for Computing Machinery},
address = {New York, NY, USA},
url = {https://doi.org/10.1145/3715275.3732198},
doi = {10.1145/3715275.3732198},
booktitle = {Proceedings of the 2025 ACM Conference on Fairness, Accountability, and Transparency},
pages = {3119–3130},
numpages = {12},
location = {
},
series = {FAccT '25}
}

@inproceedings{nguyen2026validating,
title = {Validating and Refining Generative AI Evaluations via Stakeholder Engagement},
author = {Tonya Nguyen and Jean Garcia-Gathright and Hannah Washington and Alexandra Chouldechova and Hanna Wallach and Jennifer Wortman Vaughan},
booktitle = {Proceedings of the 9th ACM Conference on Fairness, Accountability, and Transparency},
year = {2026}
}

@inbook{egede2025exploring,
author = {Egede, Lisa},
title = {Exploring Black Communities’ Perceptions and Design Approaches for Building Culturally Tailored AI Systems},
year = {2025},
isbn = {9798400714863},
publisher = {Association for Computing Machinery},
address = {New York, NY, USA},
url = {https://doi.org/10.1145/3715668.3735629},
booktitle = {Companion Publication of the 2025 ACM Designing Interactive Systems Conference},
pages = {72–76},
numpages = {5}
}

@inproceedings{taylor2024cruising,
author = {Taylor, Jordan and Simpson, Ellen and Tran, Anh-Ton and Brubaker, Jed R. and Fox, Sarah E and Zhu, Haiyi},
title = {Cruising Queer HCI on the DL: A Literature Review of LGBTQ+ People in HCI},
year = {2024},
isbn = {9798400703300},
publisher = {Association for Computing Machinery},
address = {New York, NY, USA},
url = {https://doi.org/10.1145/3613904.3642494},
doi = {10.1145/3613904.3642494},
booktitle = {Proceedings of the 2024 CHI Conference on Human Factors in Computing Systems},
articleno = {507},
numpages = {21},
location = {Honolulu, HI, USA},
series = {CHI '24}
}

@misc{adilazuarda2024measuring,
      title={Towards Measuring and Modeling "Culture" in LLMs: A Survey}, 
      author={Muhammad Farid Adilazuarda and Sagnik Mukherjee and Pradhyumna Lavania and Siddhant Singh and Alham Fikri Aji and Jacki O'Neill and Ashutosh Modi and Monojit Choudhury},
      year={2024},
      archivePrefix={arXiv},
      primaryClass={cs.CY},
      url={https://arxiv.org/abs/2403.15412}, 
}

@article{zhou2025culture,
  title={Culture is not trivia: Sociocultural theory for cultural nlp},
  author={Zhou, Naitian and Bamman, David and Bleaman, Isaac L},
  journal={arXiv preprint arXiv:2502.12057},
  year={2025}
}

@misc{romero2024cvqa,
      title={CVQA: Culturally-diverse Multilingual Visual Question Answering Benchmark}, 
      author={David Romero and Chenyang Lyu and Haryo Akbarianto Wibowo and Teresa Lynn and Injy Hamed and Aditya Nanda Kishore and Aishik Mandal and Alina Dragonetti and Artem Abzaliev and Atnafu Lambebo Tonja and Bontu Fufa Balcha and Chenxi Whitehouse and Christian Salamea and Dan John Velasco and David Ifeoluwa Adelani and David Le Meur and Emilio Villa-Cueva and Fajri Koto and Fauzan Farooqui and Frederico Belcavello and Ganzorig Batnasan and Gisela Vallejo and Grainne Caulfield and Guido Ivetta and Haiyue Song and Henok Biadglign Ademtew and Hernán Maina and Holy Lovenia and Israel Abebe Azime and Jan Christian Blaise Cruz and Jay Gala and Jiahui Geng and Jesus-German Ortiz-Barajas and Jinheon Baek and Jocelyn Dunstan and Laura Alonso Alemany and Kumaranage Ravindu Yasas Nagasinghe and Luciana Benotti and Luis Fernando D'Haro and Marcelo Viridiano and Marcos Estecha-Garitagoitia and Maria Camila Buitrago Cabrera and Mario Rodríguez-Cantelar and Mélanie Jouitteau and Mihail Mihaylov and Mohamed Fazli Mohamed Imam and Muhammad Farid Adilazuarda and Munkhjargal Gochoo and Munkh-Erdene Otgonbold and Naome Etori and Olivier Niyomugisha and Paula Mónica Silva and Pranjal Chitale and Raj Dabre and Rendi Chevi and Ruochen Zhang and Ryandito Diandaru and Samuel Cahyawijaya and Santiago Góngora and Soyeong Jeong and Sukannya Purkayastha and Tatsuki Kuribayashi and Teresa Clifford and Thanmay Jayakumar and Tiago Timponi Torrent and Toqeer Ehsan and Vladimir Araujo and Yova Kementchedjhieva and Zara Burzo and Zheng Wei Lim and Zheng Xin Yong and Oana Ignat and Joan Nwatu and Rada Mihalcea and Thamar Solorio and Alham Fikri Aji},
      year={2024},
      archivePrefix={arXiv},
      primaryClass={cs.CV},
      url={https://arxiv.org/abs/2406.05967}, 
}

@inproceedings{suresh2024participation,
  title={Participation in the Age of Foundation Models},
  author={Suresh, Harini and Tseng, Emily and Young, Meg and Gray, Mary and Pierson, Emma and Levy, Karen},
  booktitle={The 2024 ACM Conference on Fairness, Accountability, and Transparency},
  pages={1609--1621},
  year={2024}
}

@article{ParticipationScaleTensions,
	title        = {Participation versus Scale: Tensions in the Practical Demands on Participatory AI},
	author       = {Young, Meg and Ehsan, Uphol and Singh, Ranjit and Tafesse, Emnet and Gilman, Michele and Harrington, Christina and Metcalf, Jacob},
	year         = 2024,
	journal      = {First Monday},
	langid       = {english}
}

@book{walters2003all,
  title={All the rage: The story of gay visibility in America},
  author={Walters, Suzanna Danuta},
  year={2003},
  publisher={University of Chicago Press}
}

@inproceedings{chasalow2021representativeness,
  title={Representativeness in statistics, politics, and machine learning},
  author={Chasalow, Kyla and Levy, Karen},
  booktitle={Proceedings of the 2021 ACM Conference on Fairness, Accountability, and Transparency},
  pages={77--89},
  year={2021}
}

@article{adilazuarda2024culturesurvey,
	title        = {Towards Measuring and Modeling “Culture” in LLMs: A Survey},
	author       = {Adilazuarda, Muhammad Farid and Mukherjee, Sagnik and Lavania, Pradhyumna and Singh, Siddhant and Dwivedi, Ashutosh and Aji, Alham Fikri and O'Neill, Jacki and Modi, Ashutosh and Choudhury, Monojit},
	year         = 2024,
	journal      = {arXiv preprint arXiv:2403.15412}
}

@article{blake2000defining,
	title        = {On Defining the Cultural Heritage},
	author       = {Blake, Janet},
	year         = 2000,
	journal      = {International \& Comparative Law Quarterly},
	publisher    = {Cambridge University Press},
	volume       = 49,
	number       = 1,
	pages        = {61--85}
}

@article{kirk2024prism,
	title        = {The PRISM Alignment Project: What Participatory, Representative and Individualised Human Feedback Reveals About the Subjective and Multicultural Alignment of Large Language Models},
	author       = {Kirk, Hannah Rose and Whitefield, Alexander and R{\"o}ttger, Paul and Bean, Andrew and Margatina, Katerina and Ciro, Juan and Mosquera, Rafael and Bartolo, Max and Williams, Adina and He, He and others},
	year         = 2024,
	journal      = {arXiv preprint arXiv:2404.16019}
}

@article{winata2024worldcuisines,
	title        = {WorldCuisines: A Massive-Scale Benchmark for Multilingual and Multicultural Visual Question Answering on Global Cuisines},
	author       = {Winata, Genta Indra and Hudi, Frederikus and Irawan, Patrick Amadeus and Anugraha, David and Putri, Rifki Afina and Wang, Yutong and Nohejl, Adam and Prathama, Ubaidillah Ariq and Ousidhoum, Nedjma and Amriani, Afifa and others},
	year         = 2024,
	journal      = {arXiv preprint arXiv:2410.12705}
}

@article{bergman2024stela,
  title={STELA: a community-centred approach to norm elicitation for AI alignment},
  author={Bergman, Stevie and Marchal, Nahema and Mellor, John and Mohamed, Shakir and Gabriel, Iason and Isaac, William},
  journal={Scientific Reports},
  volume={14},
  number={1},
  pages={6616},
  year={2024},
  publisher={Nature Publishing Group UK London}
}

@article{Indianculturalmosaic2009,
author = {Singh, Devinder and Sharma, Manoj},
year = {2009},
month = {01},
title = {Unfolding the Indian cultural mosaic: a cross-cultural study of four regional cultures},
volume = {2},
journal = {International Journal of Indian Culture and Business Management - Int J Indian Cult Bus Manag},
doi = {10.1504/IJICBM.2009.023547}
}

@article{matias2025how,
author = {J. Nathan Matias  and Megan Price },
title = {How public involvement can improve the science of AI},
journal = {Proceedings of the National Academy of Sciences},
volume = {122},
number = {48},
pages = {e2421111122},
year = {2025},
doi = {10.1073/pnas.2421111122},
URL = {https://www.pnas.org/doi/abs/10.1073/pnas.2421111122}}

\newpage
\appendix
\clearpage
\section*{Appendix}
\addcontentsline{toc}{section}{Appendix}

This appendix contains supplementary materials for our study. We include the full evaluation rubrics for each cultural artifact (Appendix \ref{apdx:systematization-results}), our complete protocols and other study materials (Appendix \ref{apdx:complete-protocols}), and supplemental experimental results (Appendix \ref{apdx:results}). 

\startcontents[appendices]

\printcontents[appendices]{}{1}{\subsection*{Appendix Contents}}

\clearpage

\section{Complete set of evaluation rubrics}\label{apdx:systematization-results}

As described in the main text, the systematization of cultural appropriateness for each artifact is organized around binary (Y/N) criteria.
Each criteria specifies a condition that must be met in order for a depiction of the artifact to be culturally appropriate.
We say that an output is culturally appropriate if all of the criteria are satisfied.  
While the specific criteria vary by artifact, they are organized under themes that are shared across each community. 

\begin{table}[!ht]
    \centering
    \begin{tabular}{ p{4cm} | p{5cm} | p{5cm}}

     \textit{Theme}                                                                                                    & \textit{Criteria for \textbf{guide cane}} &  \textit{Criteria for \textbf{braille notetaker}}                                                                             \\ 
    \hline

\textbf{Theme 1}. The object needs to be functional as an assistive technology, and usable by someone who is blind. &  \specialcell[t]{\textbf{C1}. No deformed canes.\\ \textbf{C2}. No curved (crooked) handles. \\ \textbf{C3}. The cane must be shaped like a long (5 foot) stick. \\ \textbf{C4}. The body must have sections that are a white color.  \\ \textbf{C5}. There must be a tip at the bottom of the cane.  } & \specialcell[t]{\textbf{C1}. The device must be shaped like a thin rectangular box.\\ \\
The device must have a valid braille output, to read: \\ \textbf{C2}. The device must show braille. \\ \textbf{C3}. All depictions of braille must be tactile (embossed). No depictions of braille on electronic screens. \\\textbf{C4}. Depictions of braille must be valid: arranged in cells with 3 or 4 rows, and 2 columns. \\ \\ The device must have a valid braille input, to write: \\\textbf{C5}. The device can have a qwerty keyboard, or a braille keyboard. A braille keyboard must have 3 or 4 keys (right), a space bar, and then 3 or 4 keys (left). These keys are positioned next to each other in a straight horizontal line. \\ } \\ \hline

\textbf{Theme 2}. The object in the image should not be confused with other, more hegemonic objects, such as objects that are used predominantly by people who are sighted. &  \specialcell[t]{\textbf{C1}. No wooden walking sticks.\\ \textbf{C2}. No decorative striped patterns (candy canes).} & \specialcell[t]{\textbf{C1}. No depictions of notetaking as writing using a pen on paper.\\ \textbf{C2}. No devices that are shaped like laptops with an electronic screen output. \\ \textbf{C3}. No devices that are shaped like handheld calculators with an electronic screen output. \\\textbf{C4}. No devices that are shaped like manual typewriters. \\}  \\ \hline
\end{tabular}
\caption{\textbf{Community-informed evaluation rubrics to score AI images of a guide cane and braille notetaker.} The rubric criteria are organized under two themes. An image is defined to be culturally appropriate if \emph{all} of the criteria are met, and culturally inappropriate otherwise.}
\label{table:systematization-blv}
\end{table}

\begin{table}[H]
    \centering
    \small
    \scalebox{0.80}{
    \begin{tabular}{ p{3cm} | p{3.5cm} | p{3.5cm} | p{3.5cm} | p{3.5cm}}
     \textit{\textbf{Theme}}                &
     \textit{\textbf{Chundan Vallam}} &
     \textit{\textbf{Mridangam}} &
     \textit{\textbf{Pallanguzhi}} &
     \textit{\textbf{Kasavu saree}} \\
    \hline

\textbf{Theme 1}. The artifact must retain a culturally recognizable physical structure and must not resemble objects that are popular or associated with unrelated traditions or contexts. & 
\wrapcell[t]{\textbf{C1}. It must not resemble other passenger boats (like Kerala house boats, Chinese dragon boats, long-tailed Thai boats or ferry boats).\\
\textbf{C2}. The boat must be long and narrow.\\
\textbf{C3}. The bow of the boat must be a plain wooden extension without decorative structures.\\
\textbf{C4}. The stern of the boat must be a straight pointed tip angled slightly upward.} &
\wrapcell[t]{\textbf{C1}. It must not resemble other percussion instruments (like Tabla, Drum, Damaru, Dhol).\\
\textbf{C2}. The instrument must be long, barrel-shaped, and tapered at both ends, each ending in a rounded, double-headed form, with one end slightly larger than the other.\\
\textbf{C3}. The body of the instrument must be made out of jackwood.\\
\textbf{C4}. There must not be intricate design or detailed patterns on the body.} &

\wrapcell[t]{\textbf{C1}. It must not resemble other board games (like Monopoly, Tic Tac toe, etc.).\\
\textbf{C2}.The game board must be symmetrical along the length and consist of two or three rows of pits. The rows should have at least 5 pits.\\
\textbf{C3}. The game board must be fish or rectangular in shape. \\
\textbf{C4}. The game board must be made out of teakwood.\\
\textbf{C5}. The pits must be circular and evenly spaced.} &
\wrapcell[t]{\textbf{C1}. It must not resemble other items like a tablecloth, Kerala Mundu or curtains.\\
\textbf{C2}. The saree color must be off-white with a medium width (3--5 inch) woven gold border.\\
\textbf{C3}. The saree must be made of crisp cotton fabric throughout.\\
\textbf{C4}. The saree must not contain heavy embellishments.} \\
\hline

\textbf{Theme 2}. The artifact must preserve its intended functional or performative purpose as understood within its cultural context (e.g., as a game, garment, or tool). &
\wrapcell[t]{\textbf{C1}. Oarsmen must sit in pairs along the length of the boat. If no oarsmen are present, consider the criteria as met.\\
\textbf{C2}. Each oarsman must use only a single paddle. If no oarsmen are present, consider the criteria as met.\\
\textbf{C3}. The paddle must be longer and angled downward toward the water. If no oarsmen are present, consider the criteria as met.\\
\textbf{C4}. One person must be standing at the bow or centre position of the boat. If no oarsmen are present, consider the criteria as met.} &
\wrapcell[t]{\textbf{C1}. The heads of the instrument must be stretched goat, cow or buffalo skin.\\
\textbf{C2}. A black circular membrane must be present in the middle of both heads and must be slightly raised from the stretched skin surfaces.\\
\textbf{C3}. The black circular membrane on the smaller end must be slightly smaller than the one on the larger end.\\
\textbf{C4}. The instrument must have longitudinal leather straps lacing along its body connecting the two heads under high tension.} &
\wrapcell[t]{\textbf{C1}. The size of the tokens must not be too small. The tokens should be distributable by hand.\\
\textbf{C2}. The pits must be big enough to accommodate multiple tokens.} &
\wrapcell[t]{\textbf{C1}. The saree must be shown in a way that clearly presents its pleats and drape.} \\
\hline

\textbf{Theme 3}. The artifact should follow culturally appropriate placement or arrangement,x as practiced in traditional usage. &
\wrapcell[t]{\textbf{C1}. The oarsmen must be seated facing the stern. If no oarsmen are present, consider the criteria as met.\\
\textbf{C2}. Oarsmen must wear the same attire, typically a white traditional Kerala mundu without upper garments. If no oarsmen are present, consider the criteria as met.} &
\wrapcell[t]{\textbf{C1}. The orientation and positioning of the instrument must be horizontal, lying on its length.} &
\wrapcell[t]{\textbf{C1}. The tokens can be cowrie shells or tamarind seeds.} &
\wrapcell[t]{\emph{None}} \\
\hline
\end{tabular}}
\caption{\textbf{Community-informed evaluation rubrics to score AI images of a Chundan Vallam, Mridangam, Pallanguzhi, and Kasavu saree.} The rubric criteria are organized under three themes that are shared across the six Indian artifacts. An image is defined to be culturally appropriate if \emph{all} of the criteria are met, and culturally inappropriate otherwise.}
\label{table:systematization-indian}
\end{table}

\newpage
\section{Community workshop methodologies \& study materials}\label{apdx:complete-protocols}

In this section, we present extended methodological details for both the blind and low vision (Section \ref{apdx:protocol-blv}) and Indian (Section \ref{apdx:protocl-india}) community engagements.

\subsection{Summary of differences between protocols}\label{apdx:protocol-contrast}

Our three community engagements adopt different methods to engage community members in the process of systematization.
We conducted our study in two phases, where we adopted a shared methodology to engage both residents of Tamil Nadu and Kerala, and a different methodology to engage BLV community members in the UK. 
These different methodologies reflect different cultural contexts and best practices in creating access for each community.
Each engagement was conducted and facilitated by different members of our research team, who also experimented with small adaptations when implementing our shared protocol.
Table \ref{tab:workshop-differences} summarizes notable differences in how our study methodology differed across contexts. 
The table provides a brief justification for why we made each methodological decision.

\subsection{Extended study protocol: Assistive technologies used by the blind and low vision community}\label{apdx:protocol-blv}

\subsubsection{Creating access to images}\label{apdx:blv-accessibility}

To elicit blind and low vision community members' preferences about AI images, we needed to determine how to facilitate non-visual access to these images.  First, following a best practice from past work asking blind participants to evaluate AI images~\cite{huh2023genassist}, we created alt text for each image. To encourage consistency in the amount of detail provided for each image, a blind community member on our research team created a template of important characteristics to describe (\eg{} for each image of a guide cane, we always described its shape, material, and color).  We include example images and their alt text descriptions below.

We additionally drew on past scholarship on \emph{cross-ability collaborative work} in which a blind user and sighted partner work together to complete a task~\cite{das2019doesnt,braham2015invisible,muehlbradt2022whats,bennett2018interdependence}. While sighted strangers may misunderstand the access needs of their collaborators ~\cite{braham2015invisible}, recent studies have adopted cross-ability protocols between participants who already know each other well and have established trust and comfort working together~\cite{muehlbradt2022whats}.  We follow calls from \citet{bennett2018interdependence} to understand blind community members not as passive recipients of assistance, but to instead recognize their expertise in creating access throughout the collaboration.  When collaborating, we emphasized the unique skills of each participant: the blind community member as the expert on how the selected artifacts worked and how they would like their community to be represented, and their sighted collaborators as capable of providing a visual perspective on what is shown in images.
When responding to images, the facilitator invited participants to further discuss what is shown in each image as a pair ~\cite{muehlbradt2022whats}.

\subsubsection{Recruitment \& Workshop Activities} 
To recruit blind and low vision individuals who currently reside in the UK, we adopted a purposive sampling approach ~\cite{palinkas2012purposeful}.  We recruited participants from two email lists: an internal list of blind and low vision community members who had consented to receive information about future studies at our institution, and an open list for blind and low vision technology users in the UK.  We asked each blind or low vision participant to invite a sighted partner of their choice to the study, following ~\citet{muehlbradt2022whats}.  Relationships between community members and their partners included friends, partners, siblings, and children.  One blind community member invited a friend who is visually impaired to participate as their buddy. More information about participants is in Table~\ref{table:blv-participants}.

\begin{table}[H]
\small
\centering
\begin{tabular}{p{4.2cm} | p{5cm}  | p{5cm}}

Difference between methods & Methodology for blind and low vision in the UK & Methodology for South Indian states \\ \hline 

Prompts used to generate images & We used simple prompt templates (\eg{} ``a photo of a guide cane''), as described in Appendix \ref{apdx:blv-prompt-templates}. & Simple prompt templates that used the transliterated name for each artifact resulted in representations that were completely and totally unrelated. We used revised prompts from \dalle{} 3 that included a detailed English description of each artifact, as described in Appendix \ref{apdx:india-revised-prompts}. \\ \hline 
Composition of the generated images & Generated images that depicted artifacts in isolation (\eg{} floating in an abstract liminal space). Alt text descriptions provided to participants did not include descriptions of the surrounding scene. Discussions focused on the object in isolation. & Generated images that showed artifacts in more complex and realistic scenes, such as drums resting on the ground, or boats racing in a river. Participants who responded to images often commented on the broader scene in which the artifact appeared. \\ \hline 
Number of community members participating in each workshop & Workshops were scheduled individually with blind and low vision community members, who could invite a sighted partner, following a past practice in cross-ability research with BLV participants ~\cite{muehlbradt2022whats}. 
To understand disagreement and variance across the community, the facilitators compared findings across workshops.  & Multiple community members (4--5) participated in each workshop together. Study facilitators first asked participants to respond to images individually and then discuss their decisions as a group.  
As a result, disagreement across participants could be surfaced and discussed in real time ~\cite{bergman2024stela}. \\ \hline 

Number of images shown & Participants were shown either 5 or 10 images per artifact. Images were discussed one at a time, and presented by reading an alt text description. We took short breaks to prevent participant fatigue. & Participants were shown 16 total images of each artifact, sorted into 4 groups. Images were discussed one at a time. \\ \hline 
Rating scale & Participants were asked to provide binary judgments of cultural appropriateness:

1: This image can never be shown 

2: This image can be shown & Participants were asked to make decisions on a 3-point scale: 

1. Can be shown 

2. Needs improvement 

3. Cannot be shown 

\end{tabular}
\caption{\textbf{Summary of differences between community workshop protocols.} The table summarizes key differences between the workshop methodologies used to engage the three communities. Differences in protocols reflect differing access needs, and also small changes made at the discretion of different workshop facilitators.}
\label{tab:workshop-differences}
\end{table}

\begin{table*}[t]
\centering
\begin{tabular}{ c|c|c|c|c } 
  \textbf{IDs} & \textbf{Relationship} & \textbf{B's Age} & \textbf{B's Location} & \textbf{Objects discussed} \\ \hline
 B1/SP1 & Friends & 18--34 & Sheffield, England  & Braille notetaker  \\ 
 B2/SP2 & Friends & 18--34 & London Area, England & Guide cane, braille notetaker  \\ 
 B3/SP3  & Parent/Child  & 55--74  & Undisclosed & Guide cane  \\ 
 B4/SP4 & Parent/Child  & 35--54  & London Area, England & Guide cane, braille notetaker \\ 
 B5/VIP5 & Friends  & 35--54  & Undisclosed &  Guide cane \\ 
 B6/SP6  & Friends  & 55--74  & Glasglow, Scotland  & Guide cane, braille notetaker  \\ 
 B7/SP7  & Siblings  & 55--74  & London Area, England & Braille notetaker  \\ 
 B8/SP8  & Friends  & 55--74 & Perth, Scotland  & Guide cane, braille notetaker  \\ 
 B9/SP9  & Friends  & 55--74  & London Area, England & Guide cane  \\ \hline
\end{tabular}
\caption{\textbf{Participant information about each blind and low vision community member (B) and their sighted partners (SP).} We report each community member's age and (when disclosed) location of residence at the time of study.
One participant (B5) invited a visually impaired friend (VIP5) to participate as their buddy. 
\label{table:blv-participants}}
\end{table*}

Each pair of participants participated in their own workshop, following ~\citet{muehlbradt2022whats}, to give community members the space to openly discuss each activity with a partner they already felt comfortable with.  Workshops were conducted synchronously online between December 2024 and March 2025, were facilitated by the first author, and ranged from 45 to 90 minutes.  Participants were compensated £75, and all workshop studies were approved by our institution's ethics review board.

Workshops began by introducing the goals of the project: to help the study designers understand how community members evaluate whether an image is an appropriate representation of their culture. 
To ground participant discussions, we introduced the study activities by providing a hypothetical scenario for how the images they were shown would be used (``A media company has collected several images of a guide cane, and they need your help to understand which of these images they should show to users'').

For each artifact, we conducted two study activities.  First, we asked participant pairs to react to both the AI-generated images and real photographs selected from Step 2.  Images were presented one-at-a-time.  After providing the alt-text description of the image, the facilitator asked participants to share if they felt that the image could be shown to represent the artifact, or should never be shown, and why. 
Based on participants' responses and interests, the facilitator asked follow-up questions to prompt participants to elaborate on what exactly made a particular image a good, bad, offensive, or incorrect depiction of the artifact.  When responding to images, the facilitator invited participants to further discuss what is shown in each image as a pair ~\cite{muehlbradt2022whats}.

The last workshop activity invited participants to reflect on what they had seen so far to create a list of the most important things that need to be shown in a culturally appropriate portrayal of the object (open-ended), and provide reasons for each of their responses.  This activity encouraged participants to articulate concrete visual criteria that shaped their decisions.  The study facilitator asked clarifying questions that encouraged participants to reflect on whether a characteristic could vary between portrayals of the object or encourage participants to prioritize whether some characteristics were more important than others.

\newpage
\subsubsection{Prompt generation templates}\label{apdx:blv-prompt-templates} 
Table \ref{table:generation-prompts} shows the two prompts that we used to generate images: prompting using the artifact name, and prompting using the artifact name along with a short description written by a community member on our research team.
We qualitatively observed that providing a description of each artifact resulted in improved representations, but all of the generated images we reviewed still had at least one error (\eg{} with the arrangement of keys on a braille notetaker or the handle shape of a cane).
Images were generated using the \dalle{} 3 and Stable Diffusion 3 Medium APIs.

\begin{table*}[ht!]
\small
\centering
\begin{tabularx}{\textwidth}{>{\raggedright\arraybackslash}m{0.75\textwidth} m{0.2\textwidth}}\toprule
\textbf{Prompt template and example} & \textbf{Example image} \\
\midrule
\textbf{Artifact only}: \textit{``A photo of a guide cane''} &
\includegraphics[width=0.18\textwidth]{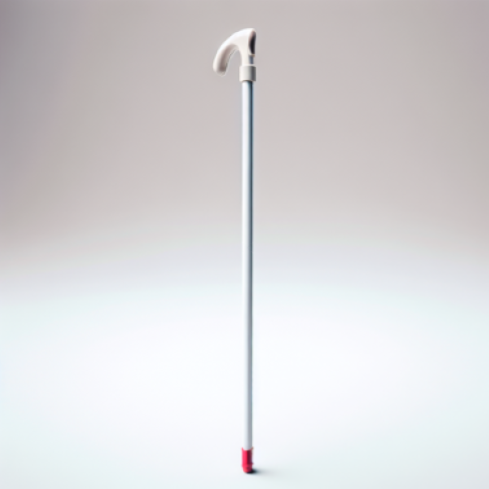} \\
\addlinespace
\textbf{Artifact + artifact description}: \textit{``A photo of a guide cane. A guide cane, or white cane, is an assistive technology used by people who are blind. It is a collapsible lightweight cane made of aluminum.''} &
\includegraphics[width=0.18\textwidth]{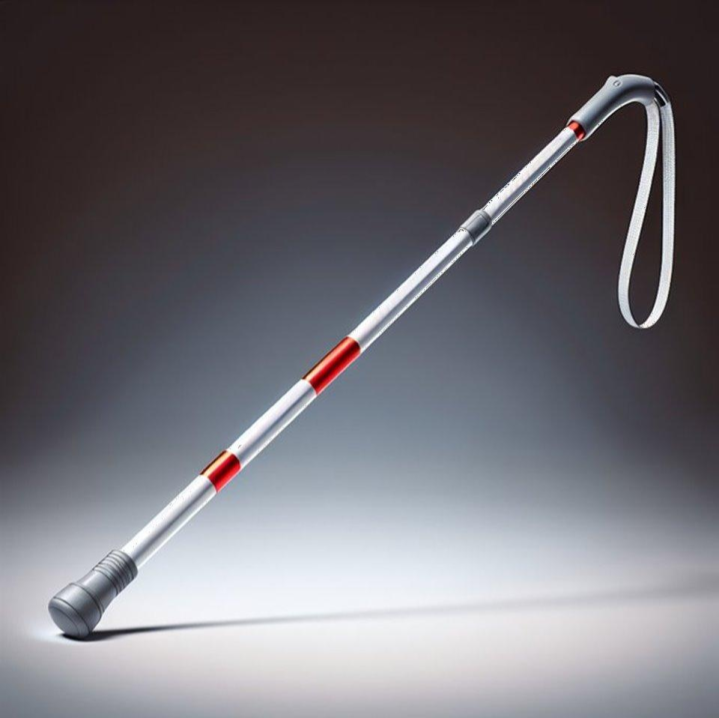} \\
\bottomrule
\end{tabularx}
\caption{\textbf{Prompt templates used to generate images to show community members.} Artifact descriptions were written by a blind community member on the research team. 
\label{table:generation-prompts}}
\end{table*}

\newpage
\subsubsection{Selected Images \& Alt Text}\label{apdx:blv-protocol-alt-text}
Below, we include the final dataset of images (including AI images and real photographs) that were shown to workshop study participants with the alt text that we provided.

\begin{table}[H]
\small
\centering
\begin{tabularx}{\textwidth}{>{\raggedright\arraybackslash} c m{0.3\textwidth} m{0.6\textwidth}}\toprule
{} & \textbf{Image} & \textbf{Alt text description} \\
\midrule
    1 & \includegraphics[width=0.7\linewidth]{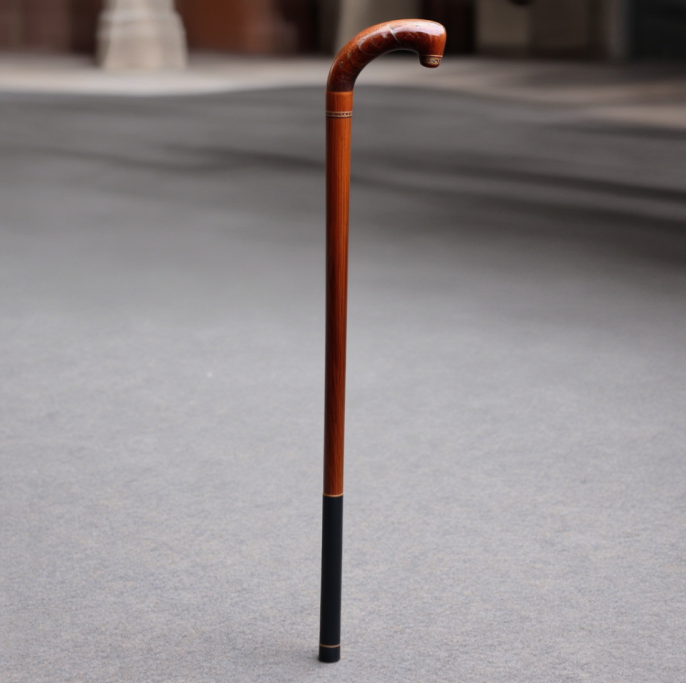} & The cane is made out of wooden material. It is a deep brown color. It has a curved grip, and straight body. The bottom part of the cane has a straight rubber tip.\\ 
    2 & \includegraphics[width=0.7\linewidth]{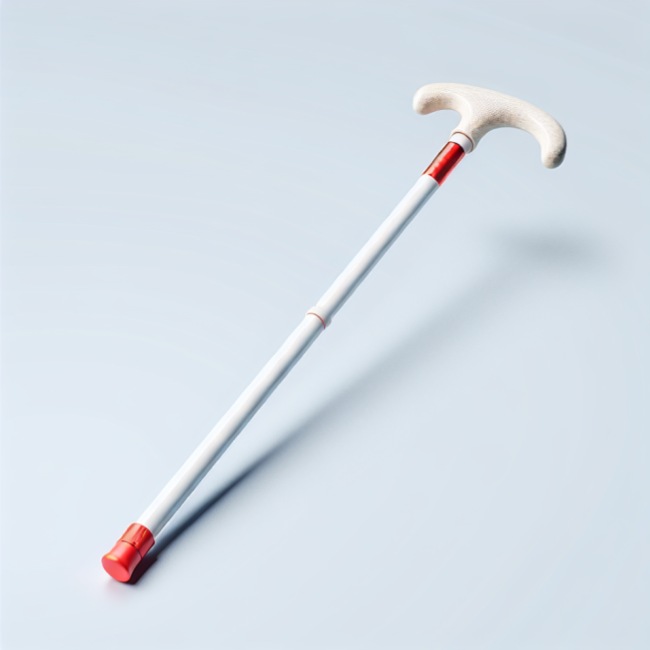} & The cane is made out of lightweight plastic material. It is a white color, with a band of red reflective tape at the top of its body. It has a curved grip, and straight body. The bottom part of the cane has a straight rubber tip.\\ 
    3 & \includegraphics[width=0.7\linewidth]{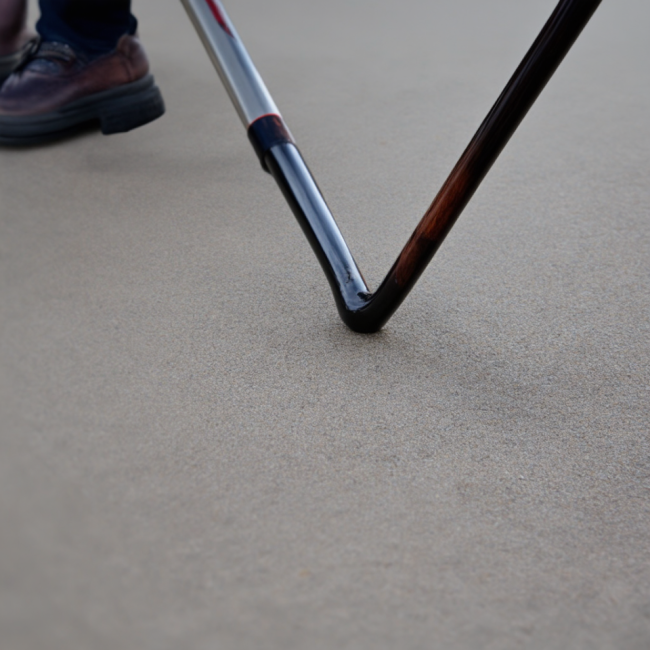} & The cane is made out of reflective metal material. It is a dark blackish brown color. The body of the cane is bent at a right angle. There is no visible handle or tip.\\ 
    4 & \includegraphics[width=0.7\linewidth]{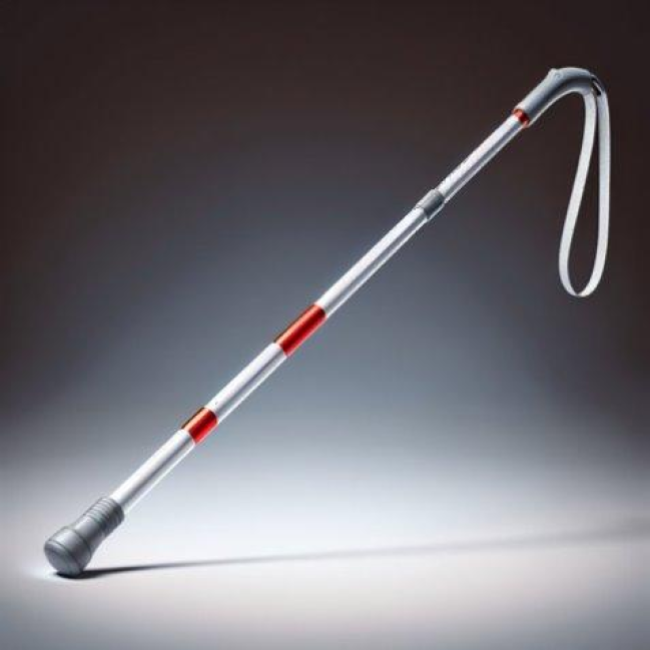} & The cane is made out of lightweight metal material. It is a light grey color, with two bands of red reflective tape. It has a straight grip, with a wrist strap. The bottom part of the cane has a round plastic marshmallow-shaped tip.\\ 

    5 & \includegraphics[width=0.7\linewidth]{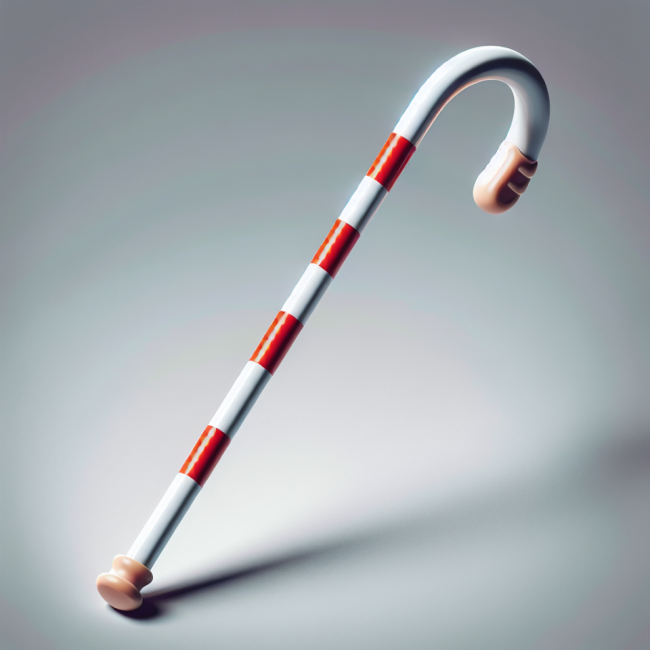} & The cane is made out of plastic material. It is a white color and has four wide bands of red reflective tape arranged like stripes. It has a curved grip, and straight body. The bottom part of the cane has a round mushroom shaped tip.\\ 
\bottomrule
\end{tabularx}
\end{table}

\newpage
\begin{table}[H]
\small
\centering
\begin{tabularx}{\textwidth}{>{\raggedright\arraybackslash} c m{0.3\textwidth} m{0.6\textwidth}}\toprule
{} & \textbf{Image} & \textbf{Alt text description} \\
\midrule
    6 & \includegraphics[width=0.6\linewidth]{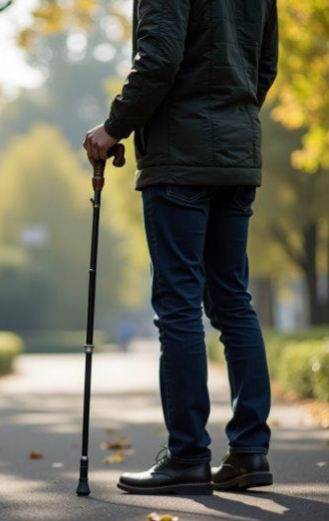} & The cane is made out of lightweight metal material. The body of the cane is black, and the handle is brown. It has a curved grip, and straight body. The bottom part of the cane has a
straight rubber tip.\\ 
    7 & \includegraphics[width=0.7\linewidth]{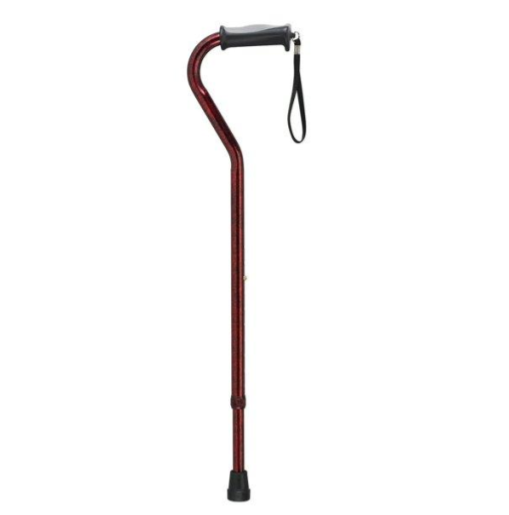} & The cane is made out of lightweight metal material. The body of the cane is a reflective red color. The cane has a crooked grip and a black handle. There is a wrist strap hanging out of the handle. The bottom part of the cane has a straight rubber tip.\\ 
    8 & \includegraphics[width=0.7\linewidth]{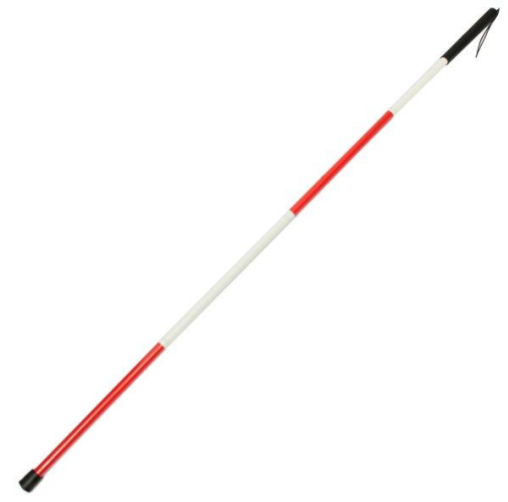} & The cane is made out of lightweight metal material. The body of the cane is white. There are two bands of red tape on the cane. The cane has a straight body, and straight black grip at the top. There is an elastic wrist strap coming out of the grip. The bottom part of the cane has a straight rubber tip.\\ 
    9 & \includegraphics[width=0.7\linewidth]{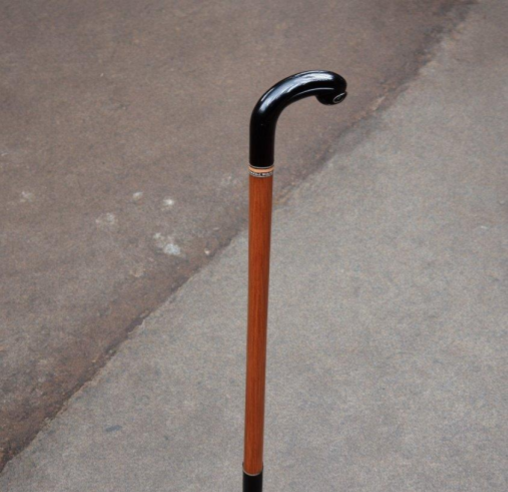} & The cane is made out of wooden material. The body of the cane is a chestnut color, and it has a black handle. It has a curved grip, and straight body. The bottom of the cane has a dark rubber tip.\\ 

    10 & \includegraphics[width=0.7\linewidth]{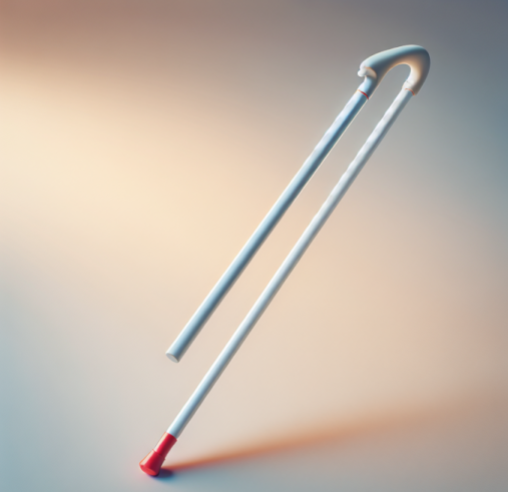} & The cane is made of lightweight plastic material. The body of the cane is curved: there is a round handle at the top, and then two long parallel sticks coming out of each end of the handle. Each of the long sticks is a white color. At the bottom of one of the long sticks is a marshmallow-shaped red tip.\\ 
\bottomrule
\end{tabularx}
\end{table}

\newpage
\begin{table}[H]
\small
\centering
\begin{tabularx}{\textwidth}{>{\raggedright\arraybackslash} c m{0.3\textwidth} m{0.6\textwidth}}\toprule
{} & \textbf{Image} & \textbf{Alt text description} \\
\midrule
    11 & \includegraphics[width=0.7\linewidth]{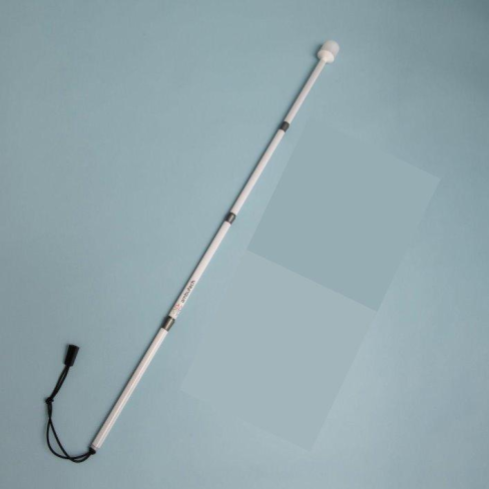} & The cane is made of lightweight metal material. The body of the cane is white. It has a straight body, and no visible handle or grip. One end of the cane has a black wrist strap. The other end of the cane has a round marshmallow-shaped tip. The body of the cane is divided into three sections by three grey joints.\\ 
    12 & \includegraphics[width=0.7\linewidth]{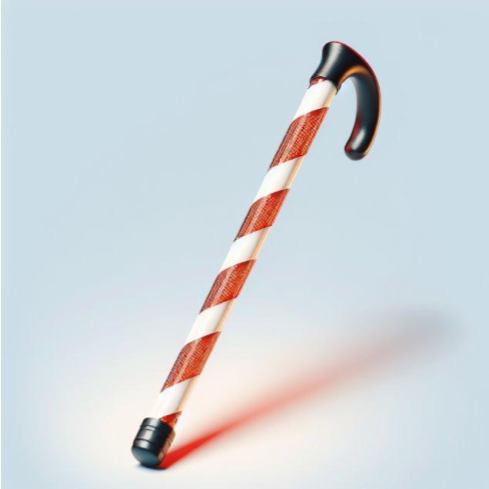} & The cane is made of lightweight plastic material. The cane's body has wide bands of red reflective tape, arranged like stripes. The cane has a straight body and curved black handle. The bottom of the cane has a straight black tip.\\ 
    13 & \includegraphics[width=0.7\linewidth]{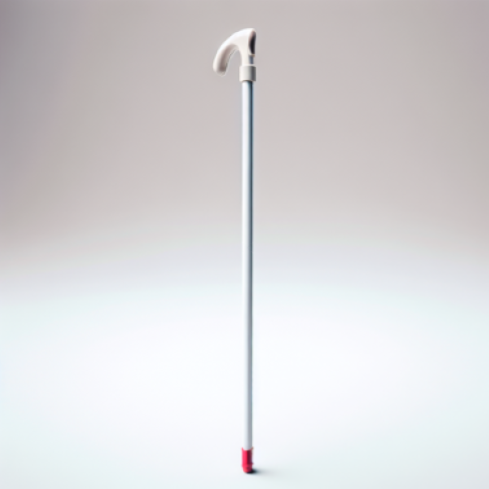} & The cane is made of lightweight plastic material. The body of the cane is white. It has a straight body, and a handle that curves downwards. The bottom of the cane has a straight red tip.\\ 
    14 & \includegraphics[width=0.7\linewidth]{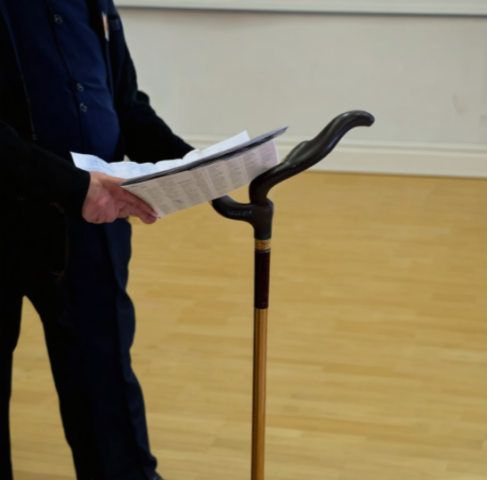} & The cane is made of lightweight plastic material. The body of the cane is a gold color. The cane has a straight body, and a black curved handle. The handle has two distinct pieces that stick up at different angles at the top of the cane. The handle is irregularly curved.\\ 

    15 & \includegraphics[width=0.6\linewidth]{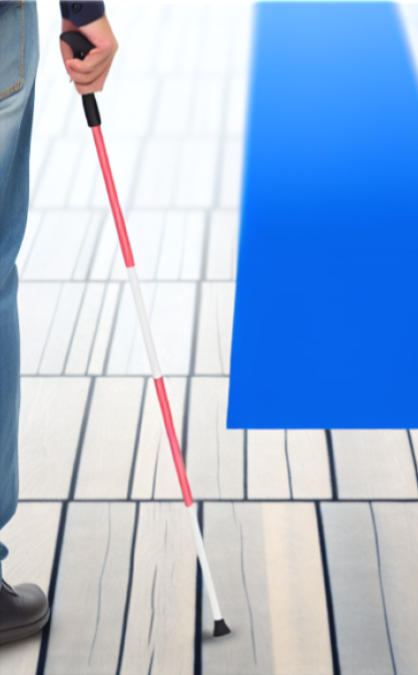} & The cane is made of lightweight metal material. The body of the cane is white, with two wide bands of red reflective tape. The top of the cane has a straight black grip.\\ 
\bottomrule
\end{tabularx}
\end{table}

\newpage
\begin{table}[H]
\small
\centering
\begin{tabularx}{\textwidth}{>{\raggedright\arraybackslash} c m{0.3\textwidth} m{0.6\textwidth}}\toprule
{} & \textbf{Image} & \textbf{Alt text description} \\
\midrule
    1 & \includegraphics[width=0.7\linewidth]{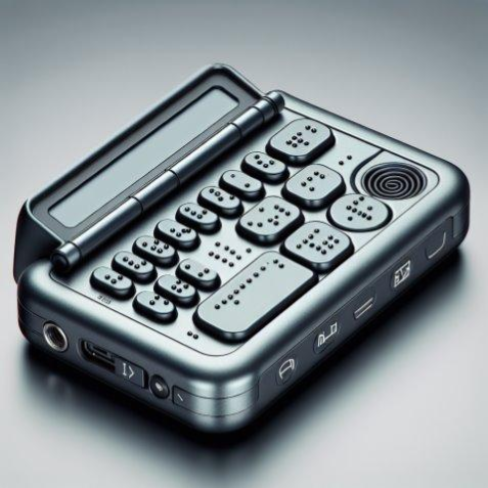} & There is a thin rectangular electronic device.
    
    The top surface of the device has a display on top, and rows of buttons below it. The display is a small electronic screen. The buttons have tactile markings on them that resemble Braille. The top surface also has a circular speaker.

    The sides of the device have additional ports and buttons.\\ 
    2 & \includegraphics[width=0.7\linewidth]{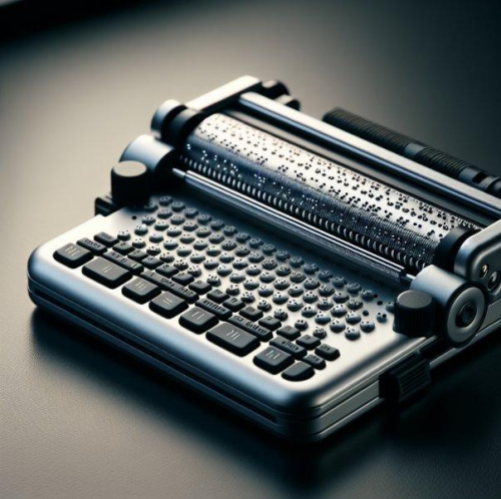} & The device is a rectangular shape, with a roller cylinder on top of its surface where paper can be inserted.
    
    The roller has tactile markings on its surface that resemble Braille. Below the roller on the top surface, there are four rows of circular keys. Each row has about 15 keys. Below that row, there are 3 rows of larger rectangular keys.\\ 
    3 & \includegraphics[width=0.7\linewidth]{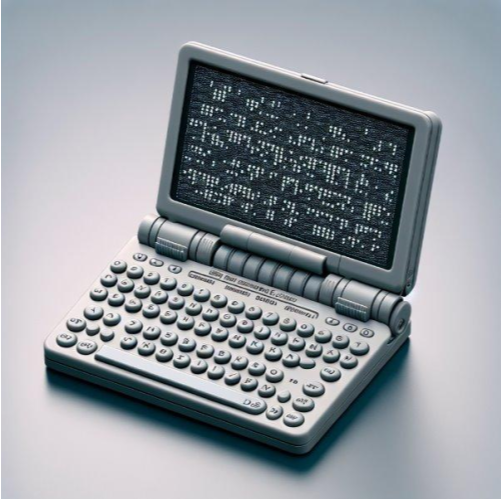} & There is a device shaped like a folding laptop computer, with an electronic screen display on top and a keyboard on the bottom.
    
    The screen of the device is displaying rows of dots that resemble Braille. There are five "lines" of Braille stacked on top of each other.
    
    The keyboard of the device resembles a qwerty keyboard. There are five rows of keys, buttons at the top, and a space bar at the bottom.\\ 
    4 & \includegraphics[width=0.7\linewidth]{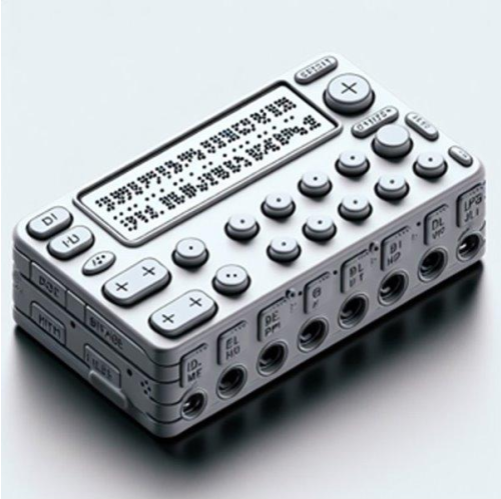} & There is a rectangular electronic device.
    
    The top surface of the device has a display on top, and rows of buttons below it. The display has tactile markings that resemble Braille. It is showing two "lines" of Braille. Below the display, there are two stacked rows of five circular buttons. The device has other buttons on its surface, for example, that resemble a volume control.
    
    The sides of the device have additional ports and buttons. For example, one side appears to have 7 circular input ports.\\ 

    5 & \includegraphics[width=0.6\linewidth]{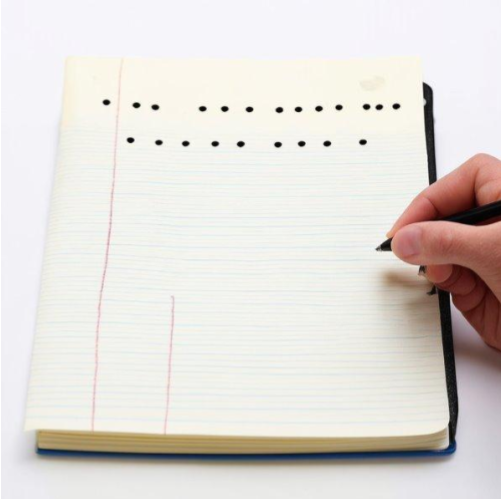} & The image shows a paper notebook and a hand holding a pen.
    
    The notebook is open to a page. Two lines of ink dots are written on the page.\\ 
\bottomrule
\end{tabularx}
\end{table}

\newpage
\begin{table}[H]
\small
\centering
\begin{tabularx}{\textwidth}{>{\raggedright\arraybackslash} c m{0.3\textwidth} m{0.6\textwidth}}\toprule
{} & \textbf{Image} & \textbf{Alt text description} \\
\midrule
    6 & \includegraphics[width=0.7\linewidth]{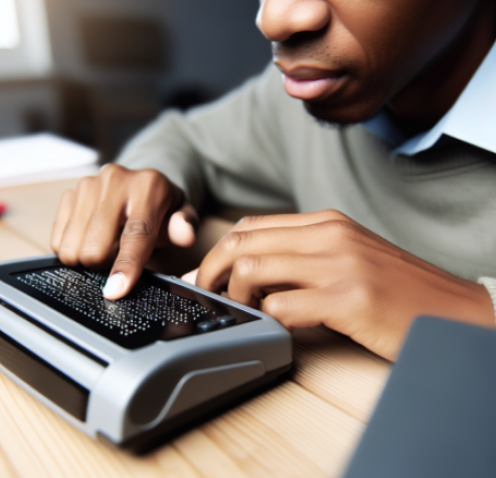} & The image shows a rectangular device.
    
    The top surface of the device has a dark display. The display has many small white tactile dots, arranged in 20 rows and 40 columns.
    
    There are no additional buttons on the sides of the device.\\ 
    7 & \includegraphics[width=0.7\linewidth]{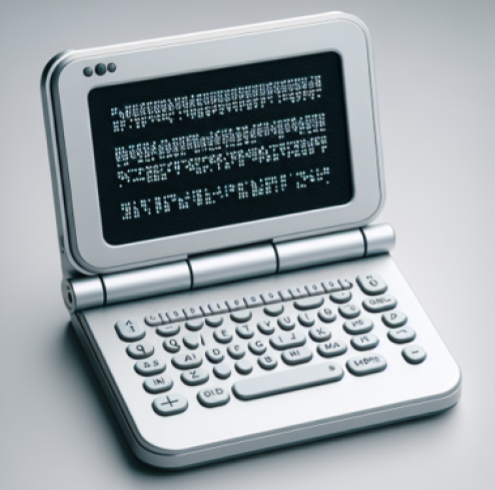} & There is a device shaped like a folding laptop computer, with an electronic screen display on top and a keyboard on the bottom.
    
    The screen of the device is displaying rows of dots that resemble Braille. There are four distinct "lines" of dots, where each "line" has about 6 rows.
    
    The keyboard of the device resembles a qwerty keyboard. There are five rows of keys, buttons at the top, and a space bar at the bottom.\\ 
    8 & \includegraphics[width=0.7\linewidth]{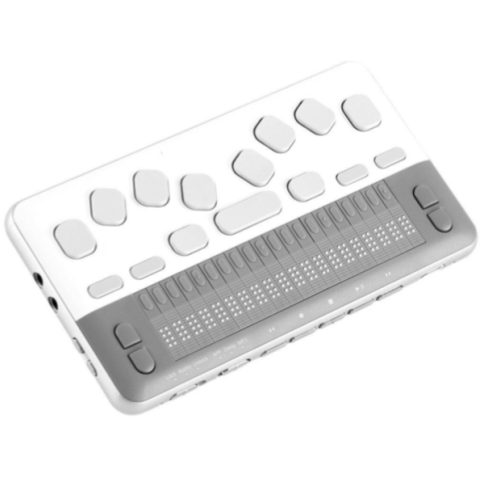} & There is a thin rectangular electronic device.
    
    The top surface of the device has buttons on top, and a tactile display at the bottom. At the top of the device, there are eight round keys arranged in a curved pattern. There are four keys on the left, four keys on the right, and a space bar in between them. There are three additional buttons on each side of the space bar. The display has 20 cells, each of which has 4 rows, and 2 columns of tactile dots. On each side of the display, there are two buttons.
    
    The side of the device has two ports and two more small buttons.\\ 
    9 & \includegraphics[width=0.7\linewidth]{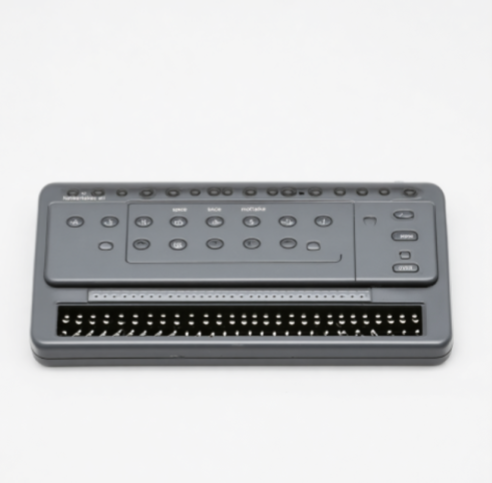} & There is a thin rectangular electronic device. The top surface of the device has several buttons on top, and a tactile display at the bottom.
    
    The top of the device has one row of about 15 small circular buttons. Below that, there are two more rows of 8 circular buttons, stacked on top of each other.
    
    The display has 3 rows, and around 25 columns, of small metal circular pins that are sticking out of the device.\\ 

    10 & \includegraphics[width=0.7\linewidth]{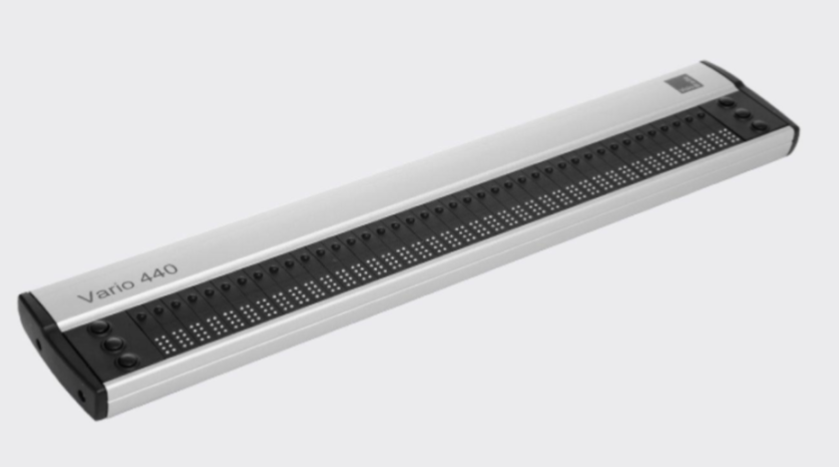} & There is a thin rectangular electronic device. It is quite wide, and not very long.
    
    The top surface of the device has a tactile display. There is no keyboard. The display has 32 cells, where each cell has 4 rows and 2 columns of small tactile dots. Each cell has a small black button above it. Next to the display, there are three circular buttons.\\ 
\bottomrule
\end{tabularx}
\end{table}

\newpage
\begin{table}[H]
\small
\centering
\begin{tabularx}{\textwidth}{>{\raggedright\arraybackslash} c m{0.3\textwidth} m{0.6\textwidth}}\toprule
{} & \textbf{Image} & \textbf{Alt text description} \\
\midrule
    11 & \includegraphics[width=0.7\linewidth]{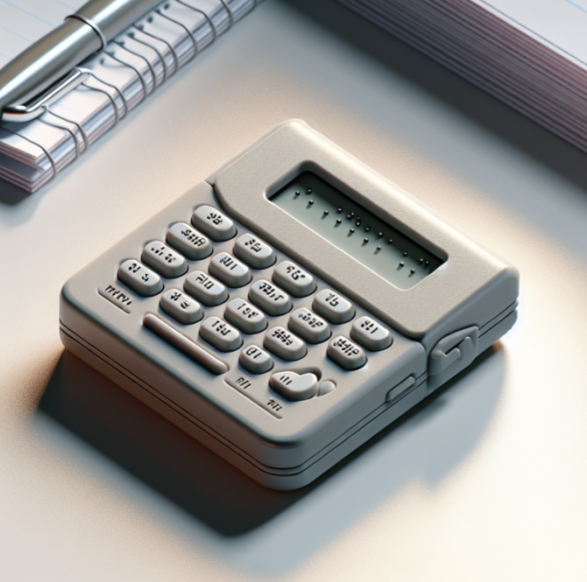} & There is a rectangular electronic device. The top surface is shaped like a handheld calculator.
    
    The top surface of the device has a tactile display on top, and rows of buttons below it. The display has small metal tactile dots. They are irregularly arranged in two rows, and around 15 columns. Below the display, there are four rows and five columns of small oval-shaped buttons.\\ 
    12 & \includegraphics[width=0.7\linewidth]{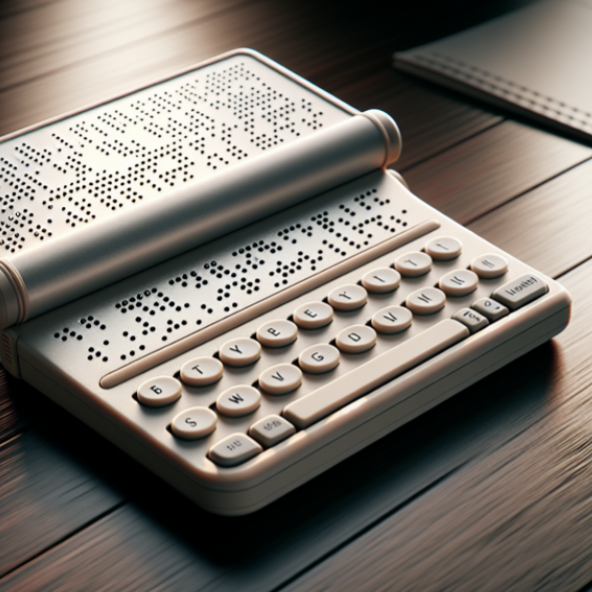} & The device is a rectangular shape, with a roller cyllinder on top of its surface.
    
    Immediately above and below the roller, the surface of the device has several tactile markings that resemble braille. Below that on the top surface, there are rows of circular keys that resemble a qwerty keyboard. There are two rows of keys marked with letters, and below that there is a space bar and additional keys.\\ 
    13 & \includegraphics[width=0.75\linewidth]{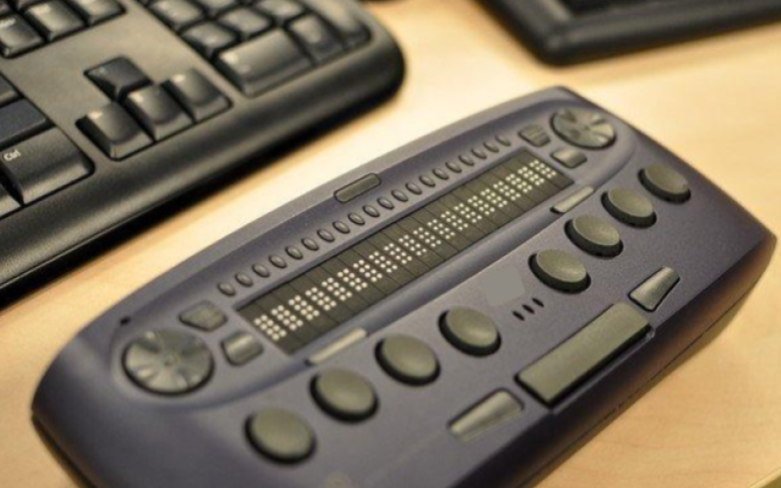} & There is a wide, thin electronic device.
    
    The top surface of the device has a braille display, and several buttons. At the top of the device, there is a tactile display. The display has 20 cells, each of which has 4 rows, and 2 columns of tactile dots. There are three round buttons to the left and right of the display. Below that, there is one row of eight round keys arranged in a curved pattern. There are four keys on the left, four keys on the right, and a space bar below them.\\ 
    14 & \includegraphics[width=0.7\linewidth]{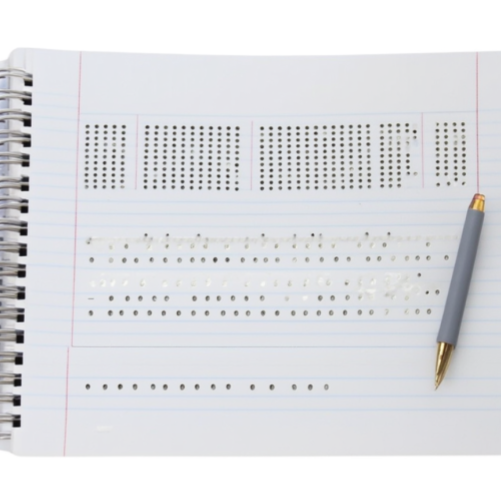} & The image shows a paper notebook with an ink pen resting on it.
    
    The notebook is open to a page. Several lines of ink dots have been written on the page. The top of the page has around 100 small dots, arranged in around 10 rows and 40 columns. The bottom of the page has four rows of larger ink dots. The page looks a bit crinkled, like several dots have been removed.\\ 

    15 & \includegraphics[width=0.75\linewidth]{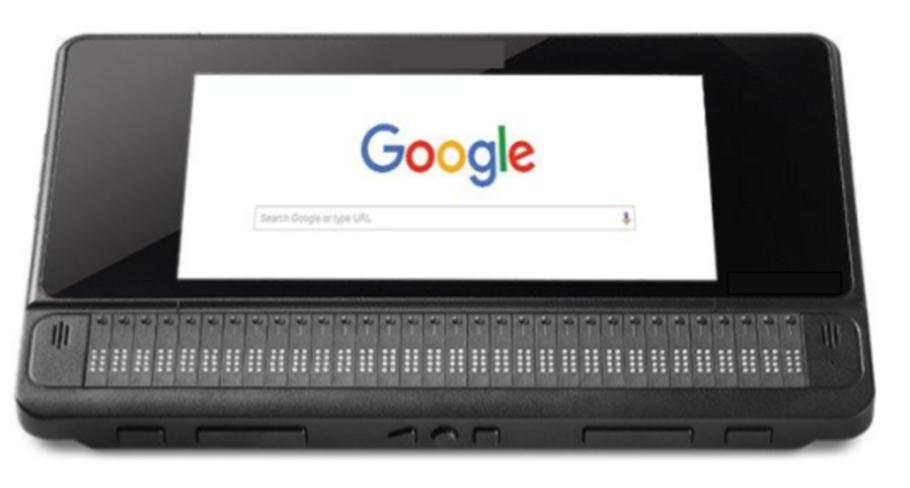} & The image shows a thin electronic device.
    
    The top surface of the device has an electronic screen, displaying the Google homepage. The bottom of the device has a tactile display. The display has 28 cells, each of which has 4 rows, and 2 columns of tactile dots. There is a button at either end of the display.\\ 
\bottomrule
\end{tabularx}
\end{table}

\newpage
\subsubsection{Workshop Study Protocol} Workshops with blind and low vision community members began by the facilitator introducing themselves, and inviting the participant pair to introduce themselves. We provide the facilitator's script below:

Today, we want to understand how you would like different assistive technologies to appear in AI images. 

Our goal as researchers is to learn from both of your expertise and past experiences with two assistive technologies as someone who uses them or as someone who has observed their usage. 

The [first/second] object we’re going to discuss, is a [OBJECT]. 
\begin{itemize}
\item If only one participant is familiar with the object: I saw from the survey that [NAME] is less familiar with [OBJECT]? Could you share more what you mean by that?     
\end{itemize}

For the activity, we’re going to discuss what you both think about some images. This activity is going to be based off a scenario, that I’ll introduce now: 

SCENARIO: ``A media company has created a service to provide images for users who are making slide decks. The service works like this: every time the user asks for an image, the company generates 10 images, and then shows the user 4. 
The media company has collected several images of a [OBJECT]. The media company needs your help to understand which of these images they should show to users, and which images should never be shown to users.''

I’ll pause: Any questions about the scenario? 

The company has gathered 10 total images of a [OBJECT] for us to discuss together today. 

\textbf{Reacting to images.} For the activity today, I am going to screen share a slide deck that has the 10 images. I’m going to go through the images one-by-one and ask you both to answer some questions for each image. I’ll begin by providing a basic description (some alt text) for each image. If anything is unclear about the images from the alt text I’ve provided, you can also ask for clarification from your partner. 

You may notice that the [OBJECT] in each image is in a different scene: for example, some images show people using a [OBJECT], versus others just depict the [OBJECT] on a plain background. When answering the questions, we’d like for you to focus only on the [OBJECT] -- not the scene.

Questions to ask for each image:
\begin{itemize}
    \item Would you tell the company that this image can be shown, or should be never shown to users? Follow-up probes:
    \begin{itemize}
        \item Why would you (not) show it? 
        \item What about this image is good/bad? 
        \item Is there something about this image that makes it an offensive or harmful depiction? 
    \end{itemize}
    \item Is this image a correct depiction of a [OBJECT]? Follow-up probes:
    \begin{itemize}
        \item What about it makes it (in)correct? 
        \item Why do you think this one is OK to show, even if it is incorrect?
    \end{itemize}
\end{itemize}

Summarizing all of the images: OK, so far we’ve selected X images that we think are not OK to show to users: [read] 
\begin{itemize}
    \item Now that we’ve discussed them all, are there any that we want to add to this final list? 
    \item Are there any images that you think actually might be OK to show, and why? 

\end{itemize}

\textbf{Identifying important charcateristics \& visual criteria.} Given everything we’ve seen, what do you think are the most important things that need to be shown for the [OBJECT] to be depicted correctly? Probes:
\begin{itemize}
    \item Do any of the things we talk about look different for different types of [OBJECT]? Is it possible for the [thing] to vary? What could you change about [this object] for it to still be correct? 
    \item What about [characteristic]? Would it be on your list at all?
    \item So far, we’ve listed [these X things] that are important to representing a [OBJECT]. Are there things on this list that are more important than others? 
\end{itemize}

Beyond the 10 images we’ve looked at today, the company is interested in understanding some general principles about what makes an image good or bad to show to a user. What advice would you give to the company? What is it that makes an image of a [OBJECT] inappropriate to show?

\newpage
\subsection{Extended study protocol: Indian cultural artifacts}\label{apdx:protocl-india}

\subsubsection{Recruitment \& Workshop Activities} We adopted a focus group methodology where we invited participants from the same state to participate in synchronous workshops together.  Past research has demonstrated the value of facilitating deliberative group discussions to surface (and clarify) points of tension and disagreement among participants ~\cite{qadri2023ai,mack2024they,qadri2023ai}.  We conducted one focus group per each artifact, where each focus group was facilitated by a community member from the relevant state.  Workshops were conducted virtually and lasted 120 minutes. Participants were compensated with an Amazon voucher worth 500 INR.

\begin{table}[H]
\centering
\begin{tabular}{ c|c|c|c|c l} 
  \textbf{State} & \textbf{Artifact} & \textbf{Number of participants} & \textbf{Gender ratio (M:F)}  &\textbf{Age Range}\\ \hline
 Tamil Nadu & Pallanguzhi & 4 & 2:2  &18--34\\ 
 Tamil Nadu & Mridangam & 5 & 1:4  &18--44\\
 Kerala & Kasavu saree & 4 & 1:3  &18--54\\
 Kerala & Chundan Vallam & 4 & 3:1  &25--44\\ \hline
 Total &  & 17 & 7:10  &18--54\\
 \hline
\end{tabular}
\caption{\textbf{Participants in workshops organized around each Indian cultural artifact.} Workshops were conducted synchronously, and a single artifact was discussed in each workshop. Eligible participants only participated in a single workshop, and were assigned to artifacts based on their availability and expertise.}
\label{table:india-participants}
\end{table}

To determine eligibility, we decided to scope our study to individuals who have spent at least several years residing in Kerala or Tamil Nadu, and would therefore be likely to be familiar with cultural artifacts of the regions.
To recruit participants for workshops, we aimed to capture a diverse sample of individuals who identified as being from each state.
Our research team disseminated a call to participate in research by disseminating a screener survey circulated on the research team's personal social networks, such as X.com and WhatsApp.
Inclusion criteria were that participants must speak English, reside in India, and have spent at least several years residing in Kerala or Tamil Nadu.
Participants also self-reported their familiarity with each cultural artifact.
We identified 23 eligible participants from our initial screener survey,  and invited 9 participants per state to participate. The final sample of participants who ultimately participated, along with their ages and gender identities, is summarized in Table \ref{table:india-participants}.

Like the BLV workshops, the focus group activities invited community members to react to images to map the broad background concept of cultural appropriateness to specific visual characteristics.  Participants were introduced to the goals of the project using a scenario where a company is exploring the use of AI-generated images in a tourism advertisement.  Participants were invited to join an online board on FigJam, a collaborative web application where multiple users can take notes. 
In the first activity, participants were presented with 16 total AI-generated images of the artifact (4 per each of the groups from Step 2) displayed on the whiteboard.
For each AI-generated image, participants were asked to rate whether they felt the image (1) cannot be shown, (2) needs improvement, or (3) can be shown as a representation of the artifact.
This three-point scale provided participants with more flexibility to note when they were uncertain about a particular image.
We first invited participants to evaluate each image individually by sharing their ratings in the meeting room chat.
Participants were then invited to share and discuss their reactions to any of the images and justifications for their rating decisions with the group.

After the first activity, the facilitator led a group discussion inviting participants to rank the groups of images, from most to least preferred.
Participants whose ratings or rankings disagreed were encouraged to understand each other's perspectives to see if they could reach a consensus, but reaching consensus was not required.

The last study activity invited participants to reflect on if there were any varieties of the artifact that they were familiar with that had not yet been discussed.
Participants were encouraged to search the Web for photographs that they felt were or were not representative of the artifact.
Participants discussed whether they felt each image was an appropriate cultural representation as a group.

\subsubsection{Generating images}

The names of each Indian cultural artifact come from the local language of each state (Tamil or Malayalam), which are written using a non-Roman script, and do not have a direct English translation. Existing text-to-image models have little to no support for these local scripts, so we followed past research and used the \emph{transliterated} name for each object, a sound-preserving transcription of the word in Roman script~\cite{knight1998machine,gupta2012mining,seth2024dosa}.
When we used basic prompt templates with these transliterations (\eg{} ``a photo of a Chundan Vallam'') to generate images, we found that many of the generated images bore very little relevance to the artifacts at hand.  For example, when we prompted Stable Diffusion 3 to create images of a Chundan Vallam (racing boat), it instead created images of South Asian homes, people, food, and scenes.  However, images created with \dalle{} 3 were more likely to create depictions that captured the essence of each artifact, due to \dalle{}'s automated prompt revision feature, which rewrote our basic input prompts to include detailed English descriptions of each artifact ~\cite{openai-promptrevision}.  To increase the quality of the Stable Diffusion images, we input these \dalle{} revised prompts instead of using basic prompt templates.  We include example \dalle{} revised prompts for each artifact in the appendix. 

As in the BLV context, community members on the research team sorted an initial set of 60 generated images (30 per model) into 4 groups that shared common visual characteristics so that we could prioritize getting participants' feedback on meaningfully different depictions.
The final set of AI-generated images differed from those shown to BLV participants in that they were more likely to picture each artifact within a detailed scene (\eg{} of rowers paddling a racing boat down a river), in contrast to the images shown to BLV participants, which showed each artifact juxtaposed on an empty or abstract background.

\subsubsection{\dalle{} revised prompts}\label{apdx:india-revised-prompts}

\begin{figure}[t]
    \centering
    \includegraphics[width=\linewidth]{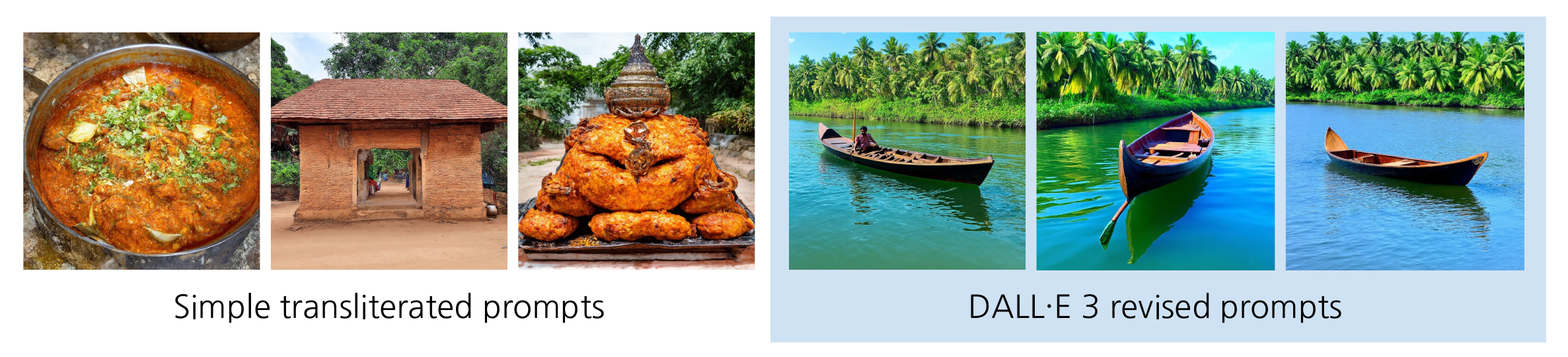}
\caption{\textbf{Stable Diffusion 3 generates depictions of unrelated cultural artifacts and scenes when given simple transliterated prompts. Depictions improve when images are generated using \dalle{} 3 revised prompts instead}. The images on the left were generated with the simple transliterated prompt ``A photo of a Chundan Vallam''. Instead of producing depictions of a boat, the generated images show unrelated depictions of foods and buildings, indicating that the models may struggle to interpret transliterated artifact names. The images on the right, which picture boats floating down a river, were generated by Stable Diffusion 3 when provided with a more descriptive revised prompt.}
\label{fig:india-example-sd-errors}
\Description[Example images generated from simple transliterated prompts (left) versus \dalle{} 3 revised prompts (right).]{The three example images generated from simple transliterated prompts show (a) a photograph of a stew or curry in a pot, ready to be served, (b) a photograph of red brick awning in an outdoor scene, and (c) a confusing image of an object that has some components that resemble food (\ie{} an orange meat dish), and other components that resemble other unrelated cultural artifacts (\eg{} an ornamental lamp). In contrast, the three generated images on the right show depictions of a brown wooden boat floating down a river.}
\end{figure}

When we used the basic ``artifact only'' prompt template from Table \ref{table:generation-prompts} to generate images of each Indian artifact, we found that Stable Diffusion 3 created images that were completely unrelated to the artifact, instead of displaying representations of unrelated objects, \eg{} of foods or buildings (Figure \ref{fig:india-example-sd-errors}).

Rather than elicit community members' feedback on images that so clearly contain errors of representation, we decided to revise our prompting strategy. 
To do this, we stored prompts created using \dalle{}'s automated prompt revision feature, which rewrote our basic input prompts to include detailed English descriptions of each artifact ~\cite{openai-promptrevision}.
To increase the quality of the Stable Diffusion images, we input these \dalle{} revised prompts instead of using basic prompt templates. 
We include example \dalle{} 3 revised prompts for each of the Indian cultural artifacts below. 

Original prompt: ``\emph{a photo of a [artifact]}''

\begin{itemize}
    \item Pallanguzhi: ``\emph{A realistic photograph of a traditional Pallanguzhi board on a rustic wooden table. The wooden board features intricate carvings, with two rows of seven circular pits, filled with vibrant, colorful seeds or shells used as counters. The scene is illuminated by warm natural lighting, which highlights the detailed craftsmanship of the board and the vivid hues of the seeds. Surrounding elements include traditional Indian decor, such as a brass lamp emitting a soft golden glow and a vibrant, patterned textile cloth draped in the background. The overall atmosphere of the image exudes nostalgia and serenity, celebrating the cultural heritage of this ancient Indian board game.}''
    \item Mridangam: ``\emph{A high-resolution image of an Indian mridangam, a traditional percussion instrument, placed on an ornately decorated cloth featuring intricate patterns in rich shades of gold and red. The wooden body of the mridangam is shown clearly, with its leather straps and detailed craftsmanship prominently visible. The setting is a warm, softly-lit indoor space with ambient light creating gentle shadows. The lighting emphasizes the mridangam as the centerpiece of the composition, exuding cultural beauty in a serene and elegant atmosphere.}''
    \item Chundan Vallam: ``\emph{A realistic photograph of a traditional Chundan Vallam, also known as a snake boat, floating on the still waters of Kerala's serene backwaters. The handcrafted, dark wooden boat is long with a slender shape and an elegantly rising, pointed bow. Intricate carvings glimmer on the boat, reflecting the bright sunlight. In the backdrop, there's a dense wall of lush, green coconut trees and tropical vegetation along the riverbanks. The sky is a clear, vibrant blue, contrasting beautifully with the greenery. The calm water creates a perfect mirror image of the boat and the surrounding landscape, enhancing the tranquil atmosphere of the scene.}''
    \item Kasavu saree: ``\emph{A serene and elegant presentation of an Indian Kasavu saree neatly folded on a rustic wooden table. The saree is predominantly white with a golden border and features intricate golden zari work embellishing its border and pallu. The scene is illuminated with warm, golden lighting, emphasizing the cultural heritage tied to Kerala. The background is softly blurred and plain to maintain a minimalist aesthetic. Near the saree, a small vase of fresh, simple flowers adds a delicate touch. The natural texture of the wooden table reinforces the timeless charm and grace of the scene, creating an artistic, warm-toned environment.}''
\end{itemize}

\subsubsection{Focus Group Protocol} Focus groups began by the facilitator introducing themselves, and inviting the group of workshop participants to introduce themselves. We provide the facilitator's script below:

We’re working on a research project that involves evaluating AI-generated images of cultural artifacts of different communities — like traditional objects, clothing, foods, etc.
While AI technologies are advancing rapidly, we have discovered that when generating images of commonly known cultural artifacts from India, examples often struggle to represent their cultural significance.
As a cultural expert from your community, your perspective is very important to us.
Our goal is to learn from your expertise and lived experiences with the cultural artifacts as someone who is from that cultural community.
We have shortlisted a cultural artifact [OBJECT] that is well known to the community.
The aim of this discussion is to understand how community members define the important visual characteristics of an artifact — those that must be included to make visual representations acceptable and respectful across the community.
This informs our evaluation of T2I (text-to-image) generated visuals.
The insights are used to improve generated visuals to ensure they reflect culturally grounded and meaningfully nuanced depictions.
Some of the AI-generated images may not represent the artifact accurately or respectfully. If an image feels inappropriate or upsetting, we sincerely apologize. Please feel free to point it out — this feedback is important to help us improve. 

\textbf{Reacting to images.} On our screen, you should see sixteen AI images of the [OBJECT]. Our team has sorted these images into four different groups that have something in common.

In this part of the study, we'll do an activity where we want you to rate each image. We'll discuss each image one-at-a-time, and are interested in knowing what everyone thinks individually before discussing each image as a group. You can message your thoughts in the chat or unmute and speak them.

There are no right or wrong answers — your unique insights matter most!

\begin{itemize}
    \item Do you think that this image can be shown, needs improvement, or cannot be shown?
    \item What exactly is good or bad about the image that influences your rating?
    \item I noticed that we aren't sure or we disagree about [image]. Can each of you discuss why you gave your rating?
\end{itemize}

\textbf{Ranking groups of images.} Now that we've seen all four of these groups of images, our next goal is to rank these four groups, from best to worst.

\begin{itemize}
    \item Which of these four groups do you think does the best job of representing the [OBJECT], and which is worst?
    \item Why is that group better or worse than the other one?
\end{itemize}

\textbf{Providing reference images.} Are there any variants or varieties of the artifacts that the model tried capturing, or didn’t have knowledge of? 
You can feel free to browse the Internet on your devices to share photos of different versions of the [OBJECT] that we haven't discussed yet.
\begin{itemize}
    \item Do you think that this version of the [OBJECT] would be okay to show? Why or why not?
\end{itemize}

\newpage
\section{Supplemental experimental results}\label{apdx:results}

This appendix presents extended descriptions of our experimental design methodology and includes supplemental experimental results. 

\begin{itemize}
    \item We begin by presenting supplemental results that analyze the content of our community-informed rubrics (RQ1), such as a lightweight validation of our final rubric criteria (Section \ref{apdx:rubric-validity}) and a qualitative analysis that compares community-informed rubrics to those generated by an LLM (Section \ref{apdx:rubric-comparisons}). 
    \item We next provide an extended description of how we \emph{operationalized} our rubrics (RQ2) to score a new dataset of images using both human and MLLM annotators (Section \ref{apdx:methods-operationalization}). 
    \item We then present extended analyses of  inter-annotator agreement among human rubric annotators (Section \ref{apdx:iaa}), qualitative trends from manually scoring T2I model outputs (Section \ref{apdx:operationalization-interpretations}), and additional results comparing human and MLLM rubric application (Section \ref{apdx:human-mllm-agreement}).
    
\end{itemize}

\subsection{Supplemental results: Validating the systematized concepts using annotations from workshops}\label{apdx:rubric-validity}

\paragraph{Methods} 

Our research team manually (by hand) applied our evaluation rubric to get final labels of cultural appropriateness for each image that was shown in workshops.
The research team (first and second authors) manually annotated whether each systematization criterion was met for each image. 
Each rubric was applied by the research team member who facilitated workshop engagements and led the creation of the rubric, to ensure its consistent application across the dataset.
Each systematization criterion corresponds directly to a concrete visual indicator in an image (\eg{} “does the guide cane have sections that are white in color?”). If an image does not contain enough detail to assess whether a criterion is met (\eg{} if an image is cropped so that a particular feature is not visible), we default to assign a positive label (\eg{} that the criterion is met) as such an image may be a potentially valid representation.

We analyze participants' ratings  to come up with a single ``majority label'' of cultural appropriateness for each image.
In workshops, each image was shown to multiple participants, who provided binary annotations of whether they felt the image was an appropriate or inappropriate depiction of each artifact.
For Indian participants, who used a three-point rating scale, we binarize ratings by treating images scored “1 — Cannot show” as inappropriate; and images scored “2 — Needs improvement” or “3 — Can show” as appropriate.
We aggregate participant labels using a majority vote. In cases of ties, we label contested images as inappropriate.

\begin{table}[H]
\small
\centering
\begin{tabular}{ cccccc}
\toprule 

{} & \textbf{Number of images} & \textbf{Agreement} & \textbf{\% contested} & \textbf{Community base rate} & \textbf{Measure base rate} \\ 
\midrule 
All BLV artifacts & 20 & 0.90 & 0.35 & {} & {} \\ 
Guide cane & 10 & 0.90 & 0.30 & 0.20 & 0.30 \\ 
Braille notetaker & 10 & 0.90 & 0.40 & 0.30 & 0.20 \\ 
\midrule 
All Indian artifacts & 64 & 0.83 & 0.32 & {} & {} \\ 
Pallanguzhi & 16 & 0.81 & 0.63 & 0.31 & 0.25 \\ 
Mridangam & 16 & 0.88 & 0.25 & 0.12 & 0.00 \\ 
Chundan Vallam & 16 & 0.81 & 0.13 & 0.25 & 0.06\\ 
Kasavu saree & 16 & 0.81 & 0.25 & 0.50 & 0.31 \\ 
\bottomrule 
\end{tabular}
\caption{\textbf{Comparing community annotations from workshops to scores obtained from applying evaluation rubrics}. We find that community annotations agree with the labels of cultural appropriateness assigned by our rubrics for over 80\% of the images shown to workshop participants. We report the ``base rate'' (the proportion of images with positive labels) for both the majority labels assigned by community members and labels assigned by our measures.}
\label{table:validity-systematizations}
\end{table}

\paragraph{Results} Table \ref{table:validity-systematizations} compares rubric-based annotations applied by the research team with cultural appropriateness ratings from workshop participants. We report agreement, defined as the proportion of images for which the rubric’s final binary label matches the participants’ majority label. 
To contextualize these agreement rates, we also report the proportion of images where participants' labels of cultural appropriateness disagreed with each other (\% contested), as well as the overall proportion of images labeled as culturally appropriate by both.

We find that our rubric-based labels agree with participants’ majority judgments for over 80\% of the images shown in workshops, suggesting close alignment with community members’ evaluations on this dataset.
At the same time, participants’ judgments themselves exhibit considerable disagreement, ranging from 13\% to 63\% of images across artifacts, underscoring the inherently subjective and contested nature of cultural representation. This level of intra-community disagreement suggests that perfect alignment with participant ratings is neither achievable nor necessarily desirable, as any formalized measure will inevitably reflect particular interpretations within a heterogeneous community rather than a single, unified standard.

While this level of agreement is in part expected given that the rubric was derived from participant feedback, the result nonetheless serves as an important validation step. In particular, it demonstrates that qualitative, free-text feedback from community discussions can be systematically translated into concrete visual criteria and recomposed into an evaluation rubric that reproduces participants’ judgments at scale. 

\clearpage
\subsection{Supplemental results: Qualitatively comparing LLM-generated vs. community-informed rubrics}\label{apdx:rubric-comparisons}

\begin{figure}[t] 
\centering 
\includegraphics[width=\linewidth]{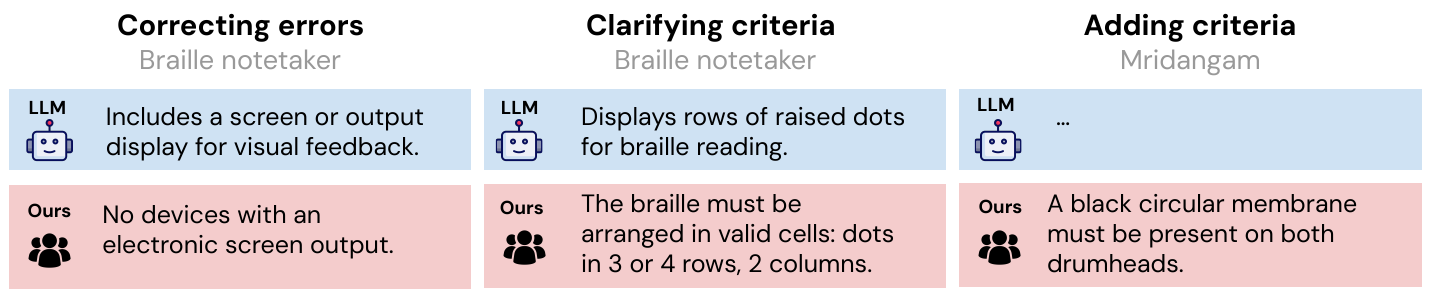} 
\caption{\textbf{Community-elicited rubrics differ meaningfully from those generated by off-the-shelf LLMs}. Our rubrics differ from LLM-generated rubrics in three ways, each illustrated using an example (Appendix \ref{apdx:rubric-comparisons}). First, LLM-generated rubrics can include factual or interpretive errors that reflect misunderstandings of the artifact (\eg{} whether a braille notetaker should have a screen). Second, our rubrics provide culturally grounded clarifications for features that LLM-generated rubrics leave underspecified. Third, our rubrics include additional criteria that LLM-generated rubrics omit, such as the black circular membrane on a Mridangam drum.} 
\label{fig:comparing-rubrics} 
\Description[Example criteria generated by LLMs vs. elicited from humans]{The figure contains example rubric criteria that illustrate three trends (organized in columns): ``correcting errors'' (left), ``clarifying criteria'' (middle), and ``adding criteria'') (right).}
\end{figure} 

\paragraph{Using an LLM to generate a rubric} We adapt the prompt that \citet{szymanski2024comparing} provide an LLM to generate rubrics, making small revisions to instruct the model to identify visual criteria that can be used to evaluate images.
We provide each prompt to GPT-4o and sample a single rubric per artifact.
The final set of generated rubrics contain 5-6 criteria that describe visual characteristics that must be present in a culturally appropriate depiction of each artifact.
 
\begin{rubricprompt}
\begin{verbatim}
System Prompt. You are a helpful and precise assistant that can create binary evaluation
criteria to evaluate images of cultural artifacts. Your task is to generate evaluation 
criteria for assessing whether an image contains a culturally appropriate depiction of an 
object. Each criterion should be a statement in which you would answer true/false. 
The criteria should describe the most important visual characteristics that should be present 
or absent in a correct depiction of the object. The criteria should not be in the form of a one 
sentence statement, not a question. You should return your final answer as a valid JSON object.

User Prompt. Create evaluation criteria for the given prompt instruction: 
"A photo of a guide cane"
\end{verbatim}
\end{rubricprompt}

\paragraph{Analysis methodology} We follow \citet{szymanski2024comparing} to conduct a simple qualitative comparison of LLM-generated and human-generated (community-informed) rubrics.
Because our dataset only contained six rubrics with a small set of criteria, our research team was able to manually review and assess the complete set of LLM-generated criteria.
To conduct the comparison, the first author annotated each LLM rubric in collaboration with community members on the research team, to highlight key differences between what was articulated in each LLM rubric, versus our own understandings of each artifact, as shaped by what we learned from workshops and also our own lived understandings of what these artifacts should be. For each criterion, we assessed if it was related to or overlapped with another criteria in our rubrics. We paid attention to places where each LLM rubric diverged from views that were articulated by the community, or our own understandings of each artifact. 
The final result of this analysis included an assessment of each individual criterion, and also broader trends in what was, or was not, captured by these LLM-generated rubrics, summarized in Figure \ref{fig:comparing-rubrics}.

\subsubsection{Results}

We share our research team's annotations of the rubric criteria and provide qualitative descriptions of how each rubric differs from our community-elicited rubrics for each artifact.

\begin{figure}[H] 
\centering 
\includegraphics[width=\linewidth]{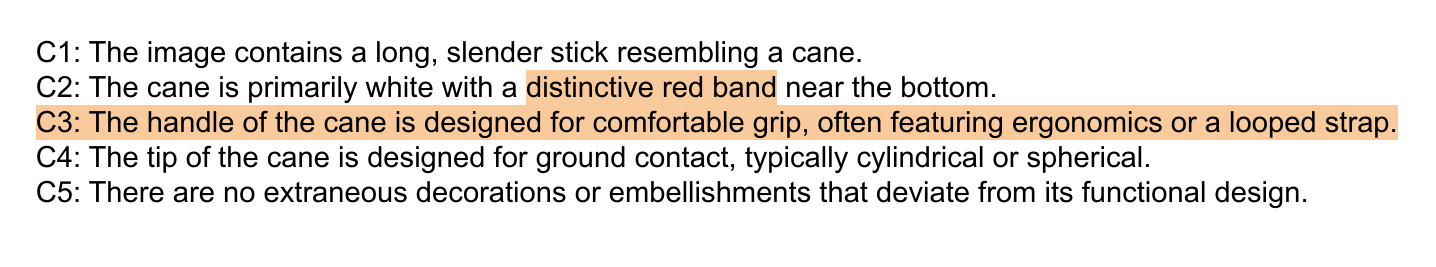} 
\caption{\textbf{Annotated LLM-generated rubric for a guide cane.} While generally providing an accurate description of a guide cane, the rubric misses several key details. The rubric does not provide a complete description of the straight handle shape of a cane (C2), a feature that is of critical importance to the community. In workshops, we learned that a band of red tape on a cane's body is often a visual signifier that the user is deaf-blind (C3), and thus should not be required for a culturally appropriate depiction.} 
\Description[Annotated rubric criteria for a guide cane]{The figure lists the LLM-generated rubric criteria, with certain criteria (C2, C3) highlighted to emphasize their misalignment with community members' understandings of each artifact.}
\label{fig:llm-rubric-guide-cane} 
\end{figure}

\begin{figure}[H] 
\centering 
\includegraphics[width=\linewidth]{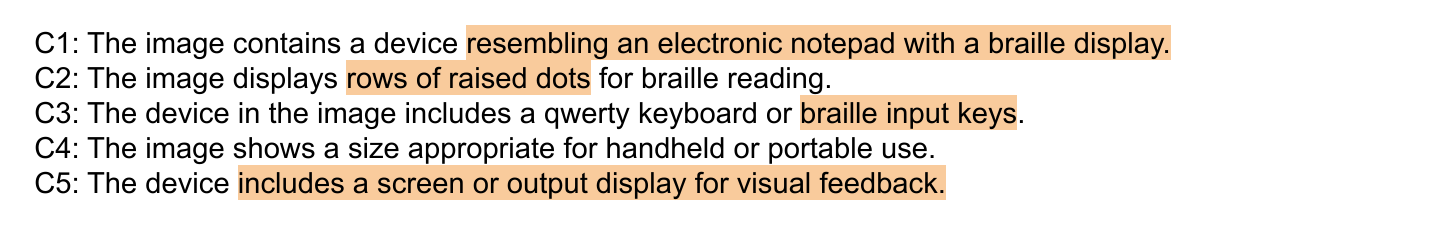} 
\caption{\textbf{Annotated LLM-generated rubric for a braille notetaker.} The rubric criteria include both inaccurate descriptions of a braille notetaking device, and do not include descriptive details about valid depictions of braille. A braille notetaker does not resemble a notepad (\eg{} it does not include a writing device such as a pen), and instead resembles a slim rectangular box (C1). The rubric does not provide a description of braille beyond ``raised dots'', which is underspecified as braille must be arranged in valid cells to be readable (C2). Similarly, the rubric does not describe the unique layout of ``braille input keys'' (C3). Many braille noteaking devices do not have a screen or visual display, as such displays are inaccessible to blind and low vision community members (C5).  } 
\Description[Annotated rubric criteria for a braille notetaker]{The figure lists the LLM-generated rubric criteria, with certain criteria (C1, C2, C3, C5) highlighted to emphasize their misalignment with community members' understandings of each artifact.}

\label{fig:llm-rubric-braille-notetaker} 
\end{figure}

\begin{figure}[H] 
\centering 
\includegraphics[width=\linewidth]{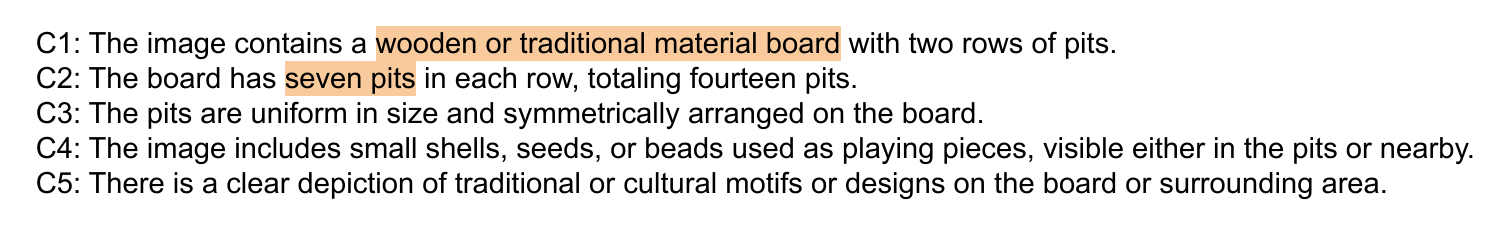} 
\caption{\textbf{Annotated LLM-generated rubric for Pallanguzhi.}  The rubric generally provides an accurate description of the most important characteristics of the a Pallanguzhi board, with two differences from the community-elicited rubric. (C1) Community members clarified that the color of the wood is important, and that Pallanguzhi baords are traditionally made of a deep-brown teakwood. (C2) The number of pits in each row of a Pallanguzhi board can vary between 5 and 7. They also emphasized the size of the playing pieces or tokens, noting that they should not be too small and should be similar to tamarind seeds or cowrie shells (C4).} 
\Description[Annotated rubric criteria for Pallanguzhi]{The figure lists the LLM-generated rubric criteria, with certain criteria (C1, C2) highlighted to emphasize their misalignment with community members' understandings of each artifact.}

\label{fig:llm-rubric-pallanguzhi} 
\end{figure}

\begin{figure}[H] 
\centering 
\includegraphics[width=\linewidth]{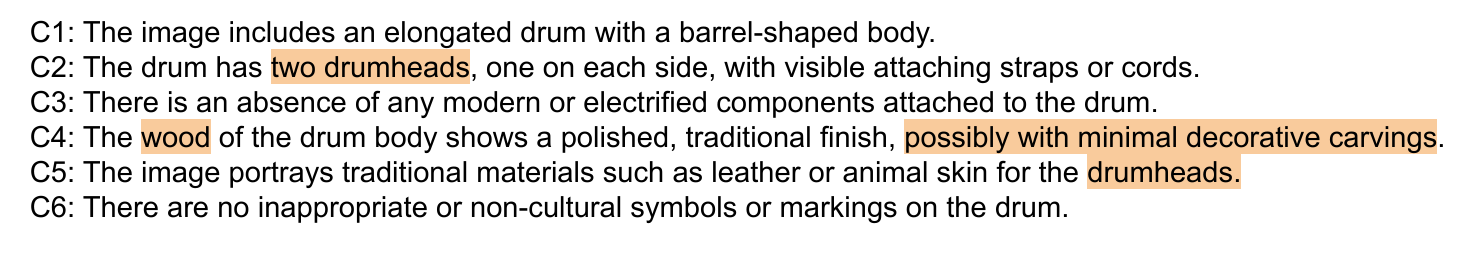} 
\caption{\textbf{Annotated LLM-generated rubric for a Mridangam.} The rubric lacks many of the critical details that distinguish the Mridangam from related drums and percussion instruments. One significant omission is the black circular membrane that must be present on both drumheads, a key feature that contributes to the timbre of the drum (C5). One drumhead is often slightly larger than the other (C2). The Mridangam should not be depicted with decorative patterns (C4). The rubric lacks details about the characteristic horizontal orientation of the drum, which must be played on its length.} 
\label{fig:llm-rubric-mridangam} 
\Description[Annotated rubric criteria for a Mridangam]{The figure lists the LLM-generated rubric criteria, with certain criteria (C2, C4, C5) highlighted to emphasize their misalignment with community members' understandings of each artifact.}

\end{figure}

\begin{figure}[H] 
\centering 
\includegraphics[width=\linewidth]{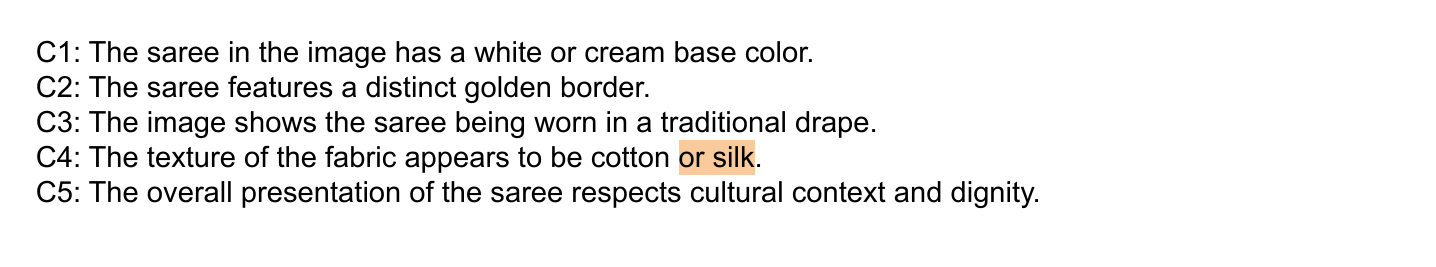} 
\caption{\textbf{Annotated LLM-generated rubric for a Kasavu saree.} The rubric generally provides an accurate description of a Kasavu saree, demonstrating substantial overlap with the community-elicited rubric. However, community members were clear that the material must be cotton and not silk (C4). } 
\label{fig:llm-rubric-kasavu-saree} 
\Description[Annotated rubric criteria for a Kasavu saree]{The figure lists the LLM-generated rubric criteria, with certain criteria (C4) highlighted to emphasize their misalignment with community members' understandings of each artifact.}

\end{figure}

\begin{figure}[H] 
\centering 
\includegraphics[width=\linewidth]{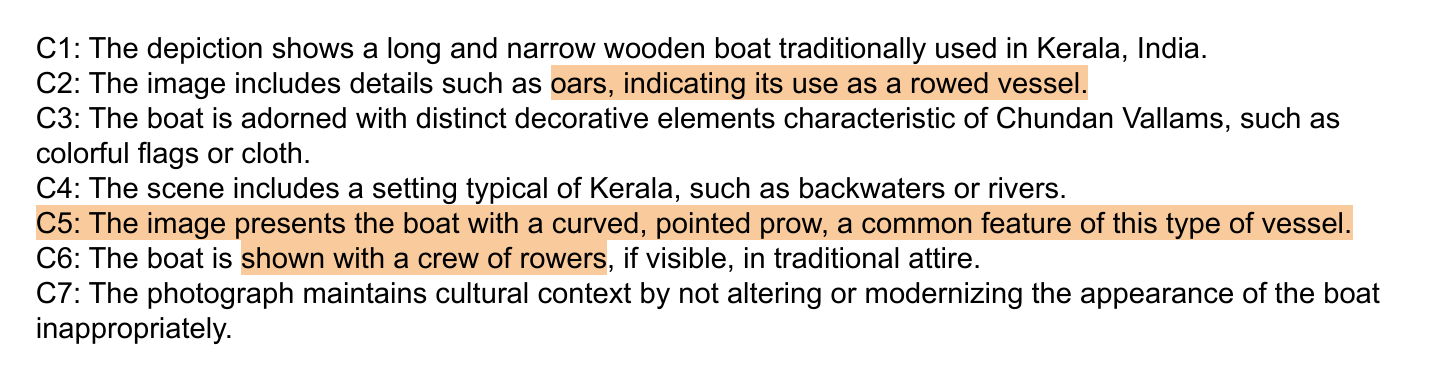} 
\caption{\textbf{Annotated LLM-generated rubric for Chundan Vallam.} {The rubric criteria cover the general structure of the Chundan Vallam but do not specify its defining features. In particular, they omit details about the oar structure and handling (C2); community members specified that the oars should be long, angled downward toward the water, and that each oarsman must use a single oar. The rubrics also do not specify the distinct characteristic of the stern being a straight, pointed tip (C5). Additionally, they lack guidance on the seating position of rowers; as emphasized by community members, when rowers are visible in images, they should be seated in pairs and face towards the stern (C6).} 
}
\Description[Annotated rubric criteria for Chundan Vallam]{The figure lists the LLM-generated rubric criteria, with certain criteria (C2, C5, C6) highlighted to emphasize their misalignment with community members' understandings of each artifact.}

\label{fig:llm-rubric-chundan-vallam} 
\end{figure}

\subsection{Extended methodology: Applying rubric criteria using human and LLM annotators }\label{apdx:methods-operationalization}

In this section, we provide an extended description of the methods that we adopted to operationalize (apply) our evaluation rubrics to score a new dataset of images in Section \ref{sec:results-operationalization}. We first describe how we generated a dataset of 50 images per each artifact. We then provide an extended description of the methods that we used to apply the rubrics manually (by collecting annotations from a member of our research team), and using a multimodal LLM-as-a-judge pipeline.

\subsubsection{Generating images}\label{apdx:new-daataset-of-images} We generated 10 images each using five image generation models that achieved state-of-the-art performance at the time of our study.
These include the two models from our case studies~\cite{openai_dalle_2022,dalle3,sd3} and three newer models, including OpenAI's latest, GPT-Image-1~\cite{gptimage1, flux2024, sd35}.
When generating images with GPT-Image-1 or \dalle{} 3, we used the simple prompt ``A photo of a \{artifact\}''\footnote{\dalle{} 3 automatically enriches this prompt with more detail.}.
For the Stable Diffusion and Flux models, we expanded this prompt with more detailed descriptions of each artifact to improve the depictions' quality as these models performed poorly out-of-the-box.
For the assistive technologies, we used prompts containing a description written by a community member on our research team (Figure \ref{table:generation-prompts}).
For the Indian cultural artifacts, we used revised \dalle{} 3 prompts as detailed in Appendix \ref{apdx:india-revised-prompts}.

\subsubsection{Manually applying the rubric criteria}\label{apdx:methods-manual-application}
To manually apply the rubric criteria, we collect human annotations from the research team member who \emph{created} each rubric by facilitating workshops with community members.  This set-up was chosen because the rubric creator would arguably be best positioned to apply the rubric consistently and in a way that is aligned with the community members' interpretations of each criterion since they have context from facilitating workshops. 
We conduct an additional exploration of inter-annotator agreement and the role that cultural expertise plays in manual rubric application in Appendix \ref{apdx:iaa}.

\subsubsection{Applying the MLLM judge}\label{apdx:gpt4o-judge}
We adopt GPT-4o as our judge model due to its demonstrated performance on vision-language understanding tasks \cite{openai2024gpt4o}.
To obtain annotations of each rubric criteria, we 
query GPT-4o by providing it with an AI image, and a user prompt written based on past LLM-as-a-judge evaluator prompt templates ~\cite{hashemi2024rubric} below.
We query GPT-4o using API calls with temperature set to the API default of $1$.
Because GPT-4o's outputs are stochastic, we report all results averaged over five runs of each LLM per AI image.

\begin{systemprompt}
\begin{verbatim}
You are an expert at assessing whether AI-generated images of a guide cane 
meet a set of evaluation criteria.
You will be provided with:
1) An image generated from the prompt "A photo of a guide cane".
2) A set of evaluation criteria, grouped by theme where each theme includes a description 
that should inform your interpretation of the criteria within that theme.

Evaluation Criteria
Theme1
The object in the image should not be confused with other, more hegemonic objects, 
such as objects that are used predominantly by people who are sighted.

C1: No wooden walking sticks.
C2: No decorative striped patterns (e.g., candy canes).

Theme2
The object needs to be functional as an assistive technology, and usable by 
someone who is blind.

C1: No deformed canes.
C2: No curved (crooked) handles.
C3: The cane must be shaped like a long (5-foot) stick.
C4: The body must have sections that are a white color.
C5: There must be a tip at the bottom of the cane.

Your Task
Determine whether the image meets each criterion, based on:

1) The generated image 
2) The text prompt.
3) Your understanding of a guide cane.
4) The description of the criteria, and the theme it belongs to.

For each criterion, respond with:

1 if the image meets the criterion.
0 if the image does not meet the criterion.

Return your evaluation as a valid JSON object with the following structure:
{
  "criteria_evaluation": {
    "Theme1": {
      "C1": 1 or 0,
      "C2": 1 or 0
    },
    "Theme2": {
      "C1": 1 or 0,
      "C2": 1 or 0,
      "C3": 1 or 0,
      "C4": 1 or 0,
      "C5": 1 or 0
    }
  },
  "overall_assessment": 1 or 0
}
The "overall_assessment" should be 1 ONLY if all criteria across all themes are met 
(i.e., all values are 1); otherwise, it should be 0.
Ensure the JSON is properly formatted and valid.
\end{verbatim}
\end{systemprompt}

\subsection{Supplemental results: Exploring inter-annotator agreement for manual rubric application}\label{apdx:iaa}

In this section, we conduct an exploration of inviting non-subject matter expert annotators to manually annotate the rubric criteria.  
As discussed in Appendix \ref{apdx:methods-manual-application}, the results presented in the main text of the paper (Section \ref{sec:results-operationalization}) use human annotations collected from the author who facilitated workshops -- the rubric designer. However, we recognize that in practice, industry practitioners often hire paid data workers that may lack subject-matter expertise to perform the data work of evaluating generative AI outputs ~\citep{weidinger2023sociotechnical,hall2024towards,zhang2025aura}. Thus, practitioners may wish to explore how they can set up an annotation task that can enable non-expert annotators to interpret and apply evaluation rubric criteria in a way that is consistent with the initial intent of the rubric designer.

To explore the possibility of asking a non-expert annotator to apply the rubric, a coauthor who did not directly engage with participants applied each rubric to label images.  These annotators were encouraged to take detailed notes throughout the annotation process, noting down when they were uncertain about how to interpret particular criteria or assess borderline cases.  They then met with the rubric creator to compare their annotations, share notes, and discuss their differing interpretations of each rubric.

\begin{table*}[h!]
\centering
{\begin{tabular}{ cc}
\toprule 

Artifact & Human Annotator Agreement  \\ \midrule 
Guide cane & 0.64 \\
Braille notetaker & 0.88 \\
Pallanguzhi & 0.92 \\
Mridangam & 0.94 \\
Kasavu saree & 0.94 \\
Chundan Vallam & 1.00 \\

\bottomrule 
\end{tabular}}
\caption{\textbf{Agreement between the final labels assigned by two human annotators}. Using 50 generated images per artifact, we report the proportion of images where the final ``cultural appropriateness'' label assigned by the primary annotator (who facilitated workshops with community members) agrees with that of a second annotator (who did not engage with the community).}
\label{table:iaa}
\end{table*}

Table \ref{table:iaa} reports the proportion of images for which the two annotators' final (binary) assessment of cultural appropriateness agreed.  We find that inter-annotator agreement varies substantially across artifacts and is particularly low for the guide cane.  
In what follows, we draw from annotators' notes and conversations to unpack some sources of disagreement and identify opportunities for future work exploring how to support the human application of evaluation rubrics.
To unpack the possible causes for inter-rater agreement, we calculated the dis-aggregated agreement rate for each individual criterion.  In conversations, annotators found that many disagreements could be explained by a lack of clarity in the language used to describe each criterion. 
These varying interpretations of the same simple rubric text reveals the difficulty of being precise when we attempt to systematize exactly what we mean, in written language \citep{wallach2025position}.  While a commonly articulated hope is that LLM-as-a-judge rubric criteria are those where ``a majority of readers should agree on whether a given model response satisfies the criterion'' \citep{akyurek2025prbench}, we found that annotators' own subjectivity and interpretations shaped how they applied our rubric criteria.

For some criteria, disagreements between annotators could potentially be resolved by simple changes to the semantics of the criterion.
For example, one annotator interpreted the rubric item ``\emph{the body (of the guide cane) must have sections that are a white color}'' (agreement = 0.88) to mean that the cane must have several (at least two) such sections, when the rubric designer's intent was for a cane with any white sections to pass.  
Such an error could be easily clarified by making a simple and direct change to the rubric text (\eg{} to ``\emph{at least one section}'').

Yet, other disagreements revealed that some criteria were underspecified, leaving room for varying annotator interpretations.
In such cases, rubric designers often drew from their memories of participants' decisions, criteria, and justifications from workshops to apply the criterion in a way that was aligned with community members' preferences.
However, this context was not held by non-expert annotators, whose interpretations occasionally strayed from the intent of the rubric designer.
For example, annotating the criterion that ``\emph{the Kasavu saree must not contain heavy embellishment}'' (agreement = 0.58) required annotators to make subjective assessments about what qualified as (in)appropriately ``heavy'' amounts of embellishment. 
When annotating, the rubric designer drew from their knowledge of what participants had described as their preferences for embellishment in workshops: in conversations, participants agreed that it was okay for the saree's woven golden border to be embellished (\eg{} woven in a mostly-solid decorative pattern), and that the criterion was written to differentiate between depictions that had undesirable additional embellishment on the saree's body. 
However, the non-expert annotator who was missing this context interpreted the criterion differently to mean that absolutely no embellishment should be permitted.
Disagreement across annotators was common for other criteria that required annotators to make subjective assessments about whether a guide cane's pattern was ``too decorative'', a Pallanguzhi looked enough like another board game, or whether a Chundan Vallam boat was appropriately long and narrow.
In such scenarios, rubrics designers could often draw upon additional knowledge of concrete visual indicators that community members used to make their assessments, and knowledge of their past ratings, that were not explicitly articulated in the rubric criteria themselves.

To address these challenges in being able to interpret and apply written annotation guidelines, practitioners can draw inspiration from past work in human computation and crowdsourcing.
For example, a lack of clarity in how to interpret and apply the written criteria might be navigated by encouraging discussion between annotators \citep{callison2009fast,chang2017revolt}, revising the rubric language to be more clear \citep{akyurek2025prbench}, or providing annotators with more context, \eg{} in the form of an annotation guide that contains example images or guidelines on how to navigate borderline examples.
This additional context might better prepare non-subject matter experts who did not participate in rubric construction to apply the rubrics in a way that is aligned with the rubric designer's intentions of synthesizing and describing the concrete visual indicators that influence community members' ratings of cultural appropriateness.
We hypothesize that iterating on the wording of criteria until higher inter-rater agreement is achieved may also yield more clearly written descriptions of criteria that lead to higher accuracy MLLM-as-a-judge implementation as well.

\newpage

\clearpage
\subsection{Supplemental results: What does manual application of rubrics reveal about cultural representation?}\label{apdx:operationalization-interpretations}

In this section, we analyze and interpret the results of \emph{manually applying} the rubric criteria, to gain insight on the types of errors made by state-of-the-art models, providing an extended discussion of the findings presented in Section \ref{sec:interpreting-rubric-results}.
Table~\ref{table:appropriateness-by-model} displays the percentage of images that were determined to be culturally appropriate by human annotators, for each artifact and model.
The results reveal that the vast majority of generated images are classified as culturally inappropriate.
This highlights that most models are failing to meet community-defined standards for cultural appropriateness, leaving few images that could be shown without reservations.
Of the five models, the most recently released, GPT-Image-1, has the highest number of images determined to be culturally appropriate.
Other models, such as Flux.1 DEV, consistently failed to produce a single appropriate image of any of the artifacts.

\begin{figure}[H]
    \centering
    \includegraphics[width=\linewidth]{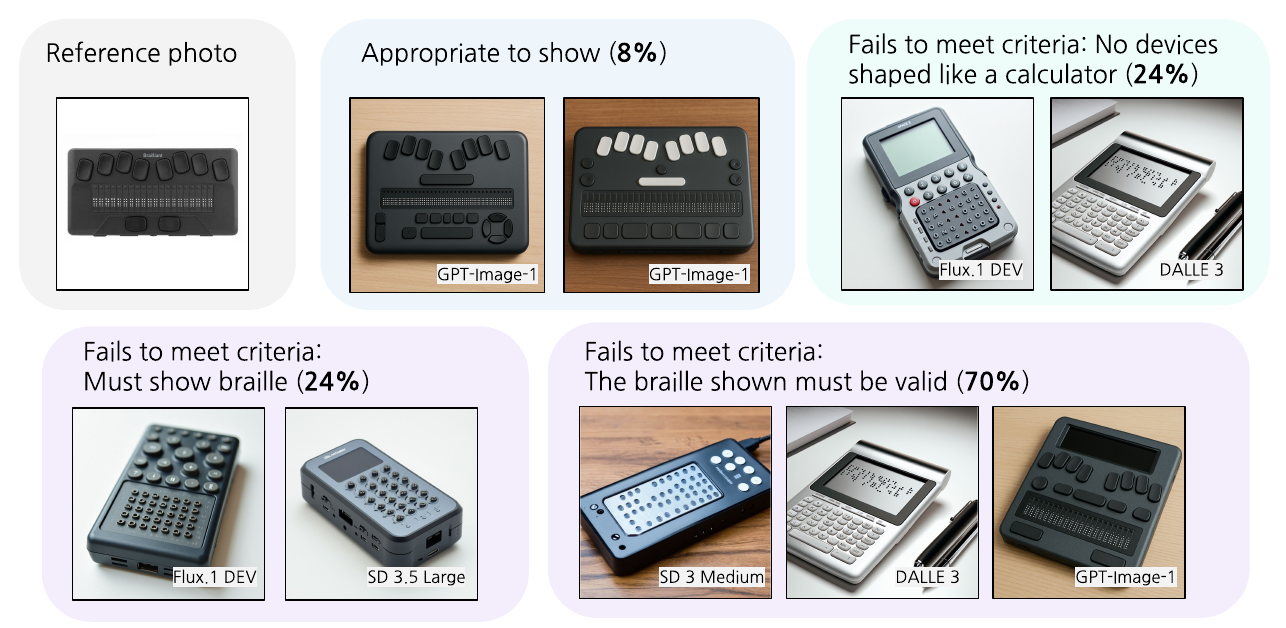}
\caption{\textbf{Criterion-level annotations provided by humans reveal the specific representational errors that make depictions of a braille notetaker inappropriate}. The figure displays a reference photo of a braille notetaker, and example AI-generated images that fall into one of four groups (as annotated by humans): (1) images that are appropriate to show (all criteria are met), (2) images that do not meet Theme 2, Criteria 2 (``No devices that are shaped like handheld calculators with an electronic screen output''), (3) images that do not meet Theme 1, Criteria 2 (``The device must show braille''), and (4) images that do not meet Theme 1, Criteria 4 (``Depictions of braille must be valid: arranged in cells with 3 or 4 rows, and 2 columns''). The figure displays the percentage of the 50 AI-generated images that fall in each group.}
\Description[Example images that fail to meet four rubric criteria for a braille notetaker]{}
\label{fig:criteria-examples}
\end{figure}

\begin{figure}[H]
    \centering
    \includegraphics[width=0.9\linewidth]{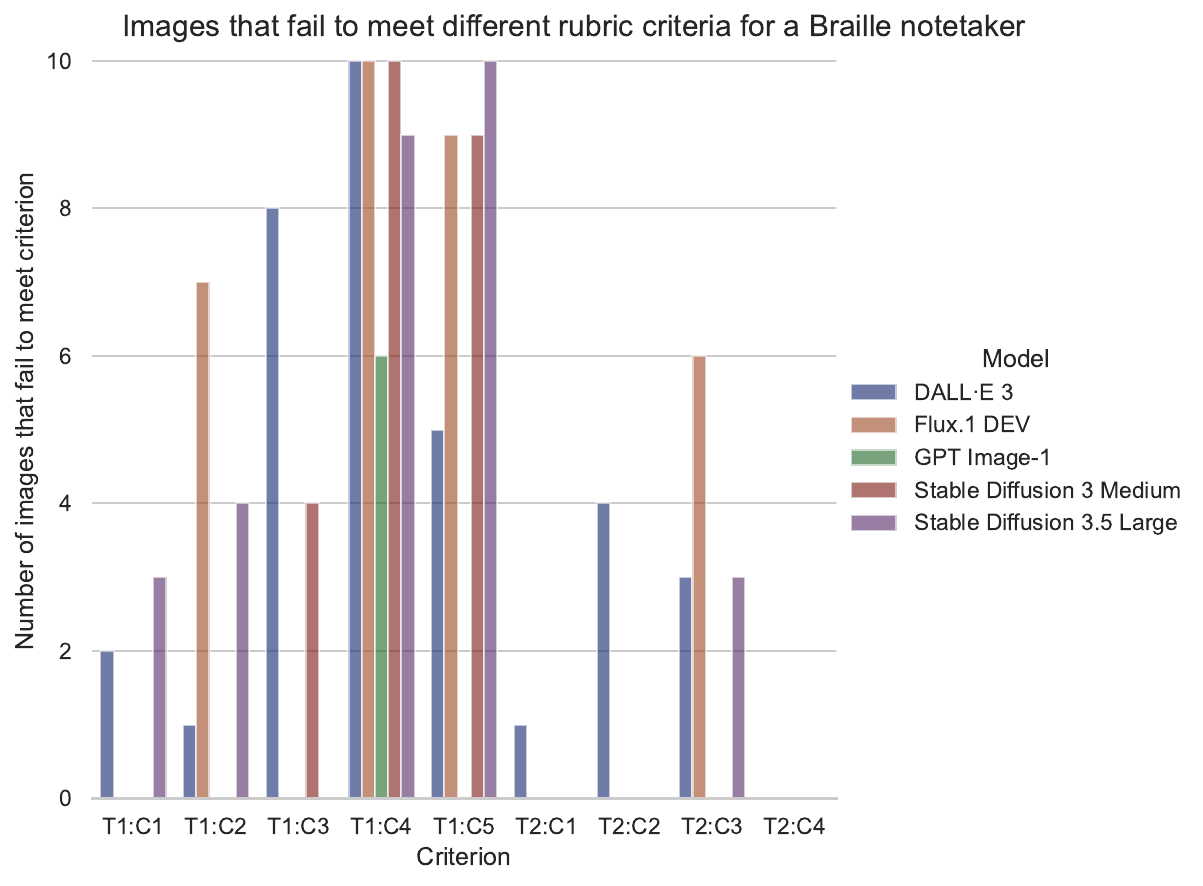}
\caption{\textbf{Comparing (manual) rubric application across models for a braille notetaker.} The frequency at which different criteria are violated (reported here using annotations provided by humans) varies across different models. For example, the GPT Image-1 images of braille notetakers that are inappropriate to show are all violate Theme 1, Criteria 4 (failing to depict valid braille). In contrast, images generated by Flux.1 DEV fail to meet a variety of different criteria, including Theme 1, Criteria 2 (failing to depict any braille) and Theme 1, Criteria 5 (failing to depict an input keyboard so that users can write). For some criteria, only some models, but not others, fail. For instance, only \dalle{} 3 fails to meet Theme 2, Criteria 2 by depicting some braille notetaker instead using a laptop device.}
\label{fig:barplot:braille}
\Description[Histogram displaying the frequency that different models (bar colors) violate different rubric criteria (bar height) for a braille notetaker.]{The histogram illustrates how different models tend to violate different criterion (see caption for more information).}

\end{figure}

\begin{figure}[H]
    \centering
    \includegraphics[width=0.9\linewidth]{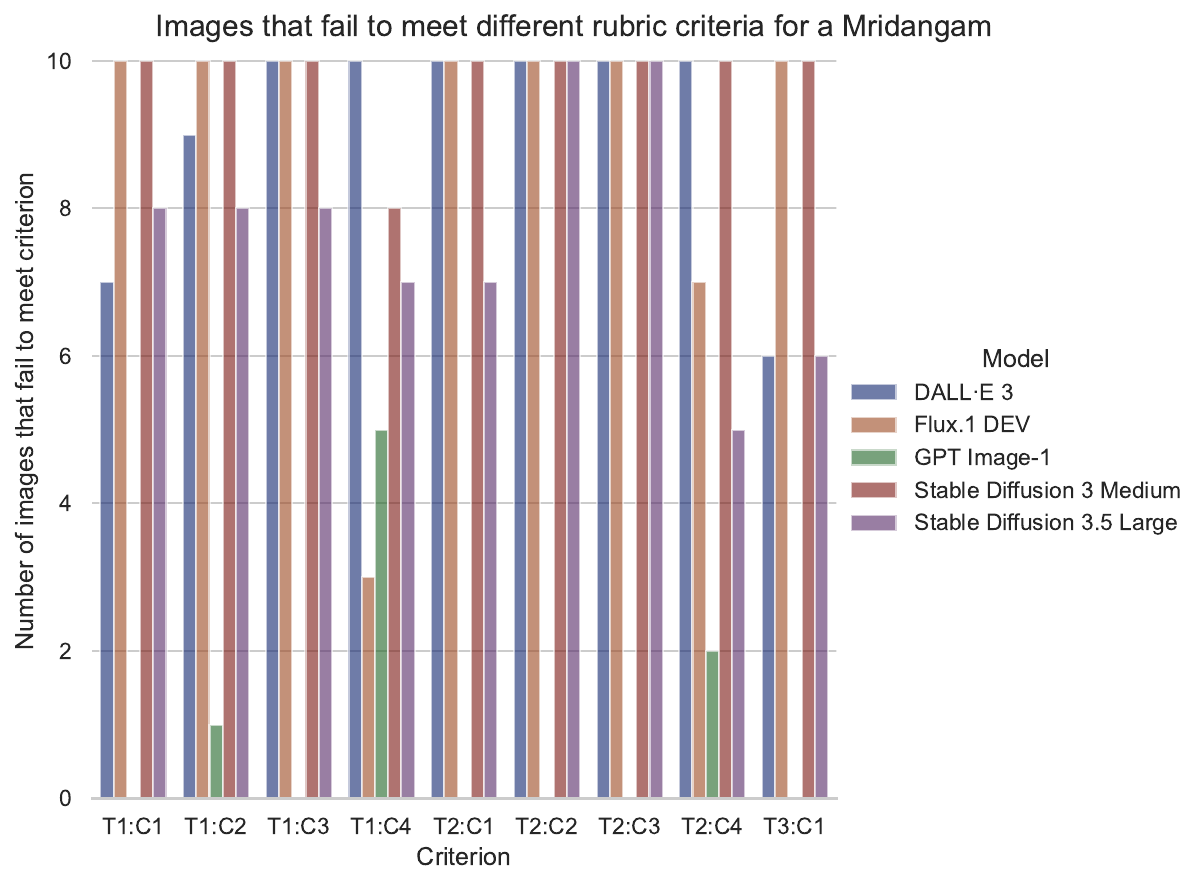}
\caption{\textbf{Comparing (manual) rubric application across models for a Mridangam.} Comparing the frequency at which different criteria are violated across models allows practitioners to draw interpretable insights about models' failure modes. With the exception of GPT Image-1, many of the models (\ie{} \dalle{} 3, Flux.1 DEV, and Stable Diffusion 3 Medium) consistently fail to meet several criteria, such as failing to depict a drum that is made of a light wooden material (Theme 1, Criteria 3), failing to depict the drum's characteristic Black circular membrane (Theme 2, Criteria 2), and excluding the long longitudinal straps that run along the drum's body (Theme 2, Criteria 4). In contrast, GPT Image-1 does frequently correctly represent many features that are characteristic to the Mridangam, but occasionally misportrays the drum by adding embellishments or detailed patterns to its body (Theme 1, Criteria 4).}
\label{fig:barplot:mridangam}
\Description[Histogram displaying the frequency that different models (bar colors) violate different rubric criteria (bar height) for a Mridangam.]{Many of the bars have high frequency counts, illustrating how the images generated by several models tend to consistently violate the Mridangam rubric criteria.}
\end{figure}

\begin{figure}[H]
    \centering
    \includegraphics[width=0.9\linewidth]{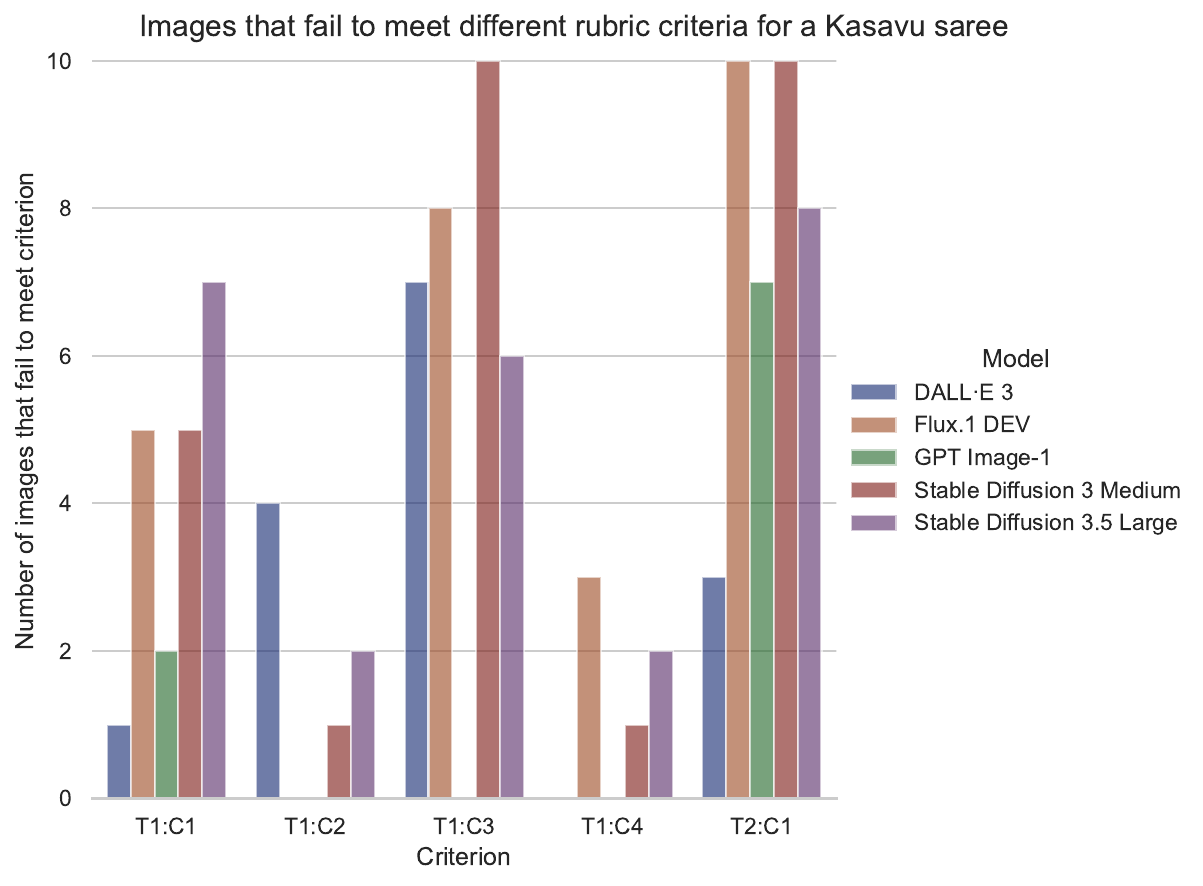}
\caption{\textbf{Comparing (manual) rubric application across models for a Kasavu Saree.} Visualizing the breakdown of criteria that are violated by different image generation models reveals interpretable insights about model behavior. For example, only Flux.1 and the Stable Diffusion models depict the saree with additional unnecessary embellishment (Theme 1, Criteria 4). The GPT Image-1 images consistently depict the saree using the correct color and material, but sometimes fail to display the saree in a way that presents it characteristics pleats and drape (Theme 2, Criteria 1).}
\label{fig:barplot:saree}
\Description[Histogram displaying the frequency that different models (bar colors) violate different rubric criteria (bar height) for a Kasavu saree.]{The histogram illustrates how different models tend to make different types of errors (see figure caption).}
\end{figure}

Beyond examining the final labels assigned to each image, practitioners can also inspect which specific criteria from the systematized concepts were not satisfied. 
Different models make different errors of representation, which are captured by different rubric criteria.
Figure \ref{fig:criteria-examples} illustrates this with example depictions of a braille notetaker: some images were deemed culturally appropriate, while others failed on particular criteria. The examples show how our systematization criteria capture meaningful representational errors in a new dataset of generated images, such as whether a depiction of a braille notetaker does not attempt to show braille at all, or the braille it shows is not valid. 
By analyzing which criteria applied, practitioners can develop a more nuanced understanding of the characteristic errors made by each model, as demonstrated in Figures \ref{fig:barplot:braille}, \ref{fig:barplot:mridangam}, and \ref{fig:barplot:saree}.

\begin{table}[htbp]
\small
\centering
\begin{tabular}{ ccccccc}
\toprule 

Artifact & \textbf{\dalle{} 3} & \textbf{Flux.1 DEV} & \textbf{GPT Image-1} & \textbf{SD 3 Medium} & \textbf{SD 3.5 Large} & Total (Appropriate) \\
Guide cane & 0.40 & \textcolor{lightgray}{0.00} & 0.90 & 0.30 & 0.40 & 0.40 \\ 
Braile notetaker & \textcolor{lightgray}{0.00} & \textcolor{lightgray}{0.00} & 0.40 & \textcolor{lightgray}{0.00} & \textcolor{lightgray}{0.00} & 0.08 \\ 
Pallanguzhi & 0.10 & \textcolor{lightgray}{0.00} & 0.80 & \textcolor{lightgray}{0.00} & \textcolor{lightgray}{0.00} & 0.18 \\ 
Mridangam & \textcolor{lightgray}{0.00} & \textcolor{lightgray}{0.00} & 0.50 & \textcolor{lightgray}{0.00} & \textcolor{lightgray}{0.00} & 0.10 \\ 
Kasavu saree & 0.20 & \textcolor{lightgray}{0.00} & 0.30 & \textcolor{lightgray}{0.00} & 0.10 & 0.12 \\ 
Chundan Vallam & \textcolor{lightgray}{0.00} & \textcolor{lightgray}{0.00} & \textcolor{lightgray}{0.00}& \textcolor{lightgray}{0.00} & \textcolor{lightgray}{0.00} & 0.00 \\ 
\midrule 

\bottomrule 
\end{tabular}
\caption{\textbf{Few images generated by state-of-the-art models are culturally appropriate when scored using our systematized concepts}. The table shows the percentage of images generated by each model that are culturally appropriate: where all of the criteria are met (annotated by our research team).}
\label{table:appropriateness-by-model}
\end{table}

\clearpage

\clearpage
\subsection{Supplemental results: Evaluating human-MLLM agreement at applying community-informed rubrics}\label{apdx:human-mllm-agreement}
We report the agreement between human and MLLM annotations for each of the rubric criteria. Columns further disaggregate and report agreement by images that the human labeled as appropriate, versus inappropriate. Note that we do not report agreement for groups that contain zero images (columns with an ``N/A'' value).
We also report the base rates of the proportion of images that the human versus the MLLM assigned positive labels to.

\begin{table}[H]
\small
\centering
\caption{Comparing human to MLLM annotations applying the rubric for a \textbf{guide cane}. See Section \ref{apdx:human-mllm-agreement} for a complete description of each column.}
\label{tab:op-agreement-guidecane}
\begin{tabular}{l C{4cm} c g c c c c}
\toprule
\textbf{Criteria} & \textbf{Description} &  \textbf{\shortstack{Human\\{\footnotesize(\% Appropriate)}}} & \textbf{\shortstack{MLLM\\{\footnotesize(\% Appropriate)}}} & \shortstack{\textbf{Agreement}\\{\footnotesize Overall}} & \shortstack{\textbf{Agreement}\\{\footnotesize Appropriate}} & \shortstack{\textbf{Agreement}\\{\footnotesize Inappropriate}} \\
\midrule

Final label & {} & 0.40 & 0.44 & 0.84 & 0.84 & 0.83 \\ \lightrule

T1, C1 & No deformed canes.	& 0.82 & 0.98 & 0.84 & 1.00 & 0.11 \\
T1, C2 & No curved (crooked) handles. & 0.58 & 0.68 & 0.87 & 0.97 & 0.72 \\
T1, C3 & The cane must be shaped like a long (5-foot) stick. & 0.68 & 0.76 & 0.72 & 0.85 & 0.44 \\
T1, C4 & The body must have sections that are a white color. & 1.00 & 0.98 & 0.98 & 0.98 & N/A \\ 
T1, C5 & There must be a tip at the bottom of the cane. & 0.94 & 0.97 & 0.95 & 0.99 & 0.33 \\ \lightrule
T2, C1 & No wooden walking sticks. & 1.00 & 0.96 & 0.96 & 0.96 & N/A \\ 
T2, C2 & No decorative striped patterns (e.g., candy canes). & 0.80 & 0.71 & 0.84 & 0.85 & 0.84 \\
\bottomrule
\end{tabular}
\end{table}

\begin{table}[H]
\small
\centering
\caption{Comparing human to MLLM annotations applying the rubric for a \textbf{Kasavu saree}. See Section \ref{apdx:human-mllm-agreement} for a complete description of each column. }
\label{tab:op-agreement-saree}
\begin{tabular}{l C{4cm} c g c c c c}
\toprule
\textbf{Criteria} & \textbf{Description} &  \textbf{\shortstack{Human\\{\footnotesize(\% Appropriate)}}} & \textbf{\shortstack{MLLM\\{\footnotesize(\% Appropriate)}}} & \shortstack{\textbf{Agreement}\\{\footnotesize Overall}} & \shortstack{\textbf{Agreement}\\{\footnotesize Appropriate}} & \shortstack{\textbf{Agreement}\\{\footnotesize Inappropriate}} \\
\midrule

Final label & {} & 0.12 & 0.21 & 0.88 & 0.87 & 0.88 \\ \lightrule
T1, C1 & It must not resemble other items like tablecloth, Kerala Mundu or curtains. & 0.60 & 0.88 & 0.60 & 0.90 & 0.14 \\
T1, C2 & The saree color must be off-white with a medium wide (3-5 inch) woven gold border. & 0.86 & 0.97 & 0.84	& 0.97 & 0.03 \\
T1, C3 & The saree must be made of crisp cotton fabric throughout. & 0.38 & 0.72 & 0.56 & 0.87 & 0.37 \\
T1, C4 & The saree must not contain heavy embellishments. & 0.88 & 0.76 & 0.80 & 0.82 & 0.67 \\ \lightrule
T2, C1 & The saree must be shown in a way that clearly presents its pleats and drape. & 0.24 & 0.28 & 0.89 & 0.87 & 0.90 \\
\bottomrule
\end{tabular}
\end{table}

\begin{table}[H]
\small
\centering
\caption{Comparing human to MLLM annotations applying the rubric for a \textbf{braille notetaker}. See Section \ref{apdx:human-mllm-agreement} for a complete description of each column. }
\label{tab:op-agreement-braillenote}
\begin{tabular}{l C{4cm} c g c c c c}
\toprule
\textbf{Criteria} & \textbf{Description} &  \textbf{\shortstack{Human\\{\footnotesize(\% Appropriate)}}} & \textbf{\shortstack{MLLM\\{\footnotesize(\% Appropriate)}}} & \shortstack{\textbf{Agreement}\\{\footnotesize Overall}} & \shortstack{\textbf{Agreement}\\{\footnotesize Appropriate}} & \shortstack{\textbf{Agreement}\\{\footnotesize Inappropriate}} \\
\midrule
Final label & {} &	0.08 & 0.20 & 0.82 & 0.65 & 0.83 \\ \lightrule
T1, C1 & The device must be shaped like a thin rectangular box. & 0.90 & 0.88 & 0.83 & 0.89 & 0.28 \\
T1, C2 & The device must show braille. & 0.76 & 0.86 & 0.78 & 0.92 & 0.35 \\
T1, C3 & All depictions of braille must be tactile (embossed). No depictions of braille on electronic screens. & 0.76 & 0.57 & 0.68 & 0.67 & 0.73 \\ 
T1, C4 & Depictions of braille must be valid: arranged in cells with 3 or 4 rows, and 2 columns. & 0.10 & 0.66 & 0.39 & 0.76 & 0.35 \\ 
T1, C5 & The device can have a qwerty keyboard, or a Braille keyboard. A braille keyboard must have 3 or 4 keys (right), space bar, 3 or 4 keys (left). These keys are positioned next to each other in a straight horizontal line. & 0.34 & 0.34 & 0.74 & 0.62 & 0.81 \\
 \lightrule
T2, C1 & No depictions of notetaking as writing (using a pen) on paper. & 0.98 & 0.89 & 0.90 & 0.91 & 0.80 \\ 
T2, C2 & No devices that are shaped like laptops with an electronic screen output. & 0.92 & 0.67 & 0.73 & 0.72 & 0.85 \\ 
T2, C3 & No devices that are shaped like handheld calculators, with an electronic screen output. & 0.76 & 0.95 & 0.77 & 0.97 & 0.13 \\
T2, C4 & No devices that are shaped like manual typewriters. & 1.00 & 1.00 & 1.00 & 1.00 & N/A \\ 
\bottomrule
\end{tabular}
\end{table}

\begin{table}[H]
\small
\centering
\caption{Comparing human to MLLM annotations applying the rubric for a \textbf{Pallanguzhi}. See Section \ref{apdx:human-mllm-agreement} for a complete description of each column. }
\label{tab:op-agreement-pallanguzhi}
\begin{tabular}{l C{4cm} c g c c c c}
\toprule
\textbf{Criteria} & \textbf{Description} &  \textbf{\shortstack{Human\\{\footnotesize(\% Appropriate)}}} & \textbf{\shortstack{MLLM\\{\footnotesize(\% Appropriate)}}} & \shortstack{\textbf{Agreement}\\{\footnotesize Overall}} & \shortstack{\textbf{Agreement}\\{\footnotesize Appropriate}} & \shortstack{\textbf{Agreement}\\{\footnotesize Inappropriate}} \\
\midrule
Final label & {} & 0.18 & 0.12 & 0.78 & 0.22 & 0.90 \\ \lightrule
T1, C1 & It should not resemble other board games (like Monopoly, Tic Tac toe, etc). & 0.86 & 0.96 & 0.89 & 0.99 & 0.26 \\ 
T1, C2 & The game board must be symmetrical along the length and consist of two or three rows of pits. The rows should have at least 5 pits. & 0.78 & 0.75 & 0.64 & 0.75 & 0.25 \\ 
T1, C3 & The game board can be fish or rectangular in shape. & 0.68 & 0.94 & 0.72 & 0.98 & 0.16 \\
T1, C4 & The game board must be made out of teakwood. & 0.54 & 0.80 & 0.56 & 0.83 & 0.24 \\ 
T1, C5 & The pits must be circular and evenly spaced. & 0.86 & 0.97 & 0.83 & 0.96 & 0.00 \\ \lightrule
T2, C1 & The size of the tokens should not be too small. The tokens should be distributable by hand. & 0.60 & 1.00 & 0.60 & 1.00 & 0.01 \\ 
T2, C2 & The pits should be big enough to accommodate multiple tokens. & 0.84 & 1.00 & 0.84 & 1.00 & 0.00 \\ \lightrule
T3, C1 & The tokens can be cowrie shells or tamarind seeds. & 0.18 & 0.18 & 0.73 & 0.27 & 0.83 \\ 
\bottomrule
\end{tabular}
\end{table}

\begin{table}[H]
\small
\centering
\caption{Comparing human to MLLM annotations applying the rubric for a \textbf{Mridangam}. See Section \ref{apdx:human-mllm-agreement} for a complete description of each column. }
\label{tab:op-agreement-mridangam}
\begin{tabular}{l C{4cm} c g c c c c}
\toprule
\textbf{Criteria} & \textbf{Description} &  \textbf{\shortstack{Human\\{\footnotesize(\% Appropriate)}}} & \textbf{\shortstack{MLLM\\{\footnotesize(\% Appropriate)}}} & \shortstack{\textbf{Agreement}\\{\footnotesize Overall}} & \shortstack{\textbf{Agreement}\\{\footnotesize Appropriate}} & \shortstack{\textbf{Agreement}\\{\footnotesize Inappropriate}} \\
\midrule

Final label & {} & 0.10 & 0.21 & 0.84 & 0.76 & 0.85 \\ \lightrule
T1, C1 & It must not resemble other percussion instruments (like Tabla, Drum, Damaru, Dhol). & 0.30 & 0.42 & 0.86 & 0.97 & 0.82 \\ 
T1, C2 & The instrument must be long, barrel-shaped, and tapered at both ends, each ending in a rounded, double-headed form, with one end slightly larger than the other. & 0.24 & 0.36 & 0.88 & 1.00 & 0.85 \\ 
T1, C3 & The body of the instrument must be made out of jackwood. & 0.24 & 0.40 & 0.72 & 0.77 & 0.71 \\ 
T1, C4 & There must not be intricate design or detailed patterns on the body. & 0.34 & 0.86 & 0.44 & 0.94 & 0.18 \\ \lightrule
T2, C1 & The heads of the instrument must be stretched goat, cow or buffalo skin. & 0.26 & 0.74 & 0.46 & 0.88 & 0.31 \\
T2, C2 & Black circular membrane must be present in the middle of both the heads and must be slightly raised from the stretched skin surfaces. & 0.20 & 0.67 & 0.53 & 1.00 & 0.41 \\
T2, C3 & Black circular membrane on the smaller end must be slightly smaller than the black circular membrane on the larger end. & 0.20 & 0.60 & 0.60 & 1.00 & 0.50 \\
T2, C4 & The instrument must have longitudinal leather straps lacing along its body connecting the two heads of the instrument under high tension. & 0.32 & 0.88 & 0.39 & 0.93 & 0.14 \\ \lightrule
T3, C1 & The orientation and positioning of the instrument must be horizontal, lying on its length. & 0.36 & 0.40 & 0.88 & 0.89 & 0.88 \\ 
\bottomrule
\end{tabular}
\end{table}

\begin{table}[H]
\small
\centering
\caption{Comparing human to MLLM annotations applying the rubric for a \textbf{Chundan Vallam}. See Section \ref{apdx:human-mllm-agreement} for a complete description of each column. We provide abbreviated descriptions of several criteria in this table to save space; for the complete criteria text, refer to Table \ref{table:systematization-indian}.}
\label{tab:op-agreement-chundan}
\begin{tabular}{l C{4cm} c g c c c c}
\toprule
\textbf{Criteria} & \textbf{Description} &  \textbf{\shortstack{Human\\{\footnotesize(\% Appropriate)}}} & \textbf{\shortstack{MLLM\\{\footnotesize(\% Appropriate)}}} & \shortstack{\textbf{Agreement}\\{\footnotesize Overall}} & \shortstack{\textbf{Agreement}\\{\footnotesize Appropriate}} & \shortstack{\textbf{Agreement}\\{\footnotesize Inappropriate}} \\
\midrule

Final label & {} & 0.00 & 0.17 & 0.83 & N/A & 0.83 \\ \lightrule
T1, C1 & (Abbreviated) It must not resemble other passenger boats. & 0.38 & 0.62 & 0.60 & 0.80 & 0.48 \\
T1, C2 & The boat must be long and narrow. & 0.62 & 0.97 & 0.65 & 1.00 & 0.07 \\ 
T1, C3 & The bow of the boat must be a plain wooden extension without decorative structures. & 0.46 & 0.24 & 0.59 & 0.31 & 0.83 \\
T1, C4 & The stern of the boat must be a straight pointed tip angled slightly upward. & 0.02 & 0.59 & 0.43 & 1.00 & 0.42 \\ \lightrule
T2, C1 & (Abbreviated) Oarsmen must sit in pairs along the length of the boat. & 0.82 & 0.94 & 0.76 & 0.93 & 0.00 \\ 
T2, C2 & (Abbreviated) Each oarsman must use only a single paddle. & 	0.90 & 0.94 & 0.84 & 0.94 & 0.00 \\ 
T2, C3 & (Abbreviated) The paddle must be longer and angled downward toward the water. & 0.94 & 0.94 & 0.88 & 0.94 & 0.00 \\
T2, C4 & (Abbreviated) One person must be standing at the bow or centre position of the boat. & 0.98 & 0.89 & 0.89 & 0.90 & 0.40 \\ \lightrule
T3, C1 & (Abbreviated) The oarsmen must be seated facing the stern. &	0.78 & 0.95 & 0.73 & 0.93 & 0.00 \\
T3, C2 & Oarsmen must wear the same attire, typically a white traditional Kerala mundu without upper garments. & 0.94 & 0.72 & 0.70 & 0.72 & 0.33 \\
\bottomrule
\end{tabular}
\end{table}

\end{document}